\documentclass[sigconf]{acmart}
\AtBeginDocument{%
  }

\copyrightyear{2026}
\acmYear{2026}
\setcopyright{cc}
\setcctype{by}
\acmConference[CIKM '26]{Proceedings of the 35th ACM International Conference on Information and Knowledge Management}{November 07--11, 2026}{Rome, Italy}
\acmBooktitle{Proceedings of the 35th ACM International Conference on Information and Knowledge Management (CIKM '26), November 07--11, 2026, Rome, Italy}
\acmDOI{10.1145/3799682.3840574}
\acmISBN{979-8-4007-2539-5/2026/11}

\usepackage[a-2b]{pdfx} 
\usepackage{fontawesome}
\usepackage{balance}
\usepackage{tcolorbox}
\usepackage{colortbl}
\definecolor{mygray}{gray}{.9}
\definecolor{mypink}{rgb}{.99,.91,.95}
\definecolor{mycyan}{cmyk}{.3,0,0,0}
\usepackage[edges]{forest}
\usepackage{tikz}
\usetikzlibrary{trees}
\tikzset{parent/.style={align=center,text width=0.5cm,rounded corners=3pt},
    child/.style={align=center,text width=1.2cm,rounded corners=3pt},
    grandchild/.style={align=center,text width=1.2cm,rounded corners=3pt}
    }
\usepackage{hhline}
\usepackage{multirow}
\usepackage{tabularx}
\usepackage{makecell}
\usepackage{diagbox}
\usepackage{subfigure}
\usepackage{cleveref}
\usepackage{enumitem}
\usepackage{threeparttable}
\usepackage{multicol}
\usepackage{amsmath} 
\usepackage[T1]{fontenc}
\usepackage{pifont}

\usepackage{algorithm}
\usepackage{algorithmicx}
\usepackage[noend]{algpseudocode}
\setitemize[1]{itemsep=0pt,partopsep=0pt,parsep=0pt,topsep=0pt,leftmargin=10pt}

\begin{document}

\title{\texttt{DICE}: Decentralized Incentive-driven Federated Learning for Efficient \& Scalable Customization of Pretrained Models in Web3}

\author{Hong Lin}
\affiliation{%
  \institution{The State Key Laboratory of Blockchain and Data Security, Zhejiang University}
  \country{Hangzhou High-Tech Zone (Binjiang) Institute of Blockchain\&Data Security}
}
\email{honglin@zju.edu.cn}
\author{Zhongle Xie}
\affiliation{%
  \institution{The State Key Laboratory of Blockchain and Data Security, Zhejiang University}
  \country{Hangzhou High-Tech Zone (Binjiang) Institute of Blockchain\&Data Security}
}
\email{xiezl@zju.edu.cn}
\author[1]{Lidan Shou}
\authornote{Lidan Shou is the corresponding author.}
\affiliation{%
  \institution{The State Key Laboratory of Blockchain and Data Security, Zhejiang University}
  \country{Hangzhou High-Tech Zone (Binjiang) Institute of Blockchain\&Data Security}
}
\email{should@zju.edu.cn}
\author{Ke Chen}
\affiliation{%
  \institution{The State Key Laboratory of Blockchain and Data Security, Zhejiang University}
  \country{Hangzhou High-Tech Zone (Binjiang) Institute of Blockchain\&Data Security}
}
\email{chenk@zju.edu.cn}
\author{Gang Chen}
\affiliation{%
  \institution{The State Key Laboratory of Blockchain and Data Security, Zhejiang University}
  \country{Hangzhou High-Tech Zone (Binjiang) Institute of Blockchain\&Data Security}
}
\email{cg@zju.edu.cn}

\renewcommand{\shortauthors}{Hong Lin, Zhongle Xie, Lidan Shou, Ke Chen, \& Gang Chen}

\begin{abstract}
Customizing Pretrained Models (PMs) for downstream tasks demands task-specific data, which burgeoning Web3 applications provide abundantly. Blockchain-powered federated learning (BCFL) offers a secure and privacy-preserving paradigm to customize PMs among \textit{self-interested} and \textit{untrusted} Web3 participants. Yet, existing BCFL algorithms face severe \textit{efficiency} and \textit{scalability} challenges when applied to large PMs, due to their substantial computational and memory requirements and the \textit{hardware constraints} of many participants. This paper introduces \texttt{DICE}, a novel BCFL algorithm that addresses the efficiency and scalability challenges of customizing large PMs in Web3 via two sets of synergistic techniques: (i) a \textit{dynamic progressive learning} scheme and a \textit{model slice verification} scheme to enable a wider range of honest participants, including those unable to deploy the full model, to engage in customizing PMs, while preserving model utility; (ii) a \textit{tree-structured committee consensus} protocol and an \textit{efficiency-driven incentive} scheme to address the dominant bottleneck that hinders existing BCFL algorithms from scaling to a larger number of data holders for customizing data-hungry large PMs, while maintaining security. Theoretical analysis and extensive experimental results on customizing vision and language PMs demonstrate \texttt{DICE}'s efficiency, effectiveness, and security. 
\end{abstract}



\maketitle

\section{Introduction} \label{sec:intro}

Pretrained AI Models (PMs) and Web3 are converging to create more intelligent, automated, and efficient decentralized systems. 
These PMs require task-specific \textbf{customization} through further pretraining or fine-tuning before serving downstream tasks \cite{schneider2024foundation,bhattacharjee2025overview,yang2023customizing}. 
Meanwhile, burgeoning Web3 applications generate vast amounts of decentralized data~\cite{shen2024artificial, bara2025promptchain, oceanprotocol2020, ZIRCONTech2025, lui2025bittensor}, which offer immense potential to customize PMs. 
For instance, platforms such as MedSynapse \cite{MedSynapse} and Mind Network \cite{WorldAIHealthHub} produce terabytes of patient health data spread across millions of peer-to-peer devices. These data are valuable for customizing pretrained BERT \cite{DBLP:journals/corr/abs-1810-04805} for disease diagnosis and treatment. 
However, aggregating private data on a single platform, whether centralized or decentralized, may violate increasingly stringent data protection regulations \cite{truong2021privacy}. 
Moreover, Web3 participants are \textit{self-interested} and \textit{untrusted}: they are reluctant to contribute resources without monetary incentives and may disrupt customization tasks if malicious behavior is profitable  \cite{10476089, Yang2024ABF,10.14778/3746405.3746429}.

\setlength{\abovecaptionskip}{7.45pt}
\begin{figure}[t]
    \centering
    \includegraphics[width=0.93\columnwidth]{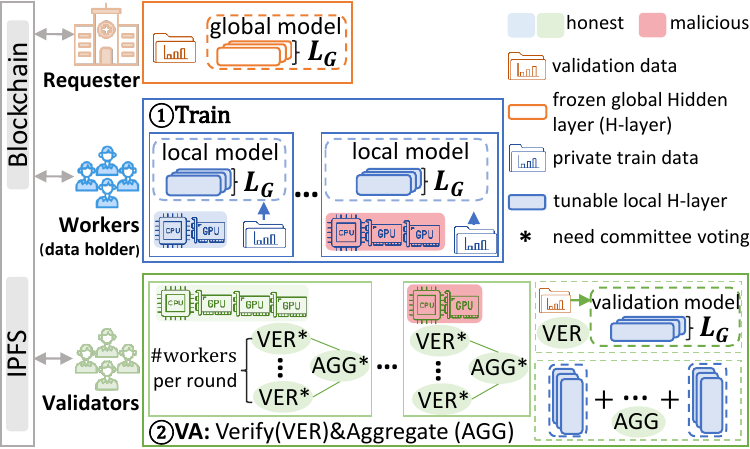} 
    \caption{A general BCFL paradigm for customizing BERT-Large (with $L_G=24$ H-layers, 340 million params), required by a medical institution for disease diagnosis, using health data in MedSynapse. }
    \label{fig:algorithm_overview_Refiner}
\end{figure}
\setlength{\tabcolsep}{4.5pt}
\setlength{\aboverulesep}{0pt}
\setlength{\belowrulesep}{0pt}
\begin{table}[t]
    \centering \small
    \caption{Per-round memory costs of a worker ($MC_w$) and a validator ($MC_v$) in Fig.~\ref{fig:algorithm_overview_Refiner} under typical algorithms \cite{10.14778/3476311.3476313,lin2024refiner}. }\label{tab:evidence_for_challenges}
    \begin{threeparttable}
        \begin{tabular}{c|c|c|c|c||c|c|c|c|c}
        \toprule
                    \textbf{Batch Size} & 8 & 16 & 32 & 64 & \textbf{\# Workers} & 10 & 40 & 70 & 100 \\
        \hline
            \textbf{$MC_{w}$ (GB)} & 11 & 17 & 30 & 59 & \textbf{$MC_{v}$ (GB)} & 18 & 71 & 125 & 170 \\ 
        \bottomrule
        \end{tabular}
        \begin{tablenotes}
        \raggedright  
        \footnotesize
        \item[\textbullet] These statistics are measured on NVIDIA A100 (PyTorch 2.6, CUDA 12.8) using the 2018n2c2 NER dataset \cite{lee2020biobert}, FP32 precision, Adam optimizer, sequence length 512, and batch size 8 unless otherwise stated. 
        \item[\textbullet] $MC_{v}$ is reported as the total GPU memory usage across devices participating in the computation and corresponds to concurrent verification of all worker updates, without memory-swapping techniques that can increase verification latency \cite{zhang2025hmi}. 
        \end{tablenotes}
    \end{threeparttable}
\end{table}

Blockchain-powered Federated Learning (BCFL)~\cite{9293091,9833437,10.1145/3524104,nguyen2021federated,zhu2023blockchain,app14209459,li2022blockchain} has emerged as a promising paradigm to support secure and privacy-preserving customization of a global model shared among Web3 participants. 
As shown in Fig.~\ref{fig:algorithm_overview_Refiner}, a typical BCFL-based customization pipeline involves three participant roles: (i) a \textit{model requester}, e.g., a medical institution, who needs a customized AI model and pays rewards to engage the other participants in customization; (ii) \textit{workers}, e.g., MedSynapse's health data holders, who train models using private data and submit \textit{updates} (model parameters updated in training); (iii) \textit{validators}, who verify and aggregate those updates. 
BCFL algorithms generally coordinate participants via a blockchain network and proceed in \textit{iterative rounds of training and verification \& aggregation (VA) steps}. 
This paradigm relies on two critical components: (i) \textit{a malicious participant handling scheme} \cite{lin2024refiner,10.14778/3476311.3476313} that requires validators to verify worker updates and execute a consensus protocol to ensure security, and (ii) \textit{an incentive scheme} \cite{9369019} that encourages self-interested entities to participate and behave honestly, typically via monetary rewards, e.g., cryptocurrency \cite{10.1145/3578581}.


However, existing BCFL algorithms are primarily designed for small models ($\leq$ tens of millions of parameters) \cite{lin2024refiner, bian2023blockchain,manoj2025blockchain}, facing critical challenges in efficiency and scalability when applied to large PMs (typically $\geq$ 100 million parameters):

\textbf{C1. Per-participant computational and memory overhead.}
Existing BCFL algorithms require every worker to perform training on the full model, and require every validator to perform multi-update verification \cite{lin2024refiner, bian2023blockchain,manoj2025blockchain,9293091,10930720}. 
As shown in Table~\ref{tab:evidence_for_challenges}, even for BERT-Large, an early-stage large pretrained model, these requirements already exceed the memory capacity of many commodity GPUs. 
Consequently, a large fraction of honest devices are excluded from effective participation.

\textbf{C2. System-level scalability bottleneck.} Since each validator in existing BCFL systems is required to verify all worker updates in every round \cite{10.1145/3524104, li2022blockchain, app14209459}, the per-validator overhead in those systems grows linearly with the worker scale (see Table~\ref{tab:evidence_for_challenges}). 
This requirement creates a critical bottleneck that can increase consensus latency and restrict participation as the worker population grows, preventing those systems from scaling to more workers to improve the performance of customizing large PMs, especially those data-hungry PMs.  


\textbf{C3. Limitations of existing efficiency techniques.}
Prior methods for efficient PM customization, such as LoRA \cite{hu2022lora}, BitFit \cite{ben-zaken-etal-2022-bitfit}, and other parameter-efficient tuning (PEFT) techniques \cite{zhang2023fedpetuning}, can reduce worker-side costs but fail to address validator-side overhead. 
While model compression techniques such as distillation \cite{hinton2015distilling}, pruning \cite{liang2021pruning}, and quantization \cite{han2015deep} can reduce validator-side costs, they might sacrifice model quality \cite{Wang2023FedBFPTAE,zhang2025hmi}. 
Critically, none of these techniques overcomes the above scalability bottleneck. 

To address these challenges, this paper proposes \texttt{DICE}, a novel \texttt{D}ecentralized \texttt{I}n\texttt{CE}ntive-driven BCFL algorithm for efficient, scalable, and secure customization of PMs in Web3. 
\texttt{DICE} supports lightweight training and verification of a large PM  by two techniques built upon insights from existing efficient customization solutions: (i) a \textit{Dynamic Progressive Learning (DPL)} scheme to reduce worker-side hardware resource requirements while maximizing the adaptability of PMs; and (ii) a \textit{Model Slice Verification (MSV)} scheme to enable honest yet hardware-constrained participants to engage as validators, increasing the pool of potential validators and thus making malicious behavior less likely to succeed. 
More importantly, \texttt{DICE} breaks the linear dependency between per-validator workload and worker scale by two core innovations: (i) a \textit{Tree-structured Committee Consensus (TCC)} protocol that decomposes the monolithic VA task into tree-structured sub-tasks and employs selective committee voting, thereby reducing the per-validator mandatory workload while maintaining security; 
and (ii) an \textit{Efficiency-Driven Incentive (EDI)} scheme that complements TCC by penalizing conflict generators, ensuring rational validators do not undermine TCC's efficiency gains. 
Jointly, TCC and EDI resolve the scalability bottleneck that prior BCFL systems cannot address even when integrated with efficient model customization techniques, while preserving robust security guarantees. 

The contributions of this paper are outlined as follows:

\textbullet{} \textbf{Systematic Problem Analysis and Novel Algorithm.} 
We systematically analyze the efficiency and scalability challenges of customizing large PMs within existing BCFL systems, and propose \texttt{DICE}, a novel algorithm that significantly lowers the entry barriers of customization tasks, enabling a wider range of honest participants to contribute effectively. 
(\S\ref{sec:algorithm})

\textbullet{} \textbf{Novel Technical Designs.} We introduce two sets of novel techniques in \texttt{DICE}: 
DPL\&MSV to support lightweight training and verification of large PMs while preserving model utility; TCC\&EDI to address the dominant bottleneck that hinders prior BCFL systems from scaling to a larger number of workers without compromising security. 
(\S\ref{ssec:slice_training}--\S\ref{ssec:consensus}) 

\textbullet{} \textbf{Theoretical Guarantees.} We provide a rigorous theoretical analysis demonstrating that \texttt{DICE} lowers the participation barrier for both workers and validators and allows the per-validator workload required for consensus to scale \textit{sub-linearly} with the worker scale, while bounding the success probabilities of malicious behaviors. (\S\ref{sec:theoretical_analysis})

\textbullet{} \textbf{Experimental Validation.} We conduct extensive experiments on customizing vision and language PMs to show that \texttt{DICE}  significantly reduces both worker-side and validator-side costs compared to baselines, delivers competitive final model performance, and maintains robust resistance against malicious actors. 
Additional experiments demonstrate that \texttt{DICE} is compatible with existing efficient customization techniques, accommodating diverse device constraints and model performance requirements without compromising security. 
(\S\ref{sec:experiment})

\section{Preliminaries}
\label{sec:preliminaries}
We begin by presenting the system model, threat model, and foundational techniques from prior BCFL work on which \texttt{DICE} builds. 

\subsection{System Model} \label{ssec:participants}

\subsubsection{Participants} Similar to existing BCFL designs \cite{lin2024refiner,10930720}, \texttt{DICE} involves three types of participants: model \textit{requesters}, \textit{workers}, and \textit{validators}, as introduced in \S\ref{sec:intro}.
Requesters need abundant data held privately by workers to customize PMs \cite{Bommasani2021OnTO} for superior performance on downstream tasks, and possess powerful hardware to perform computation on large PMs. However, workers and validators may lack the hardware capacity to deploy a full model.

\subsubsection{Coordination and Communication}  \label{sssec:cc_model}
\texttt{DICE} adopts a peer-to-peer \textit{blockchain network} to coordinate participants. 
To initialize a customization task, the requester creates a smart contract that records task specifications (e.g., global model, validation dataset, algorithm hyperparameters) and manages cryptocurrency to reward honest participants who strictly follow the algorithm's logic \cite{tian2021blockchain,ur2021trustfed,10.14778/3476311.3476313,10930720}. 
Next, workers and validators join the task by registering their unique identifiers, usually their network addresses, on this contract. 
During task runtime, all participants communicate with the smart contract to exchange data. 
For cost-efficient data exchange, we follow previous studies \cite{lin2024refiner,zhang2022study,10930720,chakraborty2022proof} to integrate a \textit{secure messaging scheme}. This scheme stores large-scale data that are too expensive to store on the blockchain (e.g., model parameters) in an additional storage system such as IPFS, persists data hashes and file handles in smart contracts, and ensures data availability and integrity by excluding participants from reward distribution if their file handles are unreachable or if their hashes, recomputed upon file retrieval, do not match the on-chain records. 
All messages are transmitted over authenticated channels using digital signatures, and messages sent repeatedly by honest participants are eventually delivered, though delays may be unbounded. 

\subsection{Threat Model} \label{ssec:threat_model}
Threats to BCFL primarily come from two aspects: the coordination and communication infrastructures and algorithm participants \cite{hallaji2024decentralized}. 

\subsubsection{Trust Assumptions} 
Following existing work \cite{lin2024refiner,10930720,ma2024towards,tbcim2025,yang2024blockchain,purefed2024}, we treat the blockchain network and IPFS as reliable underlying infrastructures and consider algorithm participants as \textit{untrusted}, \textit{self-interested}, yet \textit{rational} players. 
Rational requesters have no incentive to deviate from the algorithm, as that would undermine their goal of customizing high-quality models and engaging other participants. 
We thus focus on threats from workers and validators. 

\subsubsection{Adversary Capacities}
A malicious worker or validator may deviate from the algorithm in three forms: 
(i) \textit{Poisoning}: computing and submitting low-quality updates or corrupted VA results to degrade model performance \cite{desai2021blockfla,kalapaaking2023blockchain,lin2024refiner,chang2021blockchain,ren2024bpfl,mosquedagonzalez:hal-05002718}; 
(ii) \textit{Fraud} (free-riding): copying and submitting honest participants' updates or VA results to claim undeserved rewards at a total computational cost substantially lower than that of executing the prescribed computation, potentially with negligible impact on model performance \cite{Dai2025Byzantine,lin2024refiner, Yang2024ABF,10402107,10476089,10643358,yu2023ironforge}; 
(iii) \textit{No response}: failing to submit results due to network issues, crashes, etc., which may stall the algorithm \cite{lin2024refiner}. 
Malicious participants can launch attacks independently or collude, and may further amplify their impact via blockchain-layer malicious behaviors such as Sybil attacks \cite{10.1007/3-540-45748-8_24} (controlling multiple fake identities) or DDoS attacks \cite{hallaji2024decentralized} (disrupting message delivery).
Privacy inference attacks and their mitigation (e.g., via differential privacy) \cite{hallaji2024decentralized} are orthogonal to the algorithm-deviation threats we consider and outside the scope of this study. %

\subsection{Foundational Techniques Against Threats}\label{ssec:foundational_techniques}
\texttt{DICE} adopts a malicious participant handling scheme architecturally similar to existing BCFL systems \cite{lin2024refiner,10930720,10.14778/3476311.3476313,9293091}. 
This scheme relies on validators to execute a \textit{worker verification} step to detect active malicious workers and a \textit{cross-validator consensus} protocol to prevent active malicious validators.  
Consensus results produced by honest validators ensure that worker updates are verified and aggregated according to the prescribed rules. 
To handle non-responsive participants, this scheme integrates a \textit{timeout} mechanism. 
Below, we review these base techniques and illustrate how \texttt{DICE} adopts or innovates beyond them.

\subsubsection{Handling Malicious Workers}
Worker verification aims to detect and reject invalid updates from malicious workers. 
Existing approaches fall into two categories: scoring-based (e.g., empirical thresholds on validation performance) and cryptography-based (e.g., zero-knowledge proofs) \cite{app14209459,hallaji2024decentralized}. 
\texttt{DICE}'s MSV scheme builds upon the former category (since the latter cannot directly assess model quality) while remaining compatible with the latter. 
Unlike existing scoring-based schemes that require full model verification, MSV is specifically designed to enable validators who cannot deploy full models to engage. This increases the total resources that can be leveraged for malicious behavior prevention and raises the difficulty of successful attacks.

\subsubsection{Handling Malicious Validators} 
Consensus protocols are the core component for defending against malicious validators. 
These protocols generally fall into three paradigms \cite{zhu2023blockchain}: proof-based \cite{Nakamoto2008BitcoinAP, saleh2021blockchain}, committee-based \cite{li2020scalable, aublin2013rbft}, and hybrid. 
\texttt{DICE}'s TCC protocol builds upon the hybrid paradigm that employs \textit{economic staking} as proof of eligibility to participate and \textit{committee voting} to determine the valid results  \cite{lin2024refiner,10930720}. This paradigm offers superior resource efficiency over proof-based alternatives, and unlike committee-based paradigms deployed in permissioned environments, enables open participation in customization tasks for anyone willing to contribute data or hardware resources. 
In contrast to existing protocols where each validator must process all worker updates, TCC and associated EDI are designed to break the linear dependency between per-validator workload and worker scale, allowing \texttt{DICE} to scale to more workers for better customization performance.

Standard consensus protocols may lack inherent mechanisms that verify the authorship of the received results to prevent validator fraud attacks. 
To close this gap, a common solution \cite{lu2018enabling,lin2024refiner,toyoda2020blockchain} is to integrate a \textit{cryptographic commitment scheme} into the consensus process. 
This scheme requires validators to submit local computation results in two phases: \textit{committing} and \textit{revealing}. When committing, validators submit \textit{fingerprints} and \textit{commitments} generated by hashing raw results. Only after a consensus fingerprint is determined will validators reveal the complete raw results. 
\texttt{DICE} adopts this scheme with a slight modification to accelerate TCC. 

\subsubsection{Handling Non-Responsive Participants} \label{sssec:handling_failed_participant_preliminary}
\texttt{DICE} inherits a timeout scheme from prior work \cite{lin2024refiner} to prevent deadlock caused by non-responsive participants. 
This scheme sets a timeout for each stalling-prone task state, dynamically adjusting it based on the moving average of each state's actual execution time, and assumes that at least one honest participant who has access to the task's smart contract remains active in each BCFL task. 
When any state exceeds its timeout, any active participant can invoke the smart contract to terminate the task and let the model requester decide whether to restart the task or just acquire the global model customized so far and reclaim the remaining funds. 
The scheme is fully compatible with \texttt{DICE}'s novel designs and thus is directly incorporated.

\section{The \texttt{DICE} Algorithm}
\label{sec:algorithm}

This section presents our \texttt{DICE} algorithm. 
\S\ref{ssec:overview} first provides a high-level overview of \texttt{DICE}'s workflow. 
Then, \S\ref{ssec:slice_training}--\S\ref{ssec:consensus} introduce two sets of techniques that \texttt{DICE} utilizes to improve efficiency and scalability while maintaining effectiveness and security of model customization in Web3. 
Table~\ref{tab:notation} summarizes the main notations used in this paper.

\setlength{\tabcolsep}{2.5pt}
\begin{table}[t]\small
  \centering
  \resizebox{\columnwidth}{!}{
  \begin{threeparttable}
  \caption{Notation Summary.} 
  \label{tab:notation}
    \begin{tabular}{ll}
    \toprule
    \textbf{Symbol} & \textbf{Definition} \\
    \hline
    $r$, $R$, $\eta$ & global round index; total number of rounds; learning rate \\
    $w$, $v$ & worker's and validator's addresses on the blockchain network \tnote{$\spadesuit$} \\
    $\mathbb{W}_{all}$, $\mathbb{V}_{all}$ & the complete sets of eligible workers and eligible validators \\
    $\mathbb{AW}^r$ & workers that have submitted updates in round $r$ \\
    $\mathbb{W}^r$ & scheduled workers; replaced by $\mathbb{AW}^r$ when training completes \\
    $\mathbb{V}^r$ & validators that submit VA results in round $r$ \\
    $\theta^r$, $\theta^r_w$ & global model parameters and worker $w$'s update in round $r$ \\ 
    $\mathbb{VAR}_v$ & the set of results produced by validator $v$ for a VA sub-task\\
    $L_G$, $L_E$ &  H-layer counts in the global model and the emulator\\
    $L_w$, $L_v$ & deployable H-layer counts on worker $w$ and validator $v$ \\
    $\mathbb{M}_w^r$ & model slice selected to be trained on worker $w$ in round $r$\\
    $\mathbb{LUS}_j$ & the $j$-th sub-slice of $\bigcup_{w\in\mathbb{W}^r}\mathbb{M}_w^r$ \\
    $N_{lus}$ & number of slices $\mathbb{LUS}_{j}$ \\
    $\theta^{r-1}_{\mathbb{LUS}_{j}}$ & \makecell[l]{global model parameters corresponding to $\mathbb{LUS}_{j}$ \\at the beginning of round $r$} \\ 
    $\theta^{r-1}_{\mathbb{LUS}_{j}}[l]$ & \makecell[l]{global model parameters corresponding to layer $l$ in $\mathbb{LUS}_{j}$ \\at the beginning of round $r$} \\
    $\theta^r_{*,\mathbb{LUS}_{j}}$ &  \makecell[l]{parameters corresponding to $\mathbb{LUS}_{j}$ from any specific worker \\or set of workers} \\
    $\theta^r_{*,\mathbb{LUS}_{j}}[l]$ & \makecell[l]{parameters corresponding to layer $l$ in $\mathbb{LUS}_{j}$ \\from any specific worker or set of workers} \\
    $N_c$, $N_t$ & integer thresholds used in the TCC protocol \\
    $N_{v,t}$ & total number of distinct VA sub-tasks completed by $v$ \\
    $N_{v,t}^{correct}$ & number of distinct VA sub-tasks correctly completed by $v$ \\
    $N_{v,t}^{pre}$ & \makecell[l]{number of distinct VA sub-tasks completed by $v$ and confirmed on chain\\when the first controversial sub-task appears} \\
    $N_{v,t}^{post}$ & \makecell[l]{number of distinct VA sub-tasks correctly completed by $v$ \\ after conflict appears} \\
    $E_{total}$ & total amount of security deposit staked by all validators  \\
    $E_c$ & expected amount of stake selected for committee formation \\ 
    $\epsilon_j$, $\epsilon_c$ & model performance threshold; voting threshold  \\ %
    $WR$, $VR$ & rewards for workers and validators in each round \\ 
    $|\cdot|$, $\cdot\parallel\cdot$  &  element count in a set; hash input concatenation \\
    $\Omega(\cdot)$, $\mathcal{O}(\cdot)$  & asymptotic lower bound; asymptotic upper bound \\
    $\Theta(\cdot)$ & asymptotically tight bound\\
    $\beta_{\mathrm{hist}}$, $\delta$ & history-window length; numerical stabilizer \\
    \bottomrule
    \end{tabular}
    \begin{tablenotes}
    \raggedright  
    \footnotesize
    \item[$\spadesuit$]Each worker/validator is represented by its unique address.
    \end{tablenotes}
  \end{threeparttable}
  }
\end{table}%
\begin{figure}[t]
    \centering
    \includegraphics[width=\columnwidth]{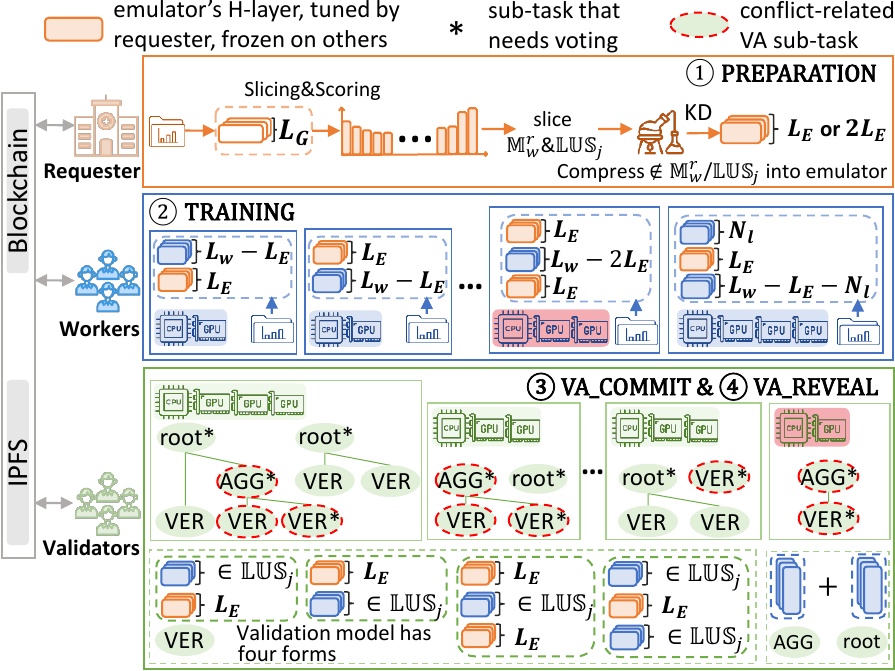} 
    \caption{The overview of \texttt{DICE}.}
    \label{fig:algorithm_overview_FMRefiner}
\end{figure}

\subsection{Overview} \label{ssec:overview}
Fig.~\ref{fig:algorithm_overview_FMRefiner} presents an overview of \texttt{DICE} for customizing the PM, which comprises an embedding layer, a sequence of intermediate hidden layers (H-layers), and an output layer. 
\texttt{DICE} executes every customization task in iterative rounds. 
Each round contains a \textit{training} step, where workers produce updates for a global model, and a \textit{VA} step, where validators verify worker updates and aggregate qualified updates to produce a new global model. 
In the training step, \texttt{DICE} uses a DPL scheme to enable hardware-constrained workers, especially those having abundant task-specific data, to engage. This scheme divides the training step into two states: \texttt{PREPARATION} and \texttt{TRAINING}. 
In the VA step, \texttt{DICE} adopts MSV, TCC, and EDI to reduce validator-side costs while maintaining security. 
TCC, integrated with a cryptographic commitment scheme, structures the VA step into \texttt{VA\_COMMIT} and \texttt{VA\_REVEAL} states.

Algorithms~\ref{alg:offchain}--\ref{alg:onchain} present core procedures of \texttt{DICE} in four states. 
In \texttt{PREPARATION}, the start state of each round, the requester prepares necessary materials for the subsequent lightweight training and verification based on our DPL and MSV schemes (Alg.~\ref{alg:offchain} lines~\ref{alg_line:requester_off_start}--\ref{alg_line:requester_off_end}, Alg.~\ref{alg:onchain} lines~\ref{alg_line:requester_on_start}--\ref{alg_line:requester_on_end}). 
\texttt{RequesterOnchainProcedure}() changes the task state to \texttt{TRAINING} upon receiving those materials. 
Then, workers produce updates for the global model cost-effectively (Alg.~\ref{alg:offchain} lines~\ref{alg_line:worker_off_start}--\ref{alg_line:worker_off_end}, Alg.~\ref{alg:onchain} lines~\ref{alg_line:worker_on_start}--\ref{alg_line:worker_on_end}). 
After a certain number of workers have submitted updates, \texttt{WorkerOnchainProcedure}() creates a VA task forest based on our TCC scheme and changes the task state to \texttt{VA\_COMMIT}. 
During \texttt{VA\_COMMIT}, validators produce and submit fingerprints and commitments of VA results (Alg.~\ref{alg:offchain} lines~\ref{alg_line:validator_offchain_start}--\ref{alg_line:offchain_commit_end}). 
\texttt{ValidatorOnchainProcedure}() checks whether validators reach consensus on the fingerprints of all VA sub-tasks (Alg.~\ref{alg:onchain} lines~\ref{alg_line:consensus_check_start}--\ref{alg_line:consensus_check_end}), and if so, transitions the task state to \texttt{VA\_REVEAL}. 
Then, validators reveal the complete VA results, which are checked by the on-chain procedure (Alg.~\ref{alg:offchain} lines~\ref{alg_line:offchain_reveal_start}--\ref{alg_line:offchain_reveal_end}, Alg.~\ref{alg:onchain} lines~\ref{alg_line:reveal_res_start}--\ref{alg_line:reveal_res_end}).
Finally, \texttt{EDI()} rewards honest participants and penalizes malicious validators (Alg.~\ref{alg:onchain} line~\ref{alg_line:onchain_reveal_end}); then the customization task proceeds to the next round (line~\ref{alg_line:onchain_reveal_end}).

\subsection{Lightweight Training and Verification powered by DPL and MSV} \label{ssec:slice_training}
This section introduces lightweight training and verification procedures powered by our DPL and MSV schemes. 
The former aims to enable constrained workers to produce updates of a large PM and customize all necessary layers of the PM on worker data to maximize the PM's adaptation ability. 
The latter aims to enable a wider range of validators, including those more hardware-constrained than workers, to verify updates, increasing the total hardware resources that can be leveraged to augment security. 
These objectives are achieved through the \textit{Slice Selection} and \textit{Large-to-Compact Model Derivation} procedures executed by the model requester. 
\S\ref{sssec:slice_selection}--\S\ref{sssec:runtime_mechanism} first illustrate how the requester executes these two procedures. 
Then, \S\ref{sssec:slice_eval} describes the lightweight training and verification process on workers and validators. 
Finally, \S\ref{sssec:discussion_on_DPL_MSV} discusses the design rationale behind our proposed schemes.

\begin{algorithm}[t] 
\small
\caption{\texttt{DICE}'s off-chain procedures, i.e., computations executed locally by each participant.} 
\label{alg:offchain} 
\begin{algorithmic}[1] 
        \Function{RequesterOffchainProcedure}{ }  \label{alg_line:requester_off_start} \Comment{\S\ref{ssec:slice_training}}
            \State $step, r, r_g \leftarrow$\texttt{ReadBlockchain}()
            \If{$step \neq $\texttt{PREPARATION} \textbf{or} $r \leq r_g$} 
                \Return
            \EndIf 
            \State $\{\mathbb{M}_w^r | w\in \mathbb{W}^r\}, \{\mathbb{LUS}_j | j\in [0,N_{lus})\} \leftarrow$SliceSelection()  \label{alg_line:select_slice_start}
            \State $\mathbb{M} \leftarrow \{\mathbb{M}_w^r | w\in \mathbb{W}^r\}; \mathbb{LUS} \leftarrow \{\mathbb{LUS}_j | j\in [0,N_{lus})\}$\label{alg_line:select_slice_end}
            \State $\{\theta^r_{\notin \mathbb{M}_w^r} | w\in \mathbb{W}^r\}$, $\{\theta^r_{\notin \mathbb{LUS}_j} | j\in [0,N_{lus})\} \leftarrow$ModelDerivation() \label{alg_line:generate_emulator}
            \State $\mathbb{EM}_{w}, \mathbb{EM}_{v}, \mathbb{GM} \leftarrow$RequesterIPFSStore() \label{alg_line:requester_submit_start}
            \State \texttt{RequesterOnchainProcedure}($\mathbb{GM}, \mathbb{EM}_{w}, \mathbb{EM}_{v}, \mathbb{M}, \mathbb{LUS}$) \label{alg_line:requester_off_end}
        \EndFunction
        \Function{WorkerOffchainProcedure}{ }  \Comment{\S\ref{ssec:slice_training}} \label{alg_line:worker_off_start}
            \State $step, r, r_w \leftarrow$\texttt{ReadBlockchain}()
            \If{$step \neq $\texttt{TRAINING} \textbf{or} $r \leq r_w$} 
                \Return
            \EndIf
            \State $\{\theta^{r}_{w,\mathbb{LUS}_j}|j\in [0,N_{lus})\} \leftarrow$ DPLtrain() \label{alg_line:DPL} 
            \State $\mathbb{LU} \leftarrow$WorkerIPFSStore(); \texttt{WorkerOnchainProcedure}($\mathbb{LU}$) \label{alg_line:worker_off_end}
        \EndFunction
        \Function{ValidatorOffchainProcedure}{ } \Comment{\S\ref{ssec:slice_training}--\ref{ssec:consensus}} \label{alg_line:validator_offchain_start}
            \State $step \leftarrow$\texttt{ReadBlockchain}()
            \If{$step \notin $\{\texttt{VA\_COMMIT},  \texttt{VA\_REVEAL}\} }
                \Return
            \EndIf
            \State $\mathbb{T}_{VA}, \mathbb{T}_{v} \leftarrow$\texttt{ReadBlockchain}();  \Comment{$\mathbb{T}_{v}$ stores sub-tasks done by $v$} \label{alg_line:get_VA_task_forest}
            \If{$step =$\texttt{VA\_COMMIT} } \label{alg_line:off_commit_start}
                \While{$step =$\texttt{VA\_COMMIT} \textbf{and} $|\mathbb{T}_{v}|< |\mathbb{T}_{VA}|$} \label{alg_line:MSV_start}
                    \State $\mathbb{RES}_v^{commit}\leftarrow$MSVorAGG() \label{alg_line:MSV}
                    \State  \texttt{ValidatorOnchainProcedure}($\mathbb{RES}_v^{commit}$)
                    \State $step, \mathbb{T}_{v} \leftarrow$\texttt{ReadBlockchain}() \label{alg_line:MSV_end}
                \EndWhile \label{alg_line:offchain_commit_end}
            \ElsIf{$|\mathbb{T}_{v}| > 0$}  
                \State $\mathbb{RES}_v^{reveal}\leftarrow$RevealResPacking() \label{alg_line:offchain_reveal_start}
                \State \texttt{ValidatorOnchainProcedure}($\mathbb{RES}_v^{reveal}$)   \label{alg_line:offchain_reveal_end}
            \EndIf
        \EndFunction
\end{algorithmic}
\end{algorithm}

\begin{algorithm}[t] 
\small
\caption{\texttt{DICE}'s on-chain procedures, programmed in the smart contract and executed on the blockchain network.} 
\label{alg:onchain} 
\begin{algorithmic}[1]
\setlength{\baselineskip}{7.7pt}  
        \Function{\texttt{RequesterOnchainProcedure}}{$\mathbb{GM}, \mathbb{EM}_{w}, \mathbb{EM}_{v}, \mathbb{M}, \mathbb{LUS}$} \label{alg_line:requester_on_start}
            \If{$step \neq $\texttt{PREPARATION} \textbf{or} $r_g \geq r$}
                \Return
            \EndIf
            \State  \texttt{RecordRequesterRes}($\mathbb{GM}, \mathbb{EM}_{w}, \mathbb{EM}_{v}, \mathbb{M}, \mathbb{LUS}$)
            \State $r_g \leftarrow r $; $step \leftarrow \texttt{TRAINING}$\label{alg_line:requester_on_end}
        \EndFunction
        \Function{\texttt{WorkerOnchainProcedure}}{$\mathbb{LU}$} \label{alg_line:worker_on_start}
            \If{$step \neq $\texttt{TRAINING} \textbf{or} $r_w \geq r$}
                \Return
            \EndIf
            \State $r_w \leftarrow r $; $\mathbb{AW}^r\leftarrow$\texttt{RecordWorkerRes}($\mathbb{LU}$); 
            \If{$\mathbb{AW}^r=\mathbb{W}^r$} \Comment{sufficient active workers}
                \State $\mathbb{W}^r \leftarrow \mathbb{AW}^r$; \texttt{CreateVATaskTree}(); $step \leftarrow$\texttt{VA\_COMMIT} \Comment{\S\ref{ssec:consensus}} \label{alg_line:worker_on_end}
            \EndIf
        \EndFunction
        \Function{\texttt{ValidatorOnchainProcedure}}{$\mathbb{RES}_v$} \label{alg_line:validator_onchain_start}
            \If{$step \notin $\{\texttt{VA\_COMMIT},  \texttt{VA\_REVEAL}\} }
                \Return
            \EndIf
            \If{$step=$\texttt{VA\_COMMIT}} 
                \texttt{OnchainCommit}($\mathbb{RES}_v$)\label{alg_line:onchain_commit}
            \Else 
                \ \texttt{OnchainReveal}($\mathbb{RES}_v$) \label{alg_line:onchain_reveal}
            \EndIf
        \EndFunction
        \Function{\texttt{OnchainCommit}}{$\mathbb{RES}_v$} \Comment{\S\ref{ssec:consensus}} \label{alg_line:onchain_commit_start}
            \State $tid $, $\mathbb{VAR}_v^{\text{commit}}$, $FP_v$, $CC_v$, $\pi_v$, $hash_v$ $\leftarrow \mathbb{RES}_v$\label{alg_line:commit_res}
            \If{not \texttt{Record}($tid $, $\mathbb{VAR}_v^{\text{commit}}$, $FP_v$, $CC_v$, $\pi_v$, $hash_v$)} \Return \EndIf \label{alg_line:consensus_check_start}
            \If{\texttt{NeedCommitteeVoting}($tid $)} \label{alg_line:voting}
                \State \texttt{Voting}($tid $, $\mathbb{VAR}_v^{\text{commit}}$, $FP_v$, $CC_v$, $\pi_v$, $hash_v$)  \label{alg_line:voting_end}
            \Else
                \State \texttt{CheckConflict}($tid $, $\mathbb{VAR}_v^{\text{commit}}$, $FP_v$, $CC_v$, $\pi_v$, $hash_v$) \label{alg_line:check_conflict}
            \EndIf
            \If{$|\mathbb{T}_{VA}|-|\mathbb{CR}|\neq0$}  \Comment{check global consensus} \label{alg_line:consensus_check1_start}
                \If{$|\{v|v\in\mathbb{V}_{all}, |\mathbb{T}_{v}\backslash\mathbb{CR}|\geq \lceil \frac{|\mathbb{T}_{VA}|-|\mathbb{CR}|}{|\mathbb{T}_{VA}|}N_t \rceil\}|<N_c$} 
                    \State \Return 
                \EndIf  \label{alg_line:consensus_check1_end} 
            \EndIf
            \If{($|\mathbb{T}_{VA}|=|\mathbb{CR}|$ \textbf{or} \texttt{SupplementaryCheck}()) \textbf{and} \\ \qquad $|\mathbb{CONSENSUS}^r|\geq |\mathbb{T}_{VA}|$}  \Comment{reach consensus} \label{alg_line:consensus_check_end}
               \State $step \leftarrow$\texttt{VA\_REVEAL}   
            \EndIf \label{alg_line:onchain_commit_end}
        \EndFunction
        \Function{\texttt{OnchainReveal}}{$\mathbb{RES}_v$} \Comment{\S\ref{ssec:consensus}} \label{alg_line:onchain_reveal_start}
            \State $\{<tid, \mathbb{VAR}_v^{\text{reveal}}>|tid \in \mathbb{T}_v\}\leftarrow \mathbb{RES}_v$\label{alg_line:reveal_res_start}
            \State \texttt{QualificationCheck}($\{<tid, \mathbb{VAR}_v^{\text{reveal}}>|tid \in \mathbb{T}_v\}$) \label{alg_line:reveal_res_end}
            \If{all($\{|\mathbb{T}_v|=0|v \in \mathbb{V}_{all}\}$)} 
                \State \texttt{EDI}(); \texttt{OnchainPreparation}(); $step \leftarrow$\texttt{PREPARATION} \label{alg_line:onchain_reveal_end}
            \EndIf 
        \EndFunction
\end{algorithmic}
\end{algorithm}

\subsubsection{Slice Selection}\label{sssec:slice_selection} In this procedure, the model requester selects a small slice of the global model for training on each worker based on the slice's importance score, and executes further-slicing for validators when necessary  (Alg.~\ref{alg:offchain} line~\ref{alg_line:select_slice_start}). 
For each worker $w$ capable of deploying the full global model, the
requester selects all H-layers and the output layer to form the slice to be trained on this worker, denoted by $\mathbb{M}^r_{w}$.

For each worker $w$ that cannot deploy the full global model, the requester divides the global model into a set of candidate slices $\mathbb{M}_{w,i}$ defined in Eq.~\ref{eq:slices} using this worker's hardware capacity $L_w$ reported at
registration. Each slice contains a small number of H-layers. 
Empty $\mathbb{M}_{w,i}$ are excluded from the subsequent slice selection process. 
\begin{equation}\label{eq:slices} \small
    \hspace{-10pt} \mathbb{M}_{w,i} =
    \begin{cases}
    \{1,\cdots, L_{w}-L_{E}\} & \text{if $i=0$ (shallow)}  \\
    \{1+L_{E}+i, \cdots, L_{w}-L_{E}+i\}& \text{if $i\in[1, N'_{\text{slice}})$ (middle)} \\
     \{L_G-(L_{w}-L_{E}-1),\cdots, L_G\} & \text{if $i=N'_{\text{slice}}$ (top)}  \\
    \{1,\cdots, L_{w}-L_{E}-(i-N'_{\text{slice}}), &  \\
    \ \ L_G-(i-N'_{\text{slice}}-1),\cdots,L_G\} & \text{if $i\in(N'_{\text{slice}},N_{\text{slice}}]$ (combined)}  
    \end{cases}
\end{equation}
Here, the numbers in $\mathbb{M}_{w,i}$ denote the indices of global H-layers. $L_G$ is the total number of H-layers in the global model. $L_E$ is the number of H-layers in an emulator. $L_w$ is the number of H-layers deployable on worker $w$, $N'_{\text{slice}}=L_G-L_w, N_{\text{slice}} = L_G-L_E-1$.

Then, the requester computes each slice's importance score denoted by $Score_{w,i}$. 
Our DPL scheme defines $Score_{w,i}$ by Eq.~\ref{eq:importance_score}, where $s^r_{\mathbb{M}_{w,i}}$ is the performance of the global model after a slice $\mathbb{M}_{w,i}$ is tuned, $div^r_{\mathbb{M}_{w,i}}=1-cos\_sim(\theta^{r-1}_{\mathbb{M}_{w,i}},\hat{\theta}^{r}_{\mathbb{M}_{w,i}})$ is the divergence of the slice parameters before and after tuning, 
$max_s=\max(\{s^r_{\mathbb{M}_{w,i}}|i \in [0,N_{\text{slice}}],\ \mathbb{M}_{w,i}\neq\emptyset\})$, $max_d=\max(\{div^r_{\mathbb{M}_{w,i}}|i\in[0,N_{\text{slice}}],\ \mathbb{M}_{w,i}\neq\emptyset\})$, and $\delta=10^{-12}$ prevents division by zero. 
Each candidate slice is evaluated independently from the same round-start global model, using the validation data. Only its selected H-layers and the output layer are tuned; all other parameters remain frozen. 
\begin{equation} \label{eq:importance_score}
    Score_{w,i} \leftarrow  \frac{s^r_{\mathbb{M}_{w,i}}}{max_s+\delta} + \frac{div^r_{\mathbb{M}_{w,i}}}{max_d+\delta}
\end{equation} 
This design prioritizes training slices that yield greater performance improvement and exhibit more substantial parameter updates. 
The model requester selects $\mathbb{M}_{w,i}$ with the highest score and combines it with the output layer to form the final slice $\mathbb{M}^r_{w}$ to be trained on worker $w$. 

\begin{figure}[t]
    \centering
    \includegraphics[width=0.95\columnwidth,height=2.7cm]{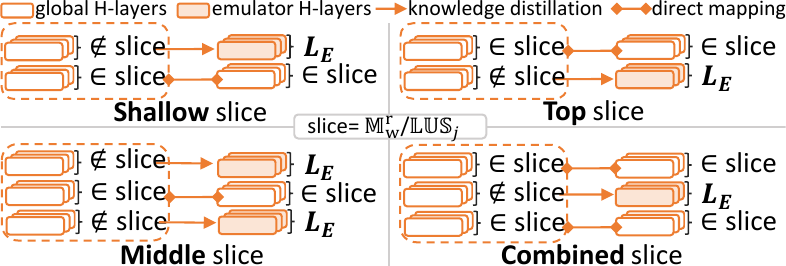} 
    \caption{Emulator generation given slice $\mathbb{M}^r_w$/$\mathbb{LUS}_{j}$.}
    \label{fig:emulator_generation}
\end{figure}

Finally, the model requester executes further-slicing based on our MSV scheme to enable more validators to participate in verifying worker updates. 
The requester checks whether the union $\mathbb{U}^r=\bigcup_{w\in\mathbb{W}^r}\mathbb{M}_w^r$ can form a validation model with at most $L_{vm}$ H-layers under the model construction scheme introduced in \S\ref{sssec:local_model}, where $L_{vm}$ is the number of H-layers that can be deployed on the most hardware-constrained validator. If so, it sets $N_{lus}=1$ and $\mathbb{LUS}_0=\mathbb{U}^r$. 
Otherwise, the requester partitions $\mathbb{U}^r$ into disjoint sub-slices $\{\mathbb{LUS}_j\}_{j=0}^{N_{lus}-1}$, where $N_{lus}$ is the number of sub-slices produced based on the following partition conditions: 
(i) The output layer forms a separate sub-slice. 
(ii) The H-layer sub-slices cover all H-layers in $\mathbb{U}^r$, and each retains a shallow, top, middle, or combined structure as shown in Fig.~\ref{fig:emulator_generation}.
(iii) Each sub-slice, together with its emulator, must form a validation model containing at most $L_{vm}$ H-layers. 
The above design requires each participant to support the deployment of at least $3L_E$ H-layers. Participants below this capacity are not admitted as workers
or validators in \texttt{DICE}.

For every slice $\mathbb{M}^{r}_w$ and $\mathbb{LUS}_{j}$, the model requester compresses the global H-layers excluded from that slice into an emulator of a compact model by ModelDerivation() (Alg.~\ref{alg:offchain} line~\ref{alg_line:generate_emulator}).

\subsubsection{Large-to-Compact Model Derivation} \label{sssec:local_model} In this procedure, the model requester derives compact models from the global model for both workers and validators  (Alg.~\ref{alg:offchain} line~\ref{alg_line:generate_emulator}). 
Each compact model retains the same embedding and output layers as the global model but contains a reduced number of H-layers. 
These compact models are designed to allow workers to efficiently generate updates for $\mathbb{M}^r_w$, and to enable validators to verify each update at a manageable cost. 
Motivated by offsite tuning \cite{xiao2023offsite}, we insert a lightweight \textit{emulator} into each compact model to ensure that each worker's update and the corresponding verification results closely approximate those obtained by training or verifying with the full model, thereby achieving performance comparable to customizing the full model. 
As presented in Fig.~\ref{fig:emulator_generation}, an emulator is obtained by compressing the global H-layers not in $\mathbb{M}^r_w$ (or $\mathbb{LUS}_j$) through knowledge distillation (KD) \cite{hinton2015distilling, xiao2023offsite} using the validation data. It contains $L_E$ H-layers by default, or $2L_E$ H-layers for a middle slice, with $L_E=1$ in our default setting. 
For $\mathbb{LUS}_j$ with only the output layer, all global H-layers are distilled into an emulator containing $L_{E}$ H-layers. For $\mathbb{M}^r_w$ or $\mathbb{LUS}_j$ containing all global H-layers, no emulator is needed. 

\subsubsection{Runtime Mechanics}\label{sssec:runtime_mechanism} 
We now describe how the model requester executes the above procedures with two runtime mechanisms incorporated in DPL and MSV to balance model performance and computation efficiency. 

(i) \textit{Periodic Slice Reselection and Emulator Regeneration.} 
Rather than executing slice selection and large-to-compact model derivation in every round, we design a more cost-effective periodic mechanism that avoids unnecessary slice reselection or emulator regeneration and decouples the update frequency of the selected slices and the emulators based on their respective roles. 

Under this periodic mechanism, the requester re-computes slice importance scores and selects new slices to train on workers only when the average performance of the global model over the latest $\epsilon_r \in \mathbb{Z}^+$ rounds does not show distinct improvement. Otherwise, the requester skips slice selection and proceeds directly to the model derivation step, reusing the same slices as in the previous round. The rationale is that slice selection determines which part of the model to train, and a performance plateau signals the exhausted utility of further training the current selected slice and the necessity of exploring a different slice.

The requester generates an emulator when a slice/sub-slice is first assigned or its structure changes.
For an unchanged slice/sub-slice, the emulator is regenerated only when the output discrepancy between the full global model and the compact model exceeds  $\epsilon_{em} \in \mathbb{R}^+$. 
The rationale is that the emulator's role is to approximate the input-output mappings of the global H-layers excluded from the compact model, and a large output discrepancy implies the ineffectiveness of the current emulator for such approximation and the necessity of emulator regeneration.

With the above designs, the model requester re-selects slices and regenerates emulators in response to performance stagnation or model drift, respectively, thereby balancing model adaptability with computational efficiency.

(ii) \textit{Adaptive Mode Switching.} 
In addition to the above periodic mechanism, DPL further incorporates an adaptive mode switching mechanism to accommodate heterogeneous hardware capabilities among workers. Under this mechanism, the requester dynamically switches between two operational modes of DPL. One is \textit{Personalized} (P-DPL), where slices are selected per worker to maximize the utilization of available hardware, as described in \S\ref{sssec:slice_selection}. The other is \textit{Unified} (U-DPL), where the same slice is selected for all workers by using a unified parameter $L_{wm}=\min(\{L_w|w \in \mathbb{W}_{all}\})$ to replace every worker's specific parameter $L_w$ in \S\ref{sssec:slice_selection}. 
U-DPL is the default mode, since it requires the slice selection to be executed only once for all workers in any round and thus is cost-effective for requesters. 
The requester activates the P-DPL mode only if there exist hardware-rich workers (with $L_w > L_{wm}$) whose updates consistently achieve a performance score higher than $\frac{L_w}{L_{wm}} \cdot s_{wm}$, where $s_{wm}$ is the best performance score that hardware-constrained workers (i.e., with $L_w = L_{wm}$) have ever achieved in all past rounds. 
Once this condition ceases to hold, the requester switches DPL back to its default mode. 

This adaptive mechanism is motivated by the observation that the benefit of personalization depends on the alignment between hardware capacity and local training quality.
Specifically, assigning larger slices to train on hardware-rich workers that fail to produce well-performing updates could lead to overfitting on those workers, reducing the generality of the global H-layers that hardware-constrained workers do not update, thereby compromising the final global model performance. 
Thus, P-DPL is preferred only when capable workers demonstrate consistently superior local performance exceeding a dynamic threshold.

Together, the two runtime mechanisms enable \texttt{DICE} to adapt the requester-side operations to both the model's learning dynamics and the participants' resource heterogeneity, balancing the requester's overhead, the engagement of more honest but hardware-constrained participants, and model performance.

\subsubsection{Lightweight Training and Verification} \label{sssec:slice_eval}
After the requester-side procedures complete, the worker constructs a compact local model for training, and the validator constructs a compact validation model for verification, as shown in Fig.~\ref{fig:algorithm_overview_FMRefiner}. 
Each worker $w$ executes DPLtrain() to produce the update for $\mathbb{M}^r_w$ and divide it into $\{\theta^{r}_{w,\mathbb{LUS}_j}|j\in [0,N_{lus})\}$ (Alg.~\ref{alg:offchain} line~\ref{alg_line:DPL}). 
Each validator $v$ executes MSVorAGG() to verify and aggregate $\theta^{r}_{w,\mathbb{LUS}_j}$ (Alg.~\ref{alg:offchain} line~\ref{alg_line:MSV}).  
An update $\theta^{r}_{w,\mathbb{LUS}_j}$ will be considered valid only if it satisfies two conditions: (i) it contains at least one layer whose norm diverges from the historical norms of updates for that layer to pass the fraud check; (ii) its corresponding validation model performs better than a dynamically adjusted threshold $\epsilon_j$ to pass the poisoning check. 
When constructing $\theta^r_{w,\mathbb{LUS}_j}$'s corresponding validation model, the parameters submitted by $w$ are used for layers trained by $w$. For H-layers in $\mathbb{LUS}_j$ that $w$ did not train, and for the output layer when it is excluded from $\mathbb{LUS}_j$, the model uses the round-start global parameters. The remaining H-layers are replaced by the corresponding emulator. 

The threshold $\epsilon_j$ is initialized by the model requester using the global model's validation score at the start of the first training round. For $r\geq2$, let $s_G^r$ denote the global model's validation score before training in round $r$, $\mathbb{S}_{\mathbb{LUS}_j}$ contain the consensus-confirmed validation scores of qualified updates for the current sub-slice $\mathbb{LUS}_j$ over the latest $\beta_{\mathrm{hist}}$ rounds, $s_{VM,\mathbb{LUS}_j}$ be the validation score of $\mathbb{LUS}_j$'s corresponding compact model at its most recent derivation, and $\overline{\Delta}^{\,r} = \frac{\sum_{q=\max(2,r-\beta_{\mathrm{hist}}+1)}^{r}(s_G^q-s_G^{q-1})}{r-\max(2,r-\beta_{\mathrm{hist}}+1)+1}$. The threshold is updated before training by Eq.~\ref{eq:epsilon_j}. This rule adapts the acceptance threshold to the progress of global model performance and the actual quality of qualified updates.
\begin{equation}
\label{eq:epsilon_j}
\epsilon_j=
\begin{cases}
\min\!\left(
s_G^r+\overline{\Delta}^{\,r},
2\operatorname{median}(\mathbb{S}_{\mathbb{LUS}_j})
-\min(\mathbb{S}_{\mathbb{LUS}_j})
\right),
& \mathbb{S}_{\mathbb{LUS}_j}\neq\emptyset\\
\min\!\left(
s_G^r+\overline{\Delta}^{\,r},
s_{VM,\mathbb{LUS}_j}
\right),
& \text{otherwise}
\end{cases}
\end{equation}

\subsubsection{Discussion}\label{sssec:discussion_on_DPL_MSV} We finally discuss the design rationales of DPL and MSV. Both schemes are motivated by the resource asymmetry between the model requester and the other participants, particularly the workers and validators who cannot deploy full models. They leverage the relatively abundant hardware resources of the requester to select slices and generate emulators for compact model execution on constrained participants, enabling those participants to effectively contribute to customizing large PMs. By integrating two runtime mechanisms, namely (i) periodic slice reselection and emulator regeneration and (ii) adaptive mode switching, they slow down the growth of requester-side costs as the number of training rounds or the participant scale increases. 

Meanwhile, the two schemes are designed to mitigate the potential biases introduced by using compact models in the training and verification stages, respectively.
DPL mitigates training bias by dynamically adjusting the slice to be trained under heterogeneous hardware constraints among workers based on a slice importance score, which quantifies the expected utility of training a specific set of layers by the potential performance improvement and the magnitude of parameter updates.  
By selecting the slice with the highest score, DPL uses information in the specific task runtime to determine which part of the PM should be customized in each round, rather than following a fixed heuristic solution. 
MSV mitigates verification bias by employing a threshold $\epsilon_j$, dynamically adjusted by the model requester based on the full global model's performance trend and actual distribution of the quality of worker updates, to compensate for systematic under- or over-estimation of worker update quality.

Furthermore, DPL and MSV jointly make poisoning attacks more difficult. 
Compared to full model training and verification, the two schemes restrict parameters that each malicious worker can modify in each round to its
assigned slice and the output layer, and each worker's update still needs to pass corresponding quality checks. 
The eventual effect of a poisoned update depends on not only the honest workers' concurrent updates but also the requester's future decisions on slice selection, emulator updates, and validation threshold ($\epsilon_j$) adjustments, all of which might not be reliably predicted at the time of attack injection. 
The tunability of $\epsilon_j$ also enables MSV to accommodate diverse and evolving adversarial behaviors, as the adjustment logic for $\epsilon_j$ can be refined when new attack strategies emerge.

\subsection{Balancing Efficiency, Scalability, and Security by TCC and EDI} \label{ssec:consensus} 
This section enhances the lightweight verification introduced in the last section with our TCC protocol and EDI scheme, two integrated techniques to further reduce per-validator workload requirement and improve the system's ability to scale to more workers while ensuring security. 
\S\ref{sssec:consensus} first presents how TCC breaks the linear dependency between each validator's workload and worker scale while generating correct consensus. 
Then, \S\ref{sssec:incentive} presents how EDI complements TCC to strengthen the balance between efficiency, scalability, and security. 
Finally, \S\ref{sssec:discussion_on_TCC_EDI} discusses the synergistic relationship between TCC and EDI.

\subsubsection{Consensus Generation} \label{sssec:consensus}
TCC generates consensus over the entire VA task by four critical steps: Task Decomposition, Result Committing, Selective Committee Voting, and Result Revealing. 

\textbf{Task Decomposition} (before \texttt{VA\_COMMIT} state). \label{sssec:VA_decompose} In this step, a monolithic VA task is decomposed into $N_{VA}$ sub-tasks in a VA task forest, where $N_{VA} \leq (2|\mathbb{W}^r|-1)N_{lus}$ and $\mathbb{W}^r$ is the set of active workers in round $r$ (Alg.~\ref{alg:onchain} line~\ref{alg_line:worker_on_end}). 
As shown in Fig.~\ref{fig:algorithm_overview_FMRefiner}, the forest comprises several task trees. The leaf node in each $Tree_j$ represents a VER sub-task verifying a worker's update of $\mathbb{LUS}_{j}$. The non-leaf node represents an AGG sub-task
aggregating qualified updates from two child nodes. 
Executing each non-leaf sub-task has a strong dependency on the results of its two child sub-tasks. For example, an AGG sub-task requires the performance scores computed in its child sub-tasks to determine each update's weight in aggregation. It may also need the file handles of the partially aggregated updates to retrieve those updates for further aggregation. 
After the VA task forest is constructed on the blockchain, result committing starts.

\textbf{Result Committing} (during \texttt{VA\_COMMIT} state). Each validator iteratively executes the VA sub-task randomly selected from the forest based on its local hardware condition and submits the fingerprint and commitment of this sub-task's results to the blockchain based on a cryptographic commitment scheme to prevent fraud (Alg.~\ref{alg:offchain} lines~\ref{alg_line:off_commit_start}--\ref{alg_line:offchain_commit_end}). 
The set of results from a validator $v$ for a VA sub-task, denoted by $\mathbb{VAR}_v$, is produced as follows. 
\begin{itemize}
    \item For any VER sub-task, the validator first verifies a worker's update $\theta^{r}_{w,\mathbb{LUS}_j}$ as illustrated in \S\ref{sssec:slice_eval}, producing the sub-task's intermediate result, i.e., the output of the validation model built upon $\theta^r_{w,\mathbb{LUS}_{j}}$ on the validation data, denoted by $V(\theta^r_{w,\mathbb{LUS}_{j}},\mathbb{D}_{val})$. If the update is qualified, the validator then computes its weight in the final aggregated update, denoted by $\alpha_{w,\mathbb{LUS}_{j}}$. We define $\alpha_{w,\mathbb{LUS}_{j}}$ by the performance of $\theta^r_{w,\mathbb{LUS}_{j}}$ on the validation data, while remaining compatible with alternative aggregation weight definitions, e.g., weights based on worker data volume \cite{McMahan17,lin2024refiner} and contribution to aggregated updates \cite{electronics14051024,10.24963/ijcai.2024/554}. Finally, the validator packs the task input version (identified by $H(\theta^r_{w,\mathbb{LUS}_{j}})$), the intermediate results $\alpha_{w,\mathbb{LUS}_{j}}$ and $V(\theta^r_{w,\mathbb{LUS}_{j}},\mathbb{D}_{val})$, the iteration index, and the sub-task identifier into the result set, where $H(\cdot)$ is a cryptographic hash function.

    \item For any AGG sub-task, the validator first aggregates the qualified model updates received from its two child nodes via Eq.~\ref{eq:aggregate_weight2}--\ref{eq:aggregate}. The coefficient $\frac{\beta_{\mathbb{W}_{c},\mathbb{LUS}_{j}}}{\tau^{l}_{c1,c2}}$ ensures that, for each model layer in $\mathbb{LUS}_{j}$ with qualified worker updates,  the root node of each task tree $Tree_j$ outputs the weighted average of the qualified worker updates containing that layer. During the aggregation process, the validator produces the sub-task's intermediate results: $H(\{(\theta^r_{\mathbb{W}_{c1},\mathbb{LUS}_{j}}[l], \theta^r_{\mathbb{W}_{c2},\mathbb{LUS}_{j}}[l], \theta^r_{\mathbb{W},\mathbb{LUS}_{j}}[l], \tau^{l}_{c1,c2})\mid l\in\mathbb{LUS}_j\}\parallel \beta_{\mathbb{W}_{c1},\mathbb{LUS}_{j}}\parallel\beta_{\mathbb{W}_{c2},\mathbb{LUS}_{j}})$ and $digest_{|\mathbb{LUS}_j|}$ defined by $digest_i=H(digest_{i-1}\parallel\theta^r_{\mathbb{W}_{c1},\mathbb{LUS}_{j}}[l_i]\parallel\theta^r_{\mathbb{W}_{c2},\mathbb{LUS}_{j}}[l_i]\parallel\theta^r_{\mathbb{W},\mathbb{LUS}_{j}}[l_i])$ for $i=1,\ldots,|\mathbb{LUS}_j|$, where $digest_0$ is a fixed initialization value and $l_i=\mathbb{LUS}_j[i]$ denotes the $i$-th layer in a predefined ordering of $\mathbb{LUS}_j$. Finally, the validator packs the task input version (identified by the two child-node update hashes $H(\theta^r_{\mathbb{W}_{c1},\mathbb{LUS}_j})$ and $H(\theta^r_{\mathbb{W}_{c2},\mathbb{LUS}_j})$ together with the corresponding $\beta_{\mathbb{W}_{c1},\mathbb{LUS}_j}$, $\beta_{\mathbb{W}_{c2},\mathbb{LUS}_j}$, and $\{\tau^{l}_{c1,c2}\mid l\in\mathbb{LUS}_j\}$), the above intermediate results, the aggregated update hash $H(\theta^r_{\mathbb{W},\mathbb{LUS}_j})$, the iteration index, and the sub-task identifier into the result set.
\end{itemize}

\begin{equation}\small\label{eq:aggregate_weight2}
\beta_{\mathbb{W}_{c},\mathbb{LUS}_{j}}
=
\begin{cases}
\alpha_{w,\mathbb{LUS}_{j}},\quad w\in\mathbb{W}_{c},
& \text{if }|\mathbb{W}_{c}|=1,\\
1,
& \text{otherwise}.
\end{cases}
\end{equation}

\begin{equation}\small\label{eq:aggregate_weight3}
\tau^{l}_{c1,c2}
=
\begin{cases}
\displaystyle
\sum_{w\in\mathbb{W}_{j,l}}
\alpha_{w,\mathbb{LUS}_{j}},
& \text{if }|\mathbb{W}_{c1}\cup\mathbb{W}_{c2}|=|\mathbb{W}_{j}|,\\
1,
& \text{otherwise}.
\end{cases}
\end{equation}

\begin{equation}\small\label{eq:aggregate}
\theta^r_{\mathbb{W},\mathbb{LUS}_{j}}[l]
=
\begin{cases}
\displaystyle
\sum_{c\in\{c1,c2\}}
\frac{\beta_{\mathbb{W}_{c},\mathbb{LUS}_{j}}}
{\tau^{l}_{c1,c2}}
\theta^r_{\mathbb{W}_{c},\mathbb{LUS}_{j}}[l],
& \tau^{l}_{c1,c2}>0,\\
\theta^{r-1}_{\mathbb{LUS}_{j}}[l],
& \text{otherwise}.
\end{cases}
\end{equation}

\noindent Here, $c1$ and $c2$ denote the two child nodes, and $\mathbb{W}_{c}$ is the set of workers whose updates are involved in any child node $c\in\{c1,c2\}$. The model parameters corresponding to child node $c$ are denoted by $\theta^r_{\mathbb{W}_{c},\mathbb{LUS}_{j}}$; they are the corresponding worker's update if $\mathbb{W}_{c}$ is a singleton, and otherwise are the partially aggregated update. Moreover, $\mathbb{W}=\mathbb{W}_{c1}\cup\mathbb{W}_{c2}$, $\mathbb{W}_{j}=\{w\in\mathbb{W}^r\mid\mathbb{M}_w^r\cap\mathbb{LUS}_{j}\neq\emptyset\}$ denotes the set of workers participating in the training of the parameters corresponding to the $j$-th sub-slice, and $\mathbb{W}_{j,l}$ is the set of workers included in $\mathbb{W}_{j}$ and with $\theta^r_{w,\mathbb{LUS}_{j}}$ containing the parameters of layer $l$. 
For any update that fails verification, its corresponding weight is set to zero. If the updates from two child nodes cover different model layers, they are aligned using zero-valued padding, and the padded entries are excluded from the normalized aggregation of the corresponding layers. If a task tree $Tree_j$ contains only one leaf node, that leaf node also serves as the root: its worker update is adopted for the layers it covers if it passes verification, while the round-start global parameters are retained for all remaining layers.

The commitment scheme in prior work \cite{lin2024refiner} defines the fingerprint and commitment of each result set $\mathbb{VAR}_v$ as $FP_v=H(\mathbb{VAR}_v)$ and $CC_v=H(\mathbb{VAR}_v\parallel v)$, respectively. TCC extends this scheme by treating the partial raw results $\mathbb{VAR}^{\text{commit}}_v$ ($\subset\mathbb{VAR}_v$), together with $FP_v$, as the sub-task fingerprint on which validators must reach consensus. For any VER sub-task, $\mathbb{VAR}^{\text{commit}}_v$ contains only the weight $\alpha_{w,\mathbb{LUS}_{j}}$ and the task input version, i.e., the worker-update hash. For any AGG sub-task, it contains only the aggregated update hash $H(\theta^r_{\mathbb{W},\mathbb{LUS}_{j}})$ and the task input version identified by the two child-node update hashes and their corresponding coefficients. TCC also requires validators executing an AGG sub-task to submit the file handle of the aggregated update together with its fingerprint and commitment. These designs enable validators to reuse available descendant results when executing non-leaf sub-tasks without re-execution, thereby reducing individual workload and global consensus latency across the VA task forest.

For any given sub-task, honest validators will generate the same valid fingerprint from the same task input version under uniform computation, encoding, and serialization rules, whereas malicious validators may submit poisoned results. \texttt{ValidatorOnchainProcedure}() feeds validator $v$'s submission into \texttt{OnchainCommit}() for fingerprint validity checking and consensus fingerprint generation (Alg.~\ref{alg:onchain} lines~\ref{alg_line:onchain_commit}, \ref{alg_line:consensus_check_start}--\ref{alg_line:consensus_check_end}). This function involves two important checks: 

(i) For any given sub-task, it checks whether the task input version declared by $v$ belongs to the available input versions recorded on-chain for that sub-task. 

(ii) For a non-leaf sub-task, it further checks the results of each descendant sub-task used by $v$: these results must match either the results previously submitted by $v$ for that descendant sub-task or its adopted consensus results. 
It also checks whether the update hashes declared in the task input version match the corresponding descendant sub-tasks' results and whether the update's coefficients are consistent with those results and Eq.~\ref{eq:aggregate_weight2}--\ref{eq:aggregate_weight3}. 
These checks enforce the sub-task input provenance and dependency consistency while allowing validators to reuse available descendant sub-tasks' results and avoid redundant computation.

The submission is rejected if any of the above checks fails. 
After the submission passes the above checks, if the blockchain has already recorded a commitment from $v$ for the same sub-task and input version, the original commitment and its corresponding fingerprint are retained, and neither replacement nor an additional submission record is accepted. Otherwise, the submitted fingerprint and commitment are recorded. The validity of the retained or newly recorded fingerprint is determined by TCC's selective committee voting strategy. 

\begin{figure}[t]
    \centering
    \includegraphics[trim=0mm 3.8mm 0mm 0mm, clip,width=0.85\columnwidth]{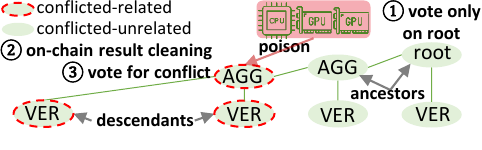} 
    \caption{Selective committee voting.}
    \label{fig:TCC}
\end{figure}

\textbf{Selective Committee Voting} (during \texttt{VA\_COMMIT} state). Fig.~\ref{fig:TCC} depicts TCC's selective committee voting strategy (Alg.~\ref{alg:onchain} lines~\ref{alg_line:voting}--\ref{alg_line:check_conflict}).  
Committees are formed only for root sub-tasks at the start of the consensus process (Fig.~\ref{fig:TCC}\ding{172}) to prevent malicious validators from prematurely satisfying the global consensus conditions by using quickly forged results without actual execution of any sub-task.   
For a non-root sub-task, a committee is formed only when it becomes controversial  (Fig.~\ref{fig:TCC}\ding{174}). 
Following prior work \cite{lin2024refiner}, our committee formation requires each validator to stake at least $E_{\min}$ units of cryptocurrency as a security deposit upon joining a customization task. 
Committee members for each sub-task are randomly selected from all validators, and each committee votes to determine the consensus fingerprint of its assigned sub-task. A validator with a larger stake has a higher probability of being selected and a larger expected number of votes.

Since the input of a root or controversial non-leaf sub-task may change with its descendant results, TCC determines whether to elect another committee for such a sub-task according to whether the identities of the existing committee members have been disclosed. If a sub-task with an existing committee needs to process a new input version and some members of the committee intended for reuse have disclosed their identities by submitting fingerprints and commitments, a new committee is formed using its corresponding election index. Otherwise, the existing committee is reused. When processing a root or controversial sub-task, each committee is associated with a specific input version and votes on the outputs generated from that version.

To prevent malicious validators from repeatedly triggering committee elections for the same sub-task, TCC records the number of committee elections for each sub-task in a single iteration and sets an upper bound $K_{\text{election}}\in\mathbb{Z}^+$, including the initial election. To distinguish different sub-tasks and different elections of the same sub-task, the election seed is generated from the current-round random seed, the sub-task identifier, and the election index of that sub-task. The on-chain procedure increments the election index from 1 to at most $K_{\text{election}}$. If another committee election is required after this limit has been reached, the current consensus process is terminated and the customization task is paused. The round-start global model parameters are retained, and the model requester determines whether to recruit validators again and restart the task.

For a root or controversial sub-task, \texttt{Voting}() is triggered (Alg.~\ref{alg:onchain} line~\ref{alg_line:voting_end}). This function first verifies validator $v$'s membership in the committee corresponding to the sub-task and calculates its number of votes using the submitted proof $\pi_v$ and hash value $hash_v$. For details of this verification process, please refer to prior work \cite{lin2024refiner,algorand}. Validator $v$'s votes are counted only if one of the following eligibility requirements is satisfied: 
(i) For a non-controversial root sub-task, $v$'s valid on-chain records must contain submissions for at least $\lceil\frac{N_{VA}-N_{CF}}{N_{VA}}N_t\rceil$ distinct conflict-unrelated sub-tasks, where $N_{CF}$ is the number of conflict-related sub-tasks. If the total number of non-root conflict-unrelated sub-tasks is smaller than this threshold, $v$ is instead required to complete and submit results for all non-root conflict-unrelated sub-tasks. 
(ii) For a controversial sub-task, $v$'s valid on-chain records must contain its submissions for all descendants of that sub-task. 
Here, valid on-chain records form a subset of the complete on-chain submission history and contain submissions accepted under the preceding rules, including submissions consistent with the consensus of non-controversial sub-tasks and submissions used for committee voting. 
The first requirement, together with the supplementary check described below, supports detection of poisoned results in conflict-unrelated sub-tasks, whereas the second forces honest committee members to execute the descendant sub-tasks on which a controversial sub-task depends, thereby producing consistent and correct results for conflict-related sub-tasks.

If $v$'s votes are eligible, \texttt{Voting}() records or updates its votes for the corresponding candidate fingerprint and updates the total votes received by each candidate. 
For each root or controversial sub-task, the highest-voted fingerprint receiving more than $E_c\epsilon_c$ votes in the committee, corresponding to this sub-task and its input version consistent with the consensus results of its dependent descendant sub-tasks, is adopted as this sub-task's consensus fingerprint. 
TCC allows validators to submit fingerprints and vote for any root or controversial sub-task before the consensus results of relevant descendant sub-tasks are available, but the consensus of any root or controversial sub-task is confirmed only after the above dependency requirement is satisfied.

For each non-root and non-controversial sub-task for which voting is unnecessary, \texttt{CheckConflict}() is invoked (Alg.~\ref{alg:onchain} line~\ref{alg_line:check_conflict}). If no conflict is detected, the received fingerprint is directly adopted as the sub-task's consensus fingerprint. Otherwise, the sub-task is marked as controversial, and the sub-task and its descendants are marked as conflict-related, triggering a two-step resolution procedure. 
Firstly, the adopted consensus fingerprints of this controversial sub-task and its ancestors are revoked, and all submissions related to these sub-tasks are removed from the valid on-chain records. The corresponding voting records, completed sub-task records, and counters related to the global consensus conditions are updated accordingly (Fig.~\ref{fig:TCC}\ding{173}). The submitted fingerprints, commitments, submitter identities, and task input versions remain in the complete on-chain submission history for duplicate-commitment checking and subsequent reward or penalty determination. If a validator's submission for a corresponding sub-task is removed from the valid on-chain records, the validator must submit it again and pass the preceding checks before it can be restored to the valid records. This step prevents a controversial sub-task from compromising the correctness of its ancestors. If a descendant's results change after conflict resolution, its ancestors should be recomputed from the updated descendant results. 
After the above step, a new committee is randomly selected to vote on the consensus fingerprint of the controversial sub-task (Fig.~\ref{fig:TCC}\ding{174}).

\textit{Global Consensus Determination.} 
To determine whether the global consensus has been reached, the on-chain procedure first checks two conditions: (i) at least $N_c$ validators each have valid on-chain records for at least $\lceil\frac{N_{VA}-N_{CF}}{N_{VA}}N_t\rceil$ distinct conflict-unrelated sub-tasks; and (ii) the consensus fingerprints of all VA sub-tasks have been confirmed. If no conflict-unrelated sub-task exists, the first condition is waived, and global consensus is reached once the second condition holds.

Otherwise, once the first condition is satisfied, a supplementary global consensus check is initiated to ensure that honest validators have submitted results for conflict-unrelated sub-tasks. The on-chain procedure selects $N_c$ validators satisfying the first condition to form the checking member set $\mathbb{V}_c$. After this set has been fixed, a newly generated random value that was unpredictable before the set was determined is used to independently and uniformly sample $\lceil\frac{N_{VA}-N_{CF}}{N_{VA}}N_t\rceil$ conflict-unrelated sub-tasks without replacement for each validator $v\in\mathbb{V}_c$, forming $v$'s supplementary checking set $\mathbb{CHECK}_v^r$. Here, $N_{CF}$ is the number of conflict-related sub-tasks at the time of sampling. Each checking member completes the sampled sub-tasks and submits the corresponding fingerprints and commitments. If the valid on-chain records already contain a submission by that member for the same input version, the on-chain procedure directly reuses the recorded submission for the supplementary check, without requiring the member to re-execute the sub-task or resubmit the result. 

Any conflict arising during the supplementary check is processed according to the preceding conflict-resolution rules, and newly conflict-related sub-tasks are removed from all supplementary checking sets. For a sub-task that remains in a checking set but whose previous submission is no longer included in the valid on-chain records due to conflict resolution, the corresponding validator checks whether the inputs on which its existing result depends are consistent with the task inputs to be used in the current execution. If they are consistent, the validator reuses the existing result; otherwise, it re-executes the sub-task. In either case, it submits the corresponding fingerprint and commitment under the preceding rules. Global consensus is reached and result revealing begins when the valid on-chain records contain fingerprints and commitments from all checking members for every sub-task in their respective $\mathbb{CHECK}_v^r$ and the second global consensus condition is satisfied.

Each supplementary check is assigned a time limit of $\lfloor \frac{T_{\text{commit}}}{K_{\text{check}}}\rfloor$, and the number of supplementary checks in each iteration is bounded by $K_{\text{check}}\in\mathbb{Z}^+$, including the initial check, where $T_{\text{commit}}$ is the total timeout of the \texttt{VA\_COMMIT} state. If a check times out before completion, new checking members are selected from the validators satisfying the first global consensus condition, and new supplementary checking sets are sampled. If a supplementary check has been completed but the second global consensus condition has not yet been satisfied, its checking sets and valid submissions are retained while the procedure waits for the second condition. If all checking attempts are exhausted without completion or the \texttt{VA\_COMMIT} state reaches its total timeout, \texttt{DICE} follows its timeout-handling rules.

\textbf{Result Revealing} (during the \texttt{VA\_REVEAL} state). In this state, each validator $v$ submits the unrevealed raw VA results
$\mathbb{VAR}^{\text{reveal}}_v
=\mathbb{VAR}_v\setminus\mathbb{VAR}^{\text{commit}}_v$
for the sub-tasks and input versions corresponding to its valid on-chain submission records (Alg.~\ref{alg:offchain} lines~\ref{alg_line:offchain_reveal_start}--\ref{alg_line:offchain_reveal_end}). These results are submitted to \texttt{OnchainReveal}(), which checks whether all of the following conditions hold:
(i) $H\big((\mathbb{VAR}^{\text{commit}}_v\cup\mathbb{VAR}^{\text{reveal}}_v)\parallel v\big)=CC_v$;
(ii) $H(\mathbb{VAR}^{\text{commit}}_v\cup\mathbb{VAR}^{\text{reveal}}_v)=FP_{\text{agreed}}$; and
(iii) $\mathbb{VAR}^{\text{commit}}_v=\mathbb{VAR}^{\text{commit}}_{\text{agreed}}$
(Alg.~\ref{alg:onchain} lines~\ref{alg_line:onchain_reveal_start}--\ref{alg_line:onchain_reveal_end}).
Here, $FP_{\text{agreed}}$ and $\mathbb{VAR}^{\text{commit}}_{\text{agreed}}$ jointly constitute the consensus fingerprint of the sub-task and input version to which $\mathbb{VAR}^{\text{reveal}}_v$ pertains. All checks are performed against records from the same iteration, sub-task, and valid input version. These checks bind each commitment to the submitter's identity and prevent the direct reuse of another validator's commitment. The formal upper bound on the probability of fraud through copying another validator's results is provided in Theorem 3 of \S\ref{ssec:security}. Result revealing ends after all validators have revealed the results corresponding to their valid on-chain submission records (line~\ref{alg_line:onchain_reveal_end}).

\subsubsection{Efficiency-Driven Incentive Scheme}
\label{sssec:incentive}

At the end of \texttt{VA\_REVEAL}, our EDI scheme rewards or penalizes workers and validators (Alg.~\ref{alg:onchain}, line~\ref{alg_line:onchain_reveal_end}) to incentivize honest behavior and encourage participants to contribute more hardware resources to accelerate customization in \texttt{DICE}. 

Rather than replacing existing incentive schemes~\cite{10.1145/3578581,9369019}, EDI augments them with workload-sensitive coefficients. The coefficient $Final_w$ for each worker $w$, defined in Eq.~\ref{workload-sensitive_coefficient_worker}, increases the rewards of honest workers that contribute higher-quality and larger-scale worker updates. The coefficients $Final_v^{\mathrm{pre}}$ and $Final_v^{\mathrm{post}}$ for each validator $v$, defined in Eq.~\ref{workload_sensitive_coefficient_validator1} and~\ref{workload_sensitive_coefficient_validator2}, respectively, increase the rewards of honest validators that correctly complete more VA sub-tasks before and after the first controversial sub-task appears. If a worker or validator is detected as malicious, its corresponding workload-sensitive coefficient is set to zero. A coefficient is also set to zero if its normalization denominator in Eq.~\ref{workload-sensitive_coefficient_worker}--\ref{workload_sensitive_coefficient_validator2} is zero.

\begin{equation} 
\label{workload-sensitive_coefficient_worker}
Final_w =
\Phi_w \left(
1+\gamma_{worker}
\frac{\sum_{j=0}^{N_{lus}-1}Contri_{w,j}}
{\sum_{w'\in\mathbb{W}^r}\sum_{j=0}^{N_{lus}-1}Contri_{w',j}}
\right),
\end{equation}

\begin{equation} 
\label{workload_sensitive_coefficient_validator1}
Final_v^{\mathrm{pre}} =
\Phi_v \left(
1+\gamma_{validator}
\frac{N_{v,t}^{\mathrm{correct}}-N_{v,t}^{\mathrm{post}}}
{\sum_{v'\in\mathbb{V}^r}N_{v',t}^{\mathrm{pre}}}
\right),
\end{equation}

\begin{equation} 
\label{workload_sensitive_coefficient_validator2}
Final_v^{\mathrm{post}} =
\Phi_v \left(
1+\gamma_{validator}
\frac{N_{v,t}^{\mathrm{post}}}
{\sum_{v'\in\mathbb{V}^r}N_{v',t}^{\mathrm{post}}}
\right).
\end{equation}

Here, $\Phi_w$ and $\Phi_v$ are the base reward weights of worker $w$ and validator $v$, respectively, which can be determined by an underlying incentive mechanism~\cite{9369019}; both are set to $1$ when no such mechanism is used. $\gamma_{worker},\gamma_{validator}\in\mathbb{R}^+$ are tunable incentive amplification factors. $Contri_{w,j}$ is worker $w$'s contribution score for the $j$-th model slice, which can be instantiated using the aggregation weight $\alpha_{w,\mathbb{LUS}_j}$ or the worker-contribution metric of the underlying incentive scheme.
$N_{v,t}^{\mathrm{correct}}$ is the number of VA sub-tasks correctly completed by validator $v$, and $N_{v,t}^{\mathrm{post}}$ is the number correctly completed after the first controversial sub-task appears, where $N_{v,t}^{\mathrm{post}}\leq N_{v,t}^{\mathrm{correct}}$. $N_{v,t}^{\mathrm{pre}}$ is the number of distinct sub-tasks completed by $v$ and confirmed on chain when the first controversial sub-task appears. If no controversial sub-task appears in the current round, we set $N_{v,t}^{\mathrm{pre}}=N_{v,t}^{\mathrm{correct}}$. 
The numbers of correctly completed sub-tasks and the corresponding rewards are determined from each validator's submissions for the input versions associated with the final consensus results. For repeated submissions of the same result for the same sub-task and input version, the earliest on-chain confirmation determines its completion time. The complete on-chain submission history is retained for commitment checks and subsequent accountability and penalties.

The rewards $R_w$ and $R_v$ for workers and validators are given by Eq.~\ref{eq:worker_reward} and~\ref{eq:validator_reward}, respectively, where $WR$ and $VR$ denote the total worker and validator reward budgets for one round, and Eq.~\ref{eq:validator_reward2} defines the normalization bases used to distribute validator rewards. When a controversial sub-task appears, $Final_{all}^{\mathrm{pre}}$ is computed using the $N_{v,t}^{\mathrm{pre}}$ values fixed when the first controversial sub-task appears. 
If $\sum_{v\in\mathbb{V}^r}N_{v,t}^{\mathrm{pre}}=0$, we set $Final_{all}^{\mathrm{pre}}=0$. Any undistributed reward remains in the task smart contract. 
The above designs provide participants with additional incentives to improve system efficiency while maintaining compatibility with existing incentive schemes.

\begin{equation} 
\label{eq:worker_reward}
R_w =
\frac{Final_w}
{\delta+\sum_{w'\in\mathbb{W}^r}Final_{w'}}
\cdot WR,
\end{equation}

\begin{equation} 
\label{eq:validator_reward2}
\begin{aligned}
Final_{all}^{\mathrm{pre}}
&=
\sum_{v\in\mathbb{V}^r}
\Phi_v\left(
1+\gamma_{validator}
\frac{N_{v,t}^{\mathrm{pre}}}
{\sum_{v'\in\mathbb{V}^r}N_{v',t}^{\mathrm{pre}}}
\right),\\
Final_{all}^{\mathrm{post}}
&=
\sum_{v\in\mathbb{V}^r}Final_v^{\mathrm{post}}.
\end{aligned}
\end{equation}

\begin{equation} 
\label{eq:validator_reward}
R_v =
\left[
\frac{Final_v^{\mathrm{pre}}}
{\delta+Final_{all}^{\mathrm{pre}}}
\left(1-\frac{N_{CF}}{N_{VA}}\right)
+
\frac{Final_v^{\mathrm{post}}}
{\delta+Final_{all}^{\mathrm{post}}}
\frac{N_{CF}}{N_{VA}}
\right]\cdot VR,
\end{equation}

Although the above design encourages active participation, it may also incentivize validators to deliberately trigger conflicts, thereby gaining additional time to complete more sub-tasks and earn higher rewards at the cost of increased consensus latency. To suppress this behavior, EDI uses the committee-confirmed consensus results of controversial sub-tasks and the complete on-chain submission history to identify and penalize submissions that deliberately trigger conflicts. A submission is deemed invalid if its declared input version matches the input version associated with the consensus result but its result fingerprint differs from the consensus fingerprint. The corresponding validator is penalized by burning part of its stake.
Eq.~\ref{eq:validator_punishment} gives the amount burned for each invalid submission in a round, where $\epsilon_{penalty}\in\mathbb{Z}^+$ is a predefined base penalty.

\begin{equation}
\label{eq:validator_punishment}
Punish = \epsilon_{penalty}+\frac{N_{CF}}{N_{VA}}\cdot VR.
\end{equation}

As shown in Eq.~\ref{eq:validator_punishment}, $Punish$ exceeds the post-conflict reward budget $\frac{N_{CF}}{N_{VA}}\cdot VR$, which upper-bounds the additional reward that any validator or colluding group can obtain by triggering conflicts. Moreover, $Punish$ increases with the number of conflict-related sub-tasks. Therefore, for any penalized validator and its colluding group, the additional reward obtained by triggering conflicts cannot offset the corresponding penalty, precluding a positive net payoff from such behavior. A sufficiently large penalty thus economically discourages validators from triggering conflicts to prolong consensus and obtain additional rewards.

\subsubsection{Discussion}
\label{sssec:discussion_on_TCC_EDI}
We finally conclude this section by discussing how TCC and EDI achieve a balance between efficiency, scalability, and security. 
TCC's core insight is that global consensus does not require every validator to verify every update. 
Through VA task decomposition and selective committee voting, TCC allows hardware-constrained honest validators to contribute by executing subsets of VA sub-tasks rather than the entire VA task.
This shifts the system scalability bottleneck from individual capabilities to the collective power of the validator set, enabling the system to scale to more workers (for larger volumes of private data to be used to improve customized model performance) as the validator scale grows rather than being limited by individual constraints. However, a standalone TCC might degenerate if malicious validators trigger excessive conflicts. 
EDI tackles this by burning the stake of each malicious validator whose submissions are identified as deliberately triggering conflicts. By setting the penalty above the additional reward obtainable from such conflicts, EDI discourages rational validators from  provoking conflict and thus limits the degradation of TCC's efficiency caused by conflict resolution. 
In \S\ref{sec:theoretical_analysis}, we will further show that, with appropriate settings of $N_t$, $N_c$, $E_c$, and $\epsilon_c$, TCC prevents the per-validator workload required to satisfy the global consensus conditions from growing linearly with the number of workers when EDI effectively deters conflicts, thereby improving participant scalability. The proposed design also bounds the success probabilities of fraud and poisoning attacks, providing formal security guarantees. 

\section{Theoretical Analysis}\label{sec:theoretical_analysis}
This section first analyzes \texttt{DICE}'s efficiency and security in \S\ref{ssec:efficiency} and \S\ref{ssec:security}, respectively, then discusses a fundamental trade-off revealed from these analyses in \S\ref{ssec:tradeoff}. 
In \S\ref{ssec:efficiency}--\S\ref{ssec:tradeoff}, we assume that all participants are active. 
\S\ref{ssec:crash_discussion} discusses the impacts of non-responsive participants. 
Finally, \S\ref{ssec:gas_cost_discussion} discusses \texttt{DICE}'s on-chain transaction costs. 

\subsection{Efficiency} \label{ssec:efficiency}
This section analyzes the efficiency of our \texttt{DICE} algorithm in terms of per-worker and per-validator computational complexities in each learning round. 
A lower per-participant complexity is preferred, as it enables a wider range of honest participants to contribute meaningfully. Moreover, it helps reduce the overall execution time due to the parallel execution of participants within each role. 

For ease of analysis, we suppose that (i) the size of the global model is $S_{GM}$; and (ii) in \texttt{DICE}, the minimum size of the local model is $\gamma_{LM} \cdot S_{GM}$, and the minimum size of the validation model is $\gamma_{VM} \cdot S_{GM}$ (where $0<\gamma_{VM}\leq\gamma_{LM}\leq1$). 
For comparison, we also analyze the current BCFL paradigm introduced in \S\ref{sec:intro} that requires each worker to train the full global model and each validator to verify all worker updates.

\textbf{Theorem 1 (Per-Participant Complexity in Current BCFL Paradigm).}
\textit{In any round $r$ of the current BCFL paradigm, the computational complexity of each honest worker is at least $\Omega(S_{GM})$, and the per-validator computational complexity required to satisfy the consensus conditions is at least $\Omega(|\mathbb{W}^r|S_{GM})$.}

\textbf{Proof.} 
The current BCFL paradigm requires every worker to train the full global model. Therefore, the computational complexity of each honest worker is at least $\Omega(S_{GM})$. Given that $|\mathbb{W}^r|$ workers submit updates of size $S_{GM}$ in round $r$, each honest validator must verify all updates and participate in cross-validator consensus to receive rewards. Since the current BCFL paradigm typically evaluates each worker update using a validation model of the same size as the global model, the per-validator computational complexity required to satisfy the consensus conditions is at least $\Omega(|\mathbb{W}^r|S_{GM})$.

We next analyze the per-participant complexity of \texttt{DICE}, focusing on the case where EDI effectively deters conflicts, i.e., $N_{CF}=0$. 

\textbf{Theorem 2 (Per-Participant Complexity in \texttt{DICE}).}
\textit{In any round $r$ of \texttt{DICE}, the computational complexity of each honest worker is at least $\Omega(\gamma_{LM}S_{GM})$, and the per-validator computational complexity required to satisfy the consensus conditions is at least $\Omega(N_t\gamma_{VM}S_{GM})$.}

\textbf{Proof.} In \texttt{DICE}, each honest worker trains a compact local model whose size is at least $\gamma_{LM}S_{GM}$. Hence, its computational complexity is at least $\Omega(\gamma_{LM}S_{GM})$. 
Given the updates submitted by workers in $\mathbb{W}^r$, \texttt{DICE}'s consensus algorithm divides the task of verifying and aggregating those updates into $N_{VA}$ sub-tasks. 
When $N_{CF}=0$, the global consensus conditions in \S\ref{sssec:consensus} require at least $N_c$ validators, each of whom submits results for at least $N_t$ distinct sub-tasks. Honest validators counted toward this requirement execute the corresponding sub-tasks according to the protocol. A VER sub-task evaluates a validation model of size at least $\gamma_{VM}S_{GM}$. An AGG sub-task processes the parameters of the corresponding sub-slice. For a given customization task with a fixed model architecture, the ratio of each nonempty sub-slice's parameter size to its corresponding validation model size has a positive constant lower bound. Thus, either type of sub-task requires $\Omega(\gamma_{VM}S_{GM})$ computation, yielding the per-validator lower bound $\Omega(N_t\gamma_{VM}S_{GM})$ required to satisfy the consensus conditions.

Theorems 1 and 2 characterize the lower bounds on the per-participant computational complexities under the current BCFL paradigm and DICE, reflecting their respective participation requirements. Compared with the current BCFL paradigm, \texttt{DICE} reduces the sizes of models that each participant must accommodate for completing a training or verification operation, thus having a greater potential to enable more honest Web3 participants to contribute to customization. 
More importantly, on the validator side, existing algorithms typically require each validator to verify all worker updates, while \texttt{DICE} allows validators to reach consensus on the complete VA results without individually executing all VA sub-tasks. By setting $N_t$ to scale sub-linearly with $|\mathbb{W}^r|$ (e.g., $N_t=\mathcal{O}(\log |\mathbb{W}^r|)$ or even $N_t=\mathcal{O}(1)$), \texttt{DICE} makes the lower bound on per-validator computational complexity scale sub-linearly with the number of workers, in contrast to the linear scaling of this bound under the current BCFL paradigm, thereby lowering the validator participation barrier and improving participant scalability. 
The choice of $N_t$ will be further discussed later in \S\ref{ssec:tradeoff}.

\textbf{Discussions.} The preceding analysis focuses on per-worker and per-validator computational complexities. Beyond these, two additional efficiency dimensions warrant consideration: the requester-side computation overhead, and the system-level communication and coordination complexities. We discuss each below.

\textit{Requester-side Computation Overhead.}
The model requesters in prior BCFL algorithms~\cite{9293091,9833437,10.1145/3524104,nguyen2021federated,zhu2023blockchain,app14209459,li2022blockchain} are typically not involved in the iterative training and VA process after deploying smart contracts for their customization tasks, since those algorithms do not consider asymmetric hardware conditions across participants. In contrast, \texttt{DICE} is designed for resource-asymmetric scenarios and aims to fully exploit requester-side resources to enable lightweight training and verification for hardware-constrained workers and validators. 

The requester in \texttt{DICE} undertakes slice selection and large-to-compact model derivation. 
During slice selection, the requester computes importance scores for candidate slices with complexity $\mathcal{O}(\sum_{g=1}^{G} N_{\text{slice}, g} \cdot S_{GM})$, where $G$ ($G=1$ under U-DPL mode) is the number of worker groups with workers in the same group sharing the same hardware capacity indicator $L_w$, and $N_{\text{slice},g}$ is the number of candidate slices for group $g$. 
During large-to-compact model derivation, the requester generates emulators for both workers and validators with complexity $\mathcal{O}(N_{\text{emulator}} \cdot S_{GM})$, where $N_{\text{emulator}}$ is the number of emulators (each corresponding to a distinct slice in $\{\mathbb{M}_w^r|w\in\mathbb{W}^r\}\cup \{\mathbb{LUS}_j | j\in [0, N_{lus})\} $). 

These requester-side costs are deliberate trade-offs for enabling broader participation in customizing large PMs in Web3. The requesters typically have sufficient hardware resources to perform the above computations, and in return, can obtain high-quality customized models and the associated economic benefits. Moreover, the requester can periodically select slices and generate emulators (as discussed in \S\ref{sssec:runtime_mechanism}), reducing computation costs over multiple rounds. Further optimization of requester-side costs via more efficient slice evaluation, model distillation, or caching strategies is a promising direction for future work.

\textit{Communication and Coordination Complexities.} 
In addition to computation, communication and coordination complexities also affect the overall efficiency of BCFL systems. \texttt{DICE} coordinates all participants via a single smart contract and stores large-scale data (e.g., individual worker updates, emulators, aggregated updates, global model parameters) in an additional distributed storage system like IPFS, as introduced in \S\ref{sssec:cc_model}. Therefore, the total communication and coordination costs comprise two parts: on-chain interactions and off-chain data transfers.

The on-chain interaction complexity depends on the data size of each submission to the smart contract, the number of submissions each participant makes per round and the number of participants.  
\texttt{DICE} controls the first factor by recording only necessary data such as cryptographic hashes and file handles on the blockchain. Therefore, our analysis below focuses on measuring on-chain interaction complexity by the latter two factors. 
The requester and each worker submit at most once per round. 
The requester publishes the model-slicing results and the file handles and hashes of the corresponding emulators, while each worker submits the file handle and hash of its worker update. Therefore, the on-chain interaction complexities of the requester and each worker are both $\mathcal{O}(1)$. 
Each validator interacts with the smart contract in two states. During \texttt{VA\_COMMIT}, a validator submits fingerprints and commitments for the sub-tasks it executes, requiring at most $\mathcal{O}(\kappa_v^rN_{v,t})$ submissions. Here, $N_{v,t}$ is the number of sub-tasks executed by validator $v$ in this state, and $\kappa_v^r$ is the maximum number of fingerprint-and-commitment submissions made by $v$ for one sub-task in round $r$, including resubmissions that reuse existing results during conflict resolution. During \texttt{VA\_REVEAL}, each validator packs its committed but unrevealed results by sub-task and input version into one submission. Therefore, the per-validator on-chain interaction complexities are $\mathcal{O}(\kappa_v^rN_{v,t})$ during \texttt{VA\_COMMIT} and $\mathcal{O}(1)$ during \texttt{VA\_REVEAL}. Overall, the total on-chain interaction complexity per round is $\mathcal{O}\left(1+|\mathbb{W}^r|+\sum_{v\in\mathbb{V}^r}(1+\kappa_v^rN_{v,t})\right)$, where the three terms correspond to the requester, all active workers, and all active validators, respectively.

Notably, different from some BCFL algorithms \cite{10.1145/3524104,11073628,9293091} in which each validator submits only once per round, validators in \texttt{DICE} submit VA results in two states: multiple times during \texttt{VA\_COMMIT} and once during \texttt{VA\_REVEAL}. This design is a deliberate trade-off: using more frequent but lightweight on-chain submissions in exchange for lower per-validator computational complexity bound and security against fraud attacks (proved in Theorem 3). This trade-off is favorable under the consideration that off-chain computation is the primary barrier for hardware-constrained validators in customizing large PMs. Moreover, validators can batch their commit-phase submissions to further reduce the on-chain interaction overhead in practice.

The off-chain data transfer costs are incurred when participants access the distributed storage system. Ignoring repeated downloads, the requester downloads the latest global model and uploads the emulators in a round when the model slicing or emulator generation is triggered, with data transfer volume $\mathcal{O}(S_{GM}+\sum_{k=1}^{N_{\mathrm{emulator}}}S_{EM,k})$, where $N_{\mathrm{emulator}}$ is the number of emulators uploaded in that round and $S_{EM,k}$ is the size of the $k$-th emulator. 
Let $S_{LM,w}$ denote the size of the local model deployed by worker $w$, $S_{v,t}$ denote the total size of the models, parameters, and off-chain VA results transferred when validator $v$ executes sub-task $t$, and $\mathbb{T}_v^r$ the set of sub-tasks processed by $v$ in round $r$. Each worker downloads the global parameters of its assigned slice and the corresponding emulator and uploads its worker update after local training, resulting in transfer volume $\mathcal{O}(S_{LM,w})$. Each validator downloads the emulators and worker updates required by its sub-tasks and uploads the corresponding VA results, resulting in transfer volume $\mathcal{O}(\sum_{t\in\mathbb{T}_v^r}\kappa_v^rS_{v,t})$. The total off-chain data transfer volume per round is therefore $\mathcal{O}\left(S_{GM}+\sum_{k=1}^{N_{\mathrm{emulator}}}S_{EM,k} +\sum_{w\in\mathbb{W}^r}S_{LM,w} +\sum_{v\in\mathbb{V}^r} \sum_{t\in\mathbb{T}_v^r}\kappa_v^rS_{v,t} \right)$. These large data objects are shared through a cost-effective distributed storage system such as IPFS, reducing on-chain storage and communication costs. In addition, reusing emulators across rounds, as discussed in \S\ref{sssec:runtime_mechanism}, can further reduce off-chain communication and coordination costs in the corresponding rounds.

\subsection{Security}
\label{ssec:security}

Similar to the existing BCFL design \cite{lin2024refiner}, \texttt{DICE} relies on validators to verify workers according to the prescribed rules, filter invalid worker updates, aggregate qualified updates, and reach consensus on the VA results used to update the global model. Security therefore requires preventing malicious validators from both copying others' VA results to fraudulently claim rewards and causing poisoned VA results to be adopted as consensus results. 
Theorems 3--6 show that TCC bounds the success probabilities of fraud and poisoning attacks while reducing the validator participation barrier. The probability analyses below focus on individual sub-tasks within a given learning round and rely on the following assumptions.

\textbf{Assumption 1 (VA Result Unpredictability and Hash Function Conditions).}
\textit{Given any probability threshold $\epsilon_p\in(0,1)$, there exists a specific design for each VA sub-task's result set and a cryptographic hash function $H(\cdot)$ satisfying $2^{-ent}+2^{-hlen}<\epsilon_p$, where $2^{-ent}$ is an upper bound on the probability that a malicious validator obtains another validator $v$'s complete VA result $\mathbb{VAR}_v$ over all computational attempts made before the end of \texttt{VA\_COMMIT}, with a total computational cost substantially lower than that of executing the prescribed computation, 
$hlen$ is the output bit length of $H(\cdot)$, and its output is modeled as a uniformly random $hlen$-bit value for any input not queried by the attacker.}

We treat $\epsilon_p$ as a security parameter fixed at each customization task initialization and use $\epsilon_p=10^{-12}$ as the target upper bound on attack success probabilities in the following analyses. 

\textbf{Designs Supporting  Assumption 1.} To support Assumption 1, each VA sub-task's complete result set includes computation results disclosed only during \texttt{VA\_REVEAL}. For VER sub-tasks, those results include validation outputs; for AGG sub-tasks, they include digests computed over the child updates and aggregated update, including a hash chain following a predefined layer order. 
Numerical VA results are represented using data types with a large representable range (e.g., uint256) and high numerical precision (e.g., 18 decimal places). Collectively, the above designs aim to make it difficult to reconstruct another validator's complete result set from available information at a substantially lower computational cost than generating it through the prescribed computation. 
Moreover, \texttt{DICE} adopts SHA-256 as the cryptographic hash function, whose 256-bit output satisfies $2^{-256}<\epsilon_p/2$ for the probability thresholds considered.

\textbf{Assumption 2 (Validator Population and Stake Conditions).}
\textit{In any learning round, the validator population and total stake satisfy
$|\mathbb{V}_{all}|>2\left\lceil \log_{1-\frac{N_t}{(2|\mathbb{W}^r|-1)N_{lus}}} \left(\frac{\epsilon_p}{K_{\text{check}}}\right) \right\rceil$ and $E_{total}\gg E_c$. The fraction of the total stake held by honest validators, denoted by $h$, satisfies $h>\frac{2}{3}$. Let $E_{\min}$ and $E_{avg}$ denote the minimum and average validator stakes, respectively; then $\frac{E_{avg}}{E_{\min}}\leq\frac{3}{2}$.}

\textbf{Designs Supporting  Assumption 2.} To support Assumption 2, \texttt{DICE} incorporates mechanisms concerning stake requirements, validator population, and economic incentives. 

First, regarding stake requirements, \texttt{DICE} admits only validators that stake between $E_{\min}$ and $\frac{3}{2}E_{\min}$ basic units of cryptocurrency as a security deposit for a PM customization task, where $E_{\min}=\max(E_c,\epsilon_{penalty}+VR)$. During task execution, any validator whose stake falls outside the range $[E_{\min},\frac{3}{2}E_{\min}]$, for example, as a result of penalties, is removed. This design enforces the bounded stake ratio $\frac{E_{avg}}{E_{\min}}\leq\frac{3}{2}$, limiting stake disparity among individual validators. It also supports deriving a lower bound on the number of honest validators from the honest stake ratio, thereby controlling the validator population required by Theorem~4.

Second, regarding the validator population, \texttt{DICE} recruits at least
$\max\!\left(100,3\left\lceil\log_{\left[1-\frac{N_t}{(2|\mathbb{W}^r|-1)N_{lus}}\right]}\frac{\epsilon_p}{K_{\text{check}}}\right\rceil\right)$ validators for each PM customization task and allows new validators to register during task execution. The above stake requirements also apply to these validators. If a validator registers in round $r$, its public key and stake weight become effective from round $r+2$ onward. This delayed activation prevents newly registered validators from affecting the secrecy and randomness of committee election and voting-power determination. 
\texttt{DICE} maintains the active validator set on-chain. This set is initialized with the validators recruited before task execution and updated at the end of each round according to the participation records, stake requirements, and activation rounds of newly registered validators. If the updated set contains fewer validators than the above threshold, \texttt{DICE} pauses the next round and reopens validator registration to replenish the set. As an exception to the above delayed-activation rule, validators registered during this pause and their stake weights become effective immediately and are eligible for committee election in the next resumed round. Once the required validator population is restored, \texttt{DICE} first fixes the public keys and stake weights eligible for committee election. The model requester then releases the election seed, after which algorithm execution resumes. Together, the stake requirements and validator population requirement ensure a sufficiently large total stake. In addition, $E_{total} \gg E_c$ and $\frac{E_{avg}}{E_{min}} \leq \frac{3}{2}$ make it more difficult for malicious validators to control more than one-third of the total stake, thus promoting $h>\frac{2}{3}$.

Finally, regarding economic incentives, \texttt{DICE} locks validators' stakes and rewards to encourage honest participation and further promote $h>\frac{2}{3}$. Specifically, suppose that validator $v$ stakes cryptocurrency or earns a positive reward in round $r$ of a PM customization task $\mathrm{T}$. \texttt{DICE} locks these tokens from round $r$ until $v$ completes at least $N_t\cdot\epsilon_{lock}$ new VA sub-tasks after round $r$, either in task $\mathrm{T}$ or in subsequent customization tasks, without being detected as malicious, where $\epsilon_{lock}\in\mathbb{Z}^{+}$. 
Under this design, validators, particularly those holding a large share of the total stake, have a stronger incentive to remain honest rather than deviate from the algorithm. Honest validators receive rewards for producing correct VA results. In contrast, deviation may cause the model requester to cease submitting new customization tasks if the resulting customized model fails to meet expectations. Previously staked tokens may consequently remain locked indefinitely without corresponding returns and may even be burned if malicious behavior is detected. These reward and withdrawal constraints increase the economic risk of deviation, thereby encouraging rational validators to participate honestly and helping sustain the condition $h>\frac{2}{3}$ in practice.

Under the above assumptions, Theorem 3 presents  \texttt{DICE}'s resistance to fraud attacks, while Theorems 4--6 and their corollaries show \texttt{DICE}'s resistance to poisoning attacks.

\textbf{Theorem 3 (Resistance to Fraud Attacks).}
\textit{Under Assumption 1, given any $\epsilon_p\in(0,1)$, the probability that, for any given sub-task and its input version, a malicious validator $v'$, operating at a total computational cost substantially lower than that of executing the prescribed computation, computes and submits $CC_{v'}=H(\mathbb{VAR}_v\parallel v')$ using results copied from another honest validator $v\neq v'$ during \texttt{VA\_COMMIT} is smaller than $\epsilon_p$, where $\mathbb{VAR}_v$ is the complete VA result produced by $v$ for that input version.}

\textbf{Proof.}
Given a VA sub-task and one of its input versions, consider a malicious validator $v'$ operating at a total computational cost substantially lower than that of executing the prescribed computation. Let $\mathcal{A}$ be the event that $v'$ obtains the corresponding complete result $\mathbb{VAR}_v$ over all computational attempts made before the end of \texttt{VA\_COMMIT}, and let $\mathcal{S}$ be the event that $v'$ submits the correct commitment. 
By Assumption 1, $\Pr(\mathcal{A})\leq2^{-ent}$. 
If $\mathcal{A}$ does not occur, the attacker needs to guess an unknown $hlen$-bit output. Since the on-chain procedure accepts only one commitment from each validator for a given sub-task and input version,
$\Pr(\mathcal{S}\cap\overline{\mathcal{A}})\leq2^{-hlen}$ under the hash output property in Assumption 1. 
Therefore, $\Pr(\mathcal{S}) = \Pr(\mathcal{S}\cap\mathcal{A}) + \Pr(\mathcal{S}\cap\overline{\mathcal{A}}) \leq2^{-ent}+2^{-hlen}<\epsilon_p$. 

Since submitting the correct commitment is necessary for copied results to pass the checks during \texttt{VA\_REVEAL}, Theorem 3 indicates that the fraud attack for any given sub-task and its input version succeeds with probability smaller than $\epsilon_p$.

\textbf{Theorem 4 (Coverage of Honest Validator Results on Conflict-Unrelated Sub-Tasks).}
\textit{Under Assumption 2, given any $\epsilon_p\in(0,1)$ and $N_t\in\mathbb{Z}^{+}$ with $N_t<N_{VA}$, let $N_c=\left\lceil\frac{|\mathbb{V}_{all}|}{2}+\log_{1-\frac{N_t}{(2|\mathbb{W}^r|-1)N_{lus}}}\left(\frac{\epsilon_p}{K_{\text{check}}}\right)\right\rceil$. Then, for any given sub-task, the probability that global consensus is reached and the sub-task is conflict-unrelated at that time, yet no honest validator has executed it according to the protocol and submitted the corresponding fingerprint and commitment, is smaller than $\epsilon_p$.}

\textbf{Proof.}
Given any VA sub-task $t$, let $\mathcal{F}_t$ denote the event that, when global consensus is reached, $t$ is conflict-unrelated but no honest validator has executed it for its current valid input version and submitted the corresponding fingerprint and commitment. Let $\mathcal{A}_i$ denote the event that the $i$-th supplementary consensus check is initiated, its conflict-unrelated sub-task sampling domain contains $t$, and none of its honest checking members samples $t$. Let $\mathcal{H}_i$ denote the execution history before this sampling, and let $\mathbb{V}_{h,i}$ denote the honest checking members. 
Under Assumption 2, the number of malicious validators satisfies $N_{\mathrm{malicious}}\leq\frac{(1-h)E_{avg}|\mathbb{V}_{all}|}{E_{\min}}<\frac{|\mathbb{V}_{all}|}{2}$. Hence, $|\mathbb{V}_{h,i}|\geq N_c-N_{\mathrm{malicious}}>N_c-\frac{|\mathbb{V}_{all}|}{2}$. 
Conditioned on $\mathcal{H}_i$, the available checking members and conflict-unrelated sub-task sampling domain have been fixed, and the members sample independently. Since $N_{VA}\leq(2|\mathbb{W}^r|-1)N_{lus}$, we have Eq.~\ref{eq:poison_attack_analysis2}. 
\begin{equation}
\label{eq:poison_attack_analysis2}
\begin{aligned}
\Pr(\mathcal{A}_i\mid\mathcal{H}_i)
&\leq
\left(
1-
\frac{
\left\lceil
\frac{N_{VA}-N_{CF}}{N_{VA}}N_t
\right\rceil
}{
N_{VA}-N_{CF}
}
\right)^{|\mathbb{V}_{h,i}|}\\
&<
\left(
1-
\frac{N_t}{(2|\mathbb{W}^r|-1)N_{lus}}
\right)^{
N_c-\frac{|\mathbb{V}_{all}|}{2}
}
\leq
\frac{\epsilon_p}{K_{\text{check}}}.
\end{aligned}
\end{equation}
This bound applies whenever the corresponding check is initiated and its conflict-unrelated sub-task sampling domain contains $t$; otherwise, $\mathcal{A}_i$ does not occur. Averaging over the execution histories therefore gives $\Pr(\mathcal{A}_i)<\frac{\epsilon_p}{K_{\text{check}}}$. 
Under the rules of \texttt{SupplementaryCheck}() introduced in \S\ref{sssec:consensus}, if $\mathcal{F}_t$ occurs, then no honest member can have sampled $t$ in the supplementary check that ultimately passed. Thus, $\mathcal{F}_t\subseteq\bigcup_{i=1}^{K_{\text{check}}}\mathcal{A}_i$. By the union bound, $\Pr(\mathcal{F}_t)\leq\sum_{i=1}^{K_{\text{check}}}\Pr(\mathcal{A}_i)<\epsilon_p$.

Theorem 4 indicates that \texttt{DICE} can provide sufficient coverage of honest validator results on conflict-unrelated sub-tasks, thereby supporting the correctness of consensus results adopted by those sub-tasks. This directly yields the following corollaries. 

\textbf{Corollary 1 (Poisoning a Non-Root Sub-Task Triggers Conflict).}
\textit{Under the same conditions as Theorem 4, for any given non-root VA sub-task, with probability greater than $1-\epsilon_p$, if global consensus is reached and a malicious validator has submitted a poisoned result for the input version ultimately adopted by that sub-task, then the sub-task has become conflict-related. Accordingly, the probability that global consensus is reached while the sub-task remains conflict-unrelated and adopts the poisoned result is smaller than $\epsilon_p$.}

\textbf{Corollary 2 (Correctness of Non-Root Conflict-Unrelated Sub-Tasks).}
\textit{Under the same conditions as Theorem 4, for any given non-root VA sub-task, with probability greater than $1-\epsilon_p$, if global consensus is reached while the sub-task remains conflict-unrelated, its adopted consensus result equals the result computed by honest validators according to the protocol for the same valid input version.}

Then, we analyze the correctness of consensus results adopted by conflict-related and root sub-tasks.

\textbf{Theorem 5 (Correctness of Controversial Sub-Tasks).}
\textit{Under Assumption 2, given any $\epsilon_p\in(0,1)$, there exist committee election parameters $E_c$ and $\epsilon_c$ such that, for any given VA sub-task, the probability that global consensus is reached, the sub-task is controversial at that time, and a poisoned result is adopted as its consensus result is smaller than $\epsilon_p$. Accordingly, with probability greater than $1-\epsilon_p$, if global consensus is reached and the sub-task is controversial, its consensus result equals the result computed by honest validators according to the protocol for the same valid input version.}

\textbf{Proof.}
Given any VA sub-task $t$, honest committee members execute $t$ and all its descendants according to \texttt{DICE}'s voting rules for controversial sub-tasks, compute the result according to the protocol for the same valid input version, and vote for the corresponding fingerprint. Let $T=\lfloor E_c\epsilon_c\rfloor+1$. For any committee election associated with $t$, adopting a poisoned result requires at least $T$ malicious votes. 

Let $e_{\mathrm{bad}}$ and $e_{\mathrm{good}}$ denote the numbers of malicious and honest votes in a committee election, respectively. According to \texttt{DICE}'s committee selection rules, the expected values of $e_{\mathrm{bad}}$ and $e_{\mathrm{good}}$ are $\mathbb{E}[e_{\mathrm{bad}}]=(1-h)E_c$ and $\mathbb{E}[e_{\mathrm{good}}]=hE_c$. Choose $1-h<\epsilon_c<h$. Then, by the Chernoff bounds for sums of independent Bernoulli random variables, we have $\Pr(e_{\mathrm{bad}}\geq T)\leq\exp\left(-\frac{E_c(\epsilon_c+h-1)^2}{\epsilon_c+1-h}\right)$ and $\Pr(e_{\mathrm{good}}<T)\leq\exp\left(-\frac{E_c(h-\epsilon_c)^2}{2h}\right)$. It suffices to choose $E_c>\ln\frac{2K_{\text{election}}}{\epsilon_p}\max\left\{\frac{\epsilon_c+1-h}{(\epsilon_c+h-1)^2},\frac{2h}{(h-\epsilon_c)^2}\right\}$. Then, both tail probabilities are smaller than $\epsilon_p/K_{\text{election}}$. 
Since $h>\frac{2}{3}$, setting $\epsilon_c=\frac{1}{2}$ gives $\frac{\epsilon_c+1-h}{(\epsilon_c+h-1)^2}<30$ and $\frac{2h}{(h-\epsilon_c)^2}<48$. Hence, one valid choice is $\epsilon_c=\frac{1}{2}$ and $E_c=\left\lfloor48\ln\frac{2K_{\text{election}}}{\epsilon_p}\right\rfloor+1$.

If global consensus is reached while $t$ is controversial and a poisoned result is adopted, the malicious vote count must reach $T$ in at least one of the committee elections associated with $t$. By a union bound over at most $K_{\text{election}}$ elections, the probability of this event is smaller than $K_{\text{election}}\cdot\bigl(\epsilon_p/K_{\text{election}}\bigr)=\epsilon_p$. Consequently, with probability greater than $1-\epsilon_p$, if global consensus is reached while $t$ is controversial, its adopted consensus result equals the result computed by honest validators according to the protocol for the same valid input version.

From Theorem 5, we can derive the correctness of consensus results adopted by conflict-related but non-controversial sub-tasks. 
Under the eligibility rules of \texttt{Voting}(), an honest committee member for a controversial sub-task must execute that sub-task and all its descendants. If the malicious vote count remains below the adoption threshold, any adopted fingerprint must receive an honest member's vote. The corresponding sub-task and its descendants therefore receive submissions from honest validators. By \texttt{CheckConflict}(), any descendant that remains non-controversial at global consensus adopts a result consistent with those submitted by honest validators for the same valid input version. 
Applying Theorem 5's committee voting analysis with threshold $\epsilon_p/N_{VA}$ to the controversial ancestors of a given sub-task, and using the fact that their number is at most $N_{VA}$, yields the following corollary.

\textbf{Corollary 3 (Correctness of Conflict-Related but Non-Controversial Sub-Tasks).}
\textit{Under the same conditions as Theorem 5, given any $\epsilon_p\in(0,1)$, there exist committee election parameters $E_c$ and $\epsilon_c$ such that, for any given VA sub-task, with probability greater than $1-\epsilon_p$, if global consensus is reached while the sub-task is conflict-related but not controversial, its adopted consensus result equals the result computed by honest validators according to the protocol for the same valid input version.}

Combining Theorems 4 and 5 with Corollary 3 yields an honest-validator result coverage guarantee for any given sub-task, whether conflict-unrelated or conflict-related. 

\textbf{Corollary 4 (Coverage of Honest Validator Results for Any Sub-task).}
\textit{Under the same conditions as Theorems 4 and 5, given any $\epsilon_p\in(0,1)$, there exist parameters $N_c$, $E_c$, and $\epsilon_c$ such that, for any given VA sub-task, with probability greater than $1-\epsilon_p$, if global consensus is reached, at least one honest validator has executed the sub-task according to the protocol and submitted the corresponding fingerprint and commitment.}

Specifically, applying Theorem 4 with threshold $\epsilon_p/2$ bounds the probability that no honest validator submits a result in the conflict-unrelated case by $\epsilon_p/2$. In the conflict-related case, applying the committee voting analysis with threshold $\epsilon_p/(2N_{VA})$ and taking a union bound over the given sub-task and its ancestors bounds the corresponding failure probability by $\epsilon_p/2$.

\textbf{Theorem 6 (Correctness of Non-Controversial Root Sub-Tasks).}
\textit{Under Assumption 2, given any $\epsilon_p\in(0,1)$, there exist committee election parameters $E_c$ and $\epsilon_c$ such that, for any given root sub-task, the probability that global consensus is reached, the root remains non-controversial, and a poisoned result is adopted as its consensus result is smaller than $\epsilon_p$. Accordingly, with probability greater than $1-\epsilon_p$, if global consensus is reached while the root remains non-controversial, its adopted consensus result follows the protocol for the corresponding valid input version.}

\textbf{Proof.}
Given any root sub-task $t$, let $\mathcal{E}_{t,i}$ denote the event that the malicious vote count reaches $\lfloor E_c\epsilon_c\rfloor+1$ in the $i$-th committee election for $t$; if that election does not occur, neither does $\mathcal{E}_{t,i}$. 
After satisfying the voting eligibility requirements, each honest validator computes the root result according to the protocol for the corresponding valid input version and votes for its fingerprint. Hence, a poisoned result can be adopted only if the malicious vote count reaches the threshold in the corresponding committee. Let $\mathcal{B}_t$ denote the event that global consensus is reached while $t$ remains non-controversial and adopts a poisoned result. Then, $\mathcal{B}_t\subseteq\bigcup_{i=1}^{K_{\text{election}}}\mathcal{E}_{t,i}$. The parameter setting in Theorem 5, i.e., $\epsilon_c=\frac{1}{2}$ and $E_c=\left\lfloor48\ln\frac{2K_{\text{election}}}{\epsilon_p}\right\rfloor+1$, gives $\Pr(\mathcal{B}_t)\leq\sum_{i=1}^{K_{\text{election}}}\Pr(\mathcal{E}_{t,i})<\epsilon_p$. 
Consequently, with probability greater than $1-\epsilon_p$, if global consensus is reached while the root remains non-controversial, its adopted consensus result follows the protocol for the corresponding valid input version.

Theorems 4--6 and the preceding corollaries together yield \texttt{DICE}'s poisoning resistance guarantee. For any given VA sub-task $t$, apply Theorem 4 with threshold $\epsilon_p/2$. Moreover, for each relevant sub-task, choose the committee election parameters so that the probability of malicious members' votes exceeding the threshold $E_c \epsilon_c$ in any of its at most $K_{\text{election}}$ committee elections is smaller than $\epsilon_p/(2N_{VA})$. 
Then, if $t$ is a conflict-unrelated non-root sub-task, Theorem 4 and Corollary 2 bound the probability of adopting poisoned results. 
If $t$ is a controversial or root sub-task, adopting a poisoned result requires malicious votes to exceed the threshold in a corresponding committee, according to Theorems 5--6.
If $t$ is conflict-related but not controversial, such an event is required to occur in a controversial ancestor of $t$, according to Corollary 3. 
Since the total number of sub-tasks comprising $t$ and its ancestors is at most $N_{VA}$, a union bound gives a probability smaller than $\epsilon_p/2+N_{VA}\epsilon_p/(2N_{VA})=\epsilon_p$ that global consensus is reached and $t$ adopts a poisoned result. 
Therefore, we have the following corollary. 

\textbf{Corollary 5 (Resistance to Poisoning Attacks).}
\textit{Under the same conditions as Theorems 4--6, given any $\epsilon_p\in(0,1)$, there exist parameters $N_c$, $E_c$, and $\epsilon_c$ such that, for any given VA sub-task in any round of \texttt{DICE}, with probability greater than $1-\epsilon_p$, if global consensus is reached, the sub-task's adopted consensus result follows the protocol for its corresponding valid input version.}

One valid parameter setting is $\epsilon_c=\frac{1}{2}$, $E_c=\left\lfloor48\ln\frac{4K_{\text{election}}N_{VA}}{\epsilon_p}\right\rfloor+1$, and $N_c=\left\lceil\frac{|\mathbb{V}_{all}|}{2}+\log_{1-\frac{N_t}{(2|\mathbb{W}^r|-1)N_{lus}}}\left(\frac{\epsilon_p}{2K_{\text{check}}}\right)\right\rceil$, with the validator population and stake requirements configured accordingly to satisfy Assumption 2 at the allocated probability thresholds. 

Collectively, the efficiency analysis in \S\ref{ssec:efficiency}, Theorem 3, and Corollary 5 show that TCC lowers the validator participation barrier while bounding the success probabilities of fraud and poisoning attacks, thereby providing security guarantees for the overall customization process.

\subsection{Per-Validator Workload Reduction and Scalability Gains versus System Overhead} \label{ssec:tradeoff}

Theorems 2 and 4 reveal a trade-off between two desirable but conflicting goals. We now discuss this trade-off and navigate it by setting $N_t$. 

\begin{itemize}
    \item \textbf{Goal 1: Reduce the per-validator workload required by the consensus conditions (favoring smaller $N_t$ as shown in Theorem 2)}. This workload affects: 
    (i) \textit{Individual validator overhead and participation barrier}: Lower workload makes it easier for honest participants to contribute as validators. 
    (ii) \textit{Cross-validator consensus latency}: Since validators work in parallel and pipeline off-chain computation of VA sub-tasks without waiting for each sub-task's on-chain confirmation, this latency is dominated by the local computation time of each validator $v$, characterized by $N_{v,t}\overline{t}_v$, where $\overline{t}_v$ is its average execution time per sub-task. Reducing the required workload can therefore shorten consensus latency. 
    (iii) \textit{System scalability}: The scalability with respect to both workers and validators can be improved significantly when decoupling the per-validator workload requirement from worker scale by setting $N_t$ to $\mathcal{O}(1)$. 
    
    \item \textbf{Goal 2: Reduce the required validator scale (favoring larger $N_t$, as quantified in Theorem 4)}. Reducing $N_t$ increases the number of validators required for security ($N_c$).  A larger $N_c$ brings three issues: 
    (i) \textit{Higher recruitment difficulty}: In ecosystems with limited participants, recruiting $N_c$ or more validators may become infeasible. 
    (ii) \textit{Increased system-level overhead}: 
    Since the complexities of on-chain interaction and off-chain data transfer in each round are both positively correlated with $|\mathbb{V}^r|$ as analyzed in \S\ref{ssec:efficiency} and $N_c$ is configured to satisfy $N_c \leq |\mathbb{V}^r|$, a larger $N_c$ could increase the system coordination and communication overhead. 
    (iii) Potential offset of latency benefit. 
    Recruiting more validators may involve slower participants with larger $\overline{t}_v$, potentially increasing the maximum $N_{v,t} \times \overline{t}_v$ among all validators and thereby offsetting some latency gains from reducing $N_t$. 
\end{itemize}

The trade-off between these two goals reflects a common pattern in distributed systems: reducing per-node workload to improve individual efficiency and support larger-scale lower-barrier participation often incurs higher system-level burden \cite{hamann2021scalability,carlini2025dynamic}. 
Prior BCFL systems did not encounter this trade-off since they retain the requirement that each validator process all worker updates~\cite{9293091,10.1145/3524104,9833437,nguyen2021federated,zhu2023blockchain,app14209459,li2022blockchain}, leaving a critical gap: how to scale customization to hundreds or more workers without assuming all validators possess sufficient hardware for processing worker updates.

\begin{figure}
    \centering
    \includegraphics[width=0.48\columnwidth]{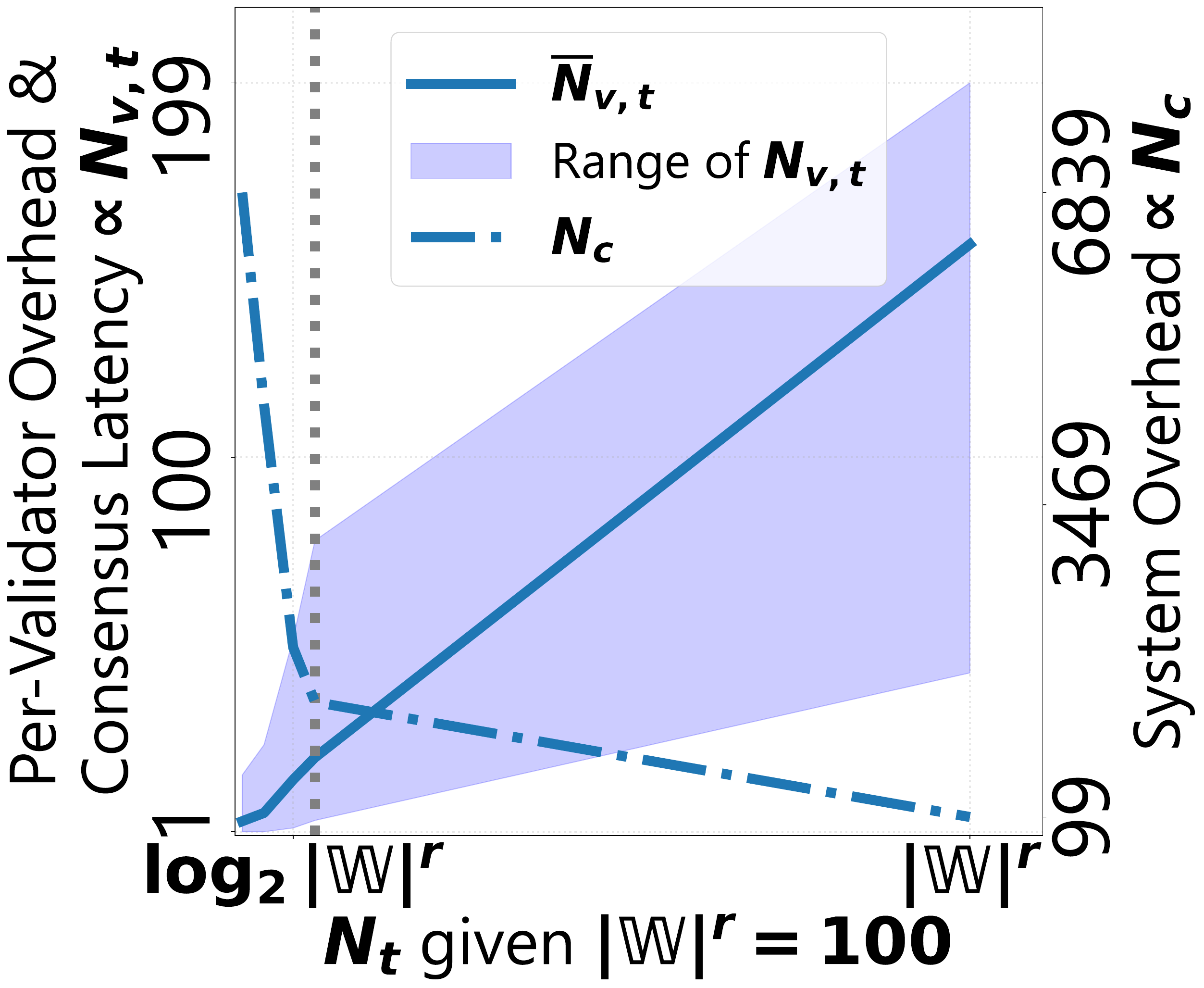}
    \includegraphics[width=0.48\columnwidth]{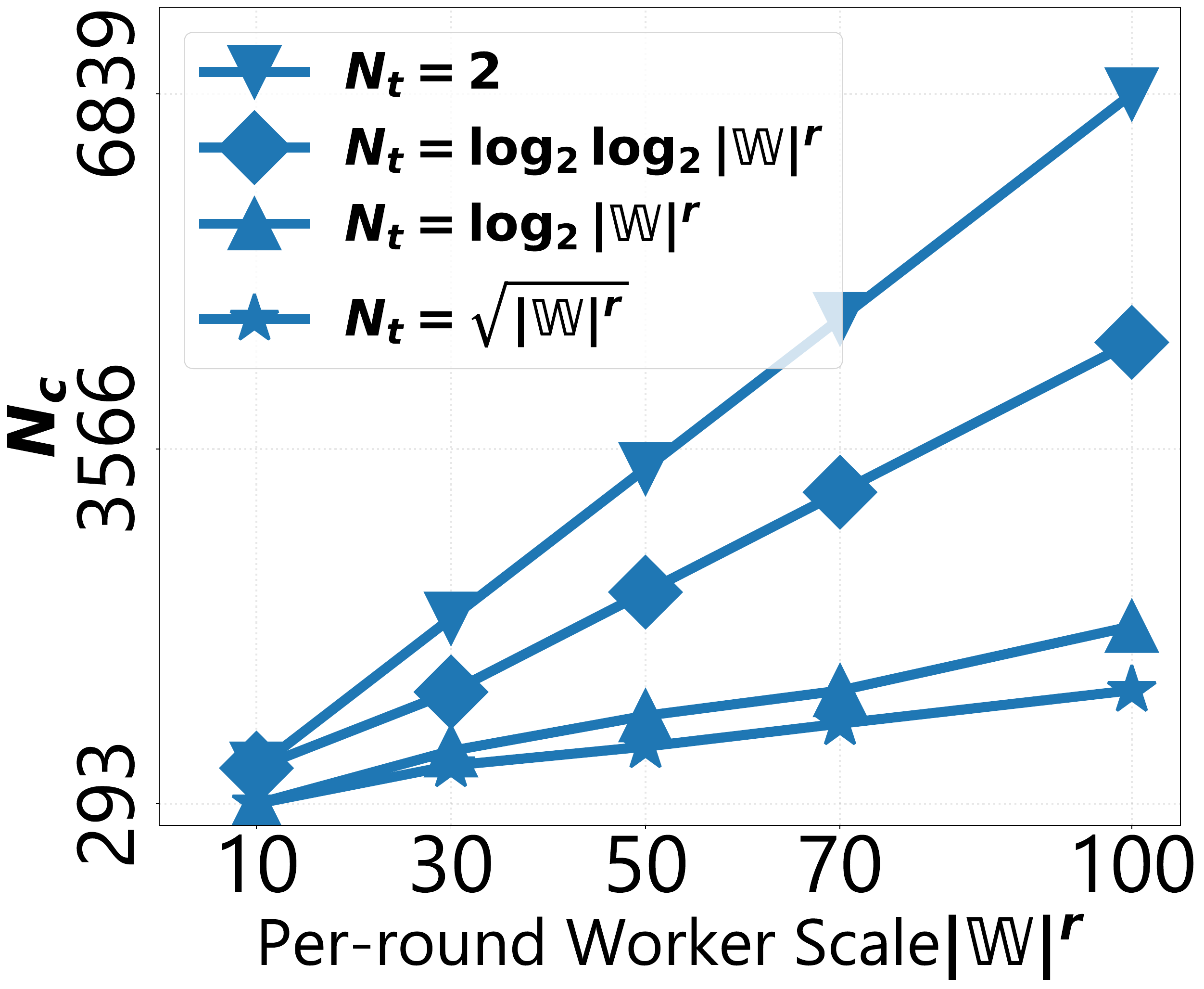}
    \caption{Effects of varying $N_t$, given $\epsilon_p=10^{-12}$, $N_{lus}=1$, $K_{\text{check}}=1$, $|\mathbb{V}_{all}|=\max(100, 3\lceil \log_{[1 - \frac{N_t}{(2|\mathbb{W}^r|-1)N_{lus}}]} \frac{\epsilon_p}{K_{\text{check}}} \rceil)$, and $N_c=\left\lceil\frac{|\mathbb{V}_{all}|}{2}+\log_{1-\frac{N_t}{(2|\mathbb{W}^r|-1)N_{lus}}}\left(\frac{\epsilon_p}{K_{\text{check}}}\right)\right\rceil$. }
    \label{fig:N_c_N_t}
\end{figure}

\texttt{DICE} navigates this trade-off by using $N_t$ as a tuning knob between the above goals.
Fig.~\ref{fig:N_c_N_t}(left) simulates the effects of varying $N_t$  under fixed worker scale. 
Fig.~\ref{fig:N_c_N_t}(right) depicts the growth of $N_c$ with increasing workers under different $N_t$. 
For most scenarios, we recommend $N_t = \lceil \log_2 |\mathbb{W}^r| \rceil$ as a balanced choice, which achieves: (i) substantially lower per-validator complexity bound than the current BCFL paradigm (from linear to \textit{sub-linear} in $|\mathbb{W}^r|$), reduced participation barrier, and improved scalability; (ii) \textit{moderate} growth rate on required validator scale and \textit{manageable} system overhead; (iii) \textit{provable} security. 
\texttt{DICE} offers users the flexibility to adapt $N_t$ to actual deployment constraints such as available participant scale, tolerable system overhead, security requirements, and individual resource budgets.

\subsection{Impact of Non-responsive Participants}\label{ssec:crash_discussion}

Below, we discuss the impact of non-responsive participants.

\subsubsection{Impacts on Liveness}
The customization task may stall in \texttt{TRAINING}, \texttt{VA\_COMMIT}, or \texttt{VA\_REVEAL} state, if too many participants become non-responsive such that the number of active workers or validators falls below the required thresholds.
\texttt{DICE} inherits a non-responsive participant handling scheme from prior work \cite{lin2024refiner}, which sets dynamically adjusted timeouts for those states, to prevent deadlock (see \S\ref{sssec:handling_failed_participant_preliminary}). 
As long as an honest participant remains active and can invoke the task's smart contract, \texttt{DICE} will leave a stalled state rather than wait indefinitely.

\subsubsection{Impacts on Efficiency and Model Utility} 
\texttt{DICE}'s overall execution time could increase when non-responsive participants slow down the update collection or the consensus process, or could decrease due to early termination invoked by the non-responsive participant handling scheme. 
\texttt{DICE}'s model performance may degrade if early termination leaves the global model insufficiently trained. 
Since \texttt{DICE} inherits the non-responsive participant handling scheme from prior work \cite{lin2024refiner}, these impacts are similar to those analyzed in that work, and we refer readers to that work for a detailed experimental evaluation of these impacts. 

We note that non-responsive participants do not affect the per-participant computational complexity bounds derived in \S\ref{ssec:efficiency}. These bounds measure individual-level efficiency performance, determined by resource constraint parameters (e.g., $L_w$, $L_v$) reported when participants join the customization task or model scale (e.g., $S_{GM}$) specified by the model requester.

\subsubsection{Impacts on Scalability}
The greater the number of non-responsive participants, the smaller the pool of available workers and validators, making it harder to meet the required participant scale. 
\texttt{DICE} mitigates this impact from two directions. 
On the one hand, \texttt{DICE}'s consensus protocol is designed to operate in permissionless blockchain environments, where any device with Internet access can join as a worker or validator without prior authorization. 
This design choice maximizes the potential participant pool and inherently improves scalability by allowing a larger and more diverse set of participants, thereby increasing resilience to non-responsive participants. 
On the other hand, as discussed in \S\ref{ssec:tradeoff}, the tunable parameter $N_t$ allows the system to adapt to the actual available participant scale, further mitigating non-responsive participant impacts.

\subsubsection{Impacts on Security}
As established in \S\ref{ssec:security}, \texttt{DICE}'s security relies on Assumptions 1--2, including the VA result unpredictability and hash function conditions and validator population and stake conditions. Non-responsive honest validators have trivial impact on the VA result unpredictability and hash function conditions, but may reduce the number of validators available for consensus. \texttt{DICE} therefore pauses and replenishes the active validator set as described in \S\ref{ssec:security}. 
When \texttt{DICE} resumes the customization process, the security guarantees continue to apply provided that Assumption 2 holds for the validators eligible for the resumed round.

\subsection{On-chain Transaction Cost}\label{ssec:gas_cost_discussion}
The overall economic costs of customizing a pretrained model in a typical BCFL system mainly comprise two parts: off-chain computation costs and on-chain transaction costs. 
\texttt{DICE} focuses on the former for hardware-constrained workers and validators, reducing their per-participant per-round computation workload requirements through DPL, MSV, TCC, and EDI, as analyzed in \S\ref{ssec:efficiency}. Although DPL and MSV involve off-chain computation for slice selection and emulator generation, this overhead is borne by the requester, who initiates the task and is typically assumed to possess sufficient hardware resources. 

For on-chain transaction costs, we use 
Ethereum \cite{Ethereum2014A} as a representative example to illustrate how \texttt{DICE} controls these costs, since BCFL systems can be deployed on various underlying blockchain platforms with different fee models. 
Ethereum uses gas as a unit to measure the computational and storage resources required to perform an on-chain procedure. The transaction fee for executing an on-chain procedure is calculated as: $fee = gas\_used \times gas\_price$, where $gas\_used$ is the total amount of gas consumed by the smart contract's EVM instructions and data storage, and $gas\_price$ is the market-determined price per unit of gas. Following previous work \cite{lin2024refiner}, \texttt{DICE} focuses on controlling $gas\_used$, as $gas\_price$ is determined by the market rather than by BCFL algorithm design. 
\texttt{DICE} reduces $gas\_used$ by two key designs. 
First, storing large-scale data, including global model parameters, worker updates and aggregated updates, in an additional cost-effective system such as IPFS. This makes $gas\_used$ largely independent of the model size, avoiding the substantial gas overhead incurred by prior BCFL systems that perform on-chain aggregation of model parameters \cite{10701002, EURECOM, wilhelmi2021blockchain}. 
Second, applying gas-efficient smart contract coding practices \cite{Nelaturu2021SmartCR,Nguyen2022GasSaverAT}, such as using low-cost EVM instructions and batching multiple operations into single transactions, to further reduce the gas costs of contract execution. According to previous work \cite{lin2024refiner}, the above designs are effective in reducing the ratio of on-chain gas fees to off-chain computation costs, which is indispensable for engaging a wider range of honest (but potentially hardware-constrained) participants and customizing an effective model.

\section{Experiments} \label{sec:experiment}
We structure a comprehensive evaluation of \texttt{DICE} into five parts to systematically assess its performance under varying conditions.  
\S\ref{ssec:exp_efficiency} evaluates \texttt{DICE} with \textit{honest but hardware-constrained} participants, against an ideal but infeasible scenario where all participants are honest and hardware-unlimited, establishing \texttt{DICE}'s performance baseline.  
\S\ref{ssec:exp_compare} compares \texttt{DICE} against the current BCFL paradigm in realistic scenarios involving \textit{both hardware-constrained and malicious} participants, demonstrating \texttt{DICE}'s holistic advantage. 
\S\ref{ssec:exp_flexibility} studies \texttt{DICE}'s flexibility in integrating with existing efficient customization techniques, which, coupled with adjustable $N_t$ discussed in \S\ref{ssec:tradeoff}, enables \texttt{DICE} to accommodate diverse device constraints and model performance requirements.  
\S\ref{ssec:otherPMs_exp} extends the evaluation to larger-scale PMs to demonstrate \texttt{DICE}'s generalizability beyond the models used in the main experiments. 
Finally, \S\ref{ssec:ablation_study} further studies the contribution of DPL and MSV schemes for effective lightweight training and verification and the necessity of using TCC with EDI for efficient consensus. 

\subsection{Experimental Settings}\label{ssec:exp_settings}

\begin{table*}[t]
    \centering \small
    \caption{Task settings. Tasks marked with $\ddag$ undergo further pretraining before fine-tuning; the others directly fine-tune the PMs.}
    \label{tab:exp_settings}
    \begin{tabular}{l|c|c}
    \toprule
         \textbf{Settings} & \textbf{Vision Tasks} & \textbf{Language Tasks} \\
    \hline
         Type & Image Classification & NER and RE \\
    \hline
         Model & \makecell[c]{ViT-base (default),\\ DINOv2-large}
               & \makecell[c]{BERT-base (default),\\ Qwen2.5-0.5B} \\
    \hline
         Dataset &
         \makecell[c]{FedISIC \cite{fedisic}$^\ddag$, CIFAR100 \cite{SHARMA2018377},\\
         DermaMNIST \cite{medmnistv1,medmnistv2}} &
         \makecell[c]{2018n2c2 \cite{lee2020biobert}$^\ddag$, BC5-disease (BC5-d) \cite{lee2020biobert}, JNLPBA \cite{kim-etal-2004-introduction}, \\
         GAD \cite{lee2020biobert}, DDI \cite{herrero2013ddi}, ChemProt \cite{taboureau2010chemprot}} \\
    \hline
         Data Partition &
         \multicolumn{2}{c}{FedISIC: $\operatorname{Dirichlet}(1.25)$; all others: $\operatorname{Dirichlet}(2.5)$} \\
    \hline
         Further Pretraining &
         MedMNIST (supervised image classification) &
         PubMed (self-supervised language modeling) \\
    \hline
         Workers / Validators ($N_c$)  & 100 / 1,173 (1,036) & 50 / 678 (599) \\
    \hline
         Total Rounds & 100 & 100 \\
    \hline
         Local Epochs / Batch Size & 3 / 50 & 3 / 64 \\
    \hline
         Learning Rate (LR) Scheduler / Optimizer & Cosine / AdamW & Linear / AdamW \\
    \hline
         LR &
         \multicolumn{2}{c}{Dynamically tuned at runtime over
         $\{7\times10^{-5},\,1\times10^{-4},\,3\times10^{-4},\,5\times10^{-4}\}$} \\
    \hline
         Emulator Distillation Epochs / LR &
         \multicolumn{2}{c}{$5$ / $1\times10^{-4}$} \\
    \hline
         Committee Election / Incentive Parameters &
         \multicolumn{2}{c}{$E_c=891$, $\epsilon_c=0.472$, $\gamma_{worker}=1$, $\gamma_{validator}=1$, $\epsilon_{penalty}=10(\times0.01\text{ETH})$} \\
    \hline
         Worker / Validator Reward Budgets &
         \multicolumn{2}{c}{$100\,(\times0.01\,\mathrm{ETH})$ / $100\,(\times0.01\,\mathrm{ETH})$} \\
    \bottomrule
    \end{tabular}
\end{table*}

Our experiments are conducted on a private Ethereum testnet deployed on our in-house private cloud to avoid the real economic costs introduced by invoking smart contracts and making transactions on the public Ethereum mainnet. We follow prior work \cite{lin2024refiner} to configure the blockchain network and refer readers to that work for more details. 
Below, we introduce the configuration of customization tasks, participants, the BCFL baselines used for comparison, and evaluation metrics.

\underline{\textit{Task Settings.}} Our experiments are conducted on nine customization tasks summarized in Table~\ref{tab:exp_settings}. 
The vision tasks customize ViT \cite{dosovitskiy2020vit} for image classification on three datasets commonly used in federated learning research \cite{10.5555/3666122.3666302,Yan2025TemporalMF,Tupper2025PersonalizedFF,Mukhtiar2024FederatedLA,lu2022distributionfreefederatedlearningconformal,Ye2025RevisitingFL,fedisic,11185620}. 
The language tasks customize BERT \cite{DBLP:journals/corr/abs-1810-04805, Wang2023FedBFPTAE} for named entity recognition or relation extraction on six biomedical domain datasets from prior work \cite{DBLP:journals/npjdm/PengLZCXSZ24, peng2023a, lee2020biobert,PubMedBERT}. 
The extended experiments in \S\ref{ssec:otherPMs_exp} use larger-scale models, namely DINOv2~\cite{oquab2023dinov2} for vision tasks and Qwen2.5~\cite{qwen2.5} for language tasks. 
For both vision and language tasks, we select one biomedical dataset to evaluate the performance of fine-tuning the PM on that dataset with a prior further-pretraining step. 
We use MedMNIST \cite{medmnistv2,medmnistv1}, a large-scale collection of standardized biomedical images, and PubMed \cite{PubMedBERT}, a large-scale biomedical text dataset, for further-pretraining in vision and language tasks, respectively. 

\underline{\textit{Participant Configuration.}} 
We deploy 100 workers and 1173 validators for vision customization tasks, and 50 workers and 678 validators for language customization tasks, along with a model requester who submits all customization tasks. 
These worker scale settings follow prior work \cite{lin2024refiner,shayan2020biscotti,carney2025fidelis,ren2024scalable}. 
For language tasks, we use a smaller number of workers due to the higher computational costs of customizing BERT and Qwen2.5 compared to ViT and DINOv2 under our experiments. 
Given worker scale settings, we compute the required number of validators using  $|\mathbb{V}_{all}|=\max(100, 3\lceil \log_{[1 - \frac{N_t}{(2|\mathbb{W}^r|-1)N_{lus}}]} \frac{\epsilon_p}{K_{\text{check}}} \rceil)$ with the $N_t=\lceil\log_2|\mathbb{W}^r|\rceil$ setting recommended in \S\ref{ssec:tradeoff}, $N_{lus}=3$ (the maximum $N_{lus}$ in our experiments), $K_{\text{election}}=3$, $K_{\mathrm{check}}=1$, and a relaxed security threshold $\epsilon_p = 10^{-2}$ that can reduce the required validator scale to achieve a balance between experimental costs and reproducibility. 
Using these settings and the analysis of Corollary 5, we choose $N_c$, $E_c$, and $\epsilon_c$ to satisfy the corresponding probability bounds.

Each participant runs in a dedicated virtual machine hosted in our in-house private cloud. We associate each participant with a unique Ethereum testnet account. 
Each worker is allocated a subset of training samples from the training datasets of the customization tasks. To simulate the non-IID data distribution prevalent in real-world cases, we allocate sample sizes across workers using a Dirichlet distribution and differentiate label distributions across workers based on the number of unique labels in each dataset. 
Validators have access to a held-out validation dataset to evaluate worker updates. For datasets that include dedicated validation samples, we use them directly. In cases where such samples are unavailable, we split the test set in half, using one half for validation and the other for final model evaluation.

\begin{table}[t]
    \centering \small
    \caption{
    Worker and validator memory overhead ($MC_w$, $MC_v$ in GB) for model training and multi-update ($|\mathbb{W}^r|$-update) verification when customizing Qwen2.5-0.5B. 
    $S_{LM}/S_{VM}$ denotes local/validation model size in GB. 
    Measurement environment follows Table~\ref{tab:evidence_for_challenges}. Hardware Ref. denotes the reference of required hardware. 
    }
    \resizebox{\columnwidth}{!}{
    \begin{tabular}{c|c|c|c|c|c}
    \toprule
    \textbf{$L_w/L_v$} & \textbf{$S_{LM}/S_{VM}$} & \textbf{$MC_w$} & \textbf{Hardware Ref.} & \textbf{$MC_v$} & \textbf{Hardware Ref.} \\
    \midrule
    24  & 1.8  & 16.4 & RTX 4090 24GB $\uparrow$ & 122.2 & 2$\times$ H100 80GB  $\uparrow$  \\ 
    20  & 1.6  & 13.8 & RX 9070 XT 16GB $\uparrow$ & 108.3 & 2 $\times$ H100 80GB  $\uparrow$  \\ 
    16  & 1.4  & 11.2 & RTX 5070 12GB $\uparrow$ & 94.5  & RTX Pro 6000 96GB  $\uparrow$  \\ 
    12  & 1.2  & 8.6  & ARC B570 10GB $\uparrow$ &  80.6  & 2$\times$RTX A6000 48GB $\uparrow$  \\
    \bottomrule
    \end{tabular}
    }
    \label{tab:fm_overhead}
\end{table}

Unless otherwise stated, participants in each experiment are heterogeneous in hardware resources and potentially malicious.  

We simulate hardware constraints by limiting the number of H-layers deployable on workers and validators. 
The number of deployable H-layers on each worker $w$ is set as $L_w = \lfloor \frac{L_G}{2} + \frac{L_G}{2} \cdot \left(1 - \frac{|\mathbb{D}_w| - D_{min}}{D_{max} - D_{min}}\right)\rceil$, where $D_{max}$ and $D_{min}$ are the maximum and minimum sizes of a worker's training dataset, respectively. This simulates a scenario where workers with abundant task-specific data cannot deploy a full model due to hardware limitations. 
For validators, we simulate hardware heterogeneity by sampling their deployable H-layer counts from a normal distribution: $L_v \sim \mathcal{N}\left(\frac{3L_G}{4}, \frac{L_G^2}{144}\right)$, clipped to the range $[L_G/2, L_G]$. This reflects a common heterogeneous scenario where some validators can deploy nearly full models, while others are constrained and even unable to deploy a worker's complete update when $L_{vm}<L_{wm}$. 
All $L_w$ and $L_v$ values are rounded to the nearest integer. 
Table~\ref{tab:fm_overhead} takes Qwen2.5-0.5B, the largest PM used in our experiments, as an example to illustrate the mapping from the total number of H-layers in a pretrained model to actual GPU memory overhead and hardware requirements. 
This table indicates that configuring deployable H-layers provides a controllable and reproducible proxy scheme to simulate real-world hardware constraints and heterogeneity.

Regarding adversarial settings, we designate 20\% of workers and 20\% of validators as malicious, with the remaining participants being honest. 
An exception is the experiment in \S\ref{ssec:exp_efficiency}, where we configure all participants as honest to establish \texttt{DICE}'s performance baseline. 
Each malicious participant randomly chooses between poisoning and fraud in each round. 
For any malicious worker, if it chooses to launch fraud attacks, it copies an honest worker's historical update to illegitimately claim rewards. Otherwise, it submits updates trained on mislabeled examples for poisoning attacks \cite{Bagdasaryan2020HowTB,Bhagoji2019AnalyzingFL}. 
For any malicious validator, if it chooses to launch fraud attacks, it copies then submits another validator's fingerprint and commitment and later submits the copied raw result. 
Otherwise, it submits poisoned VA results, generated by assigning random scores above the discard threshold to each malicious worker's update. 
Unless otherwise stated, all validators stake the same amount of security deposit ($E_{\min}\times0.01\text{ETH}$) in each experimental task, and stake-burning penalties are disabled to keep the stake distribution fixed. Therefore, the fraction of total stake controlled by malicious validators equals their share of the total validator count. 

\underline{\textit{BCFL baseline Configuration.}} 
We implement the current BCFL paradigm introduced in \S\ref{sec:intro} based on a representative system, namely Refiner \cite{10.14778/3476311.3476313,lin2024refiner}. 
This choice is justified for two reasons. 
First, Refiner shares the same participant resource requirements as mainstream BCFL algorithms \cite{9293091, desai2021blockfla,9399813,ren2024bpfl,10.1109/TDSC.2019.2952332}, namely requiring each worker to train a full model and each validator to fully verify every worker's update. This allows Refiner to faithfully represent the computational burden of the current BCFL paradigm. 
Second, Refiner is designed under the same threat model as \texttt{DICE}, addressing both poisoning and fraud attacks. 
In contrast, many BCFL algorithms \cite{9293091, Liu2020, 9284684, 10.1109/TDSC.2019.2952332, desai2021blockfla,kalapaaking2023blockchain,9399813,chang2021blockchain,ren2024bpfl,mosquedagonzalez:hal-05002718} neglect fraud attacks that have a negligible impact on model performance, making them unsuitable for a direct comparison under our broader threat model. 
Owing to the above characteristics, Refiner enables a fair and meaningful comparison of the current BCFL paradigm with \texttt{DICE}.

Our experiments also implement several enhanced variants of the current BCFL paradigm powered by four efficient customization techniques. (i) \textit{LoRA} \cite{hu2022lora}, the most popular PEFT method adopted in federated learning research \cite{zhang2023fedpetuning,xu2025dp,shen-etal-2025-pfedgpt}, which exemplifies the family of methods that insert small trainable modules while freezing the backbone of the original pretrained model. (ii) \textit{BitFit} \cite{ben-zaken-etal-2022-bitfit}, a PEFT approach that does not alter the model architecture but only tunes a subset of selected parameters. Together with LoRA, BitFit covers two fundamentally different types of PEFT techniques discussed in \S\ref{sec:related_work} without redundancy. (iii) \textit{Prune} \cite{lu2024reassessing,xiao2023offsite,hu2025trimllm} and (iv) \textit{KD} \cite{hinton2015distilling,sanh2019distilbert,jiao2020tinybert,wu2022tinyvit}, mainstream model compression techniques applicable to both model training and inference scenarios. 
For the KD-enhanced variants, we use DeiT \cite{touvron2021training} as the compressed version of ViT and DistilBERT \cite{sanh2019distilbert} as the compressed version of BERT, respectively. 
In all baselines, the embedding layers of PMs are frozen.


\begin{table*}[t] 
    \centering \small
    \caption{Results of handling constrained workers (upper half) and constrained validators (lower half). Top-1 and top-2 results are marked in bold and underlined, respectively. }
    \resizebox{\textwidth}{!}{
    \begin{tabular}{l|c|c|c|ccc||c|c|ccc||c|c|ccc}
    \toprule
         \multirow{2}*{\textbf{Methods}} & \multirow{2}*{\textbf{$\mathbb{W}_{all}$}}  & \textbf{$S_{LM} \downarrow$}  &  $|\theta_w^r| \downarrow$ & \multicolumn{3}{c||}{\textbf{ACC $\uparrow$}} & \textbf{$S_{LM} \downarrow$}  &  $|\theta_w^r| \downarrow$  & \multicolumn{3}{c||}{\textbf{F1 $\uparrow$}} & \textbf{$S_{LM} \downarrow$}  &  $|\theta_w^r| \downarrow$  & \multicolumn{3}{c}{\textbf{F1 $\uparrow$}} \\
         \cline{3-7} \cline{8-12} \cline{13-17}
         & & \multicolumn{2}{c|}{\textbf{Vision Tasks}} & \footnotesize{\textbf{FedISIC }} & \footnotesize{\textbf{CIFAR100}} & \footnotesize{\textbf{DermaMNIST}} & \multicolumn{2}{c|}{\textbf{NER Tasks}} & \footnotesize{\textbf{n2c2}} & \footnotesize{\textbf{JNLPBA}}  & \footnotesize{\textbf{BC5-d}} & \multicolumn{2}{c|}{\textbf{RE Tasks}} &  \footnotesize{\textbf{DDI}} & \footnotesize{\textbf{GAD}} & \footnotesize{\textbf{ChemProt}}\\
    \hline
         BCFL &  \multirow{3}*{\ding{55}}  & 100.0\% & 100.0\% & 79.5\% & 94.0\% & 82.9\% & 100.0\% & 100.0\% & 90.7\% & 98.5\% & 89.5\% & 100.0\% & 100.0\% & 44.3\% & 92.7\% & 13.8\% \\
        \, +LoRA & & 101.6\% & \underline{1.7\%} & 75.7\% & 91.5\% & 78.3\% & 101.2\% & \underline{1.6\%} & 86.4\% & 93.5\% & 85.2\% & 101.2\% & \underline{2.2\%} & 43.8\% & 87.4\% & 13.1\% \\
        \, +BitFit & & 100.0\% & \textbf{0.2\%} & 71.5\% & \underline{94.8\%} & 75.2\% & 100.0\% & \textbf{0.1\%} & 81.6\% & 59.3\% & 84.5\% & 100.0\% & \textbf{0.8\%} & 42.2\% & 76.3\% & 23.0\%  \\
    \cline{2-17}
        \, +Prune & \multirow{2}*{\ding{51}} & \underline{50.4\%}  & 50.0\%   & \underline{95.9\%} & 90.9\% & \underline{99.6\%} & \multirow{2}*{\textbf{\underline{60.9\%}}} & \multirow{2}*{50.0\%} & \underline{97.4\%} & 98.2\% & 98.5\% & \multirow{2}*{\textbf{\underline{61.1\%}}} & \multirow{2}*{50.4\%}  & \underline{93.3\%} & 99.6\% & 78.8\% \\
        \, +KD &  & \textbf{3.3\%} & 3.1\% & 91.7\% & 80.3\% & 91.2\% &  &  & 97.1\% & \underline{99.0\%} & \textbf{99.7\%} & & & 92.8\% & \textbf{99.9\%} & \underline{97.6\%} \\
    \hline
         \texttt{DICE} & \ding{51} & \underline{50.4\%}  & 41.6\% & \textbf{98.5\%} & \textbf{98.5\%} & \textbf{99.8\%} & \textbf{\underline{60.9\%}} & 41.7\% & \textbf{98.5\%} & \textbf{99.2\%} & \underline{98.8\%} & \textbf{\underline{61.1\%}} & 42.1\% & \textbf{96.6\%} & \underline{99.7\%} & \textbf{99.5\%} \\
    \hline \hline
        \multirow{2}*{\textbf{Methods}} & \multirow{2}*{\textbf{$\mathbb{V}_{all}$}}  & \textbf{$S_{VM} \downarrow$}  &  \textbf{$N_{VM}\downarrow$} & \multicolumn{3}{c||}{\textbf{ACC $\uparrow$}} & \textbf{$S_{VM} \downarrow$}  &  \textbf{$N_{VM}\downarrow$} & \multicolumn{3}{c||}{\textbf{F1 $\uparrow$}} & \textbf{$S_{VM} \downarrow$}  &  \textbf{$N_{VM}\downarrow$}  & \multicolumn{3}{c}{\textbf{F1 $\uparrow$}} \\
         \cline{3-7} \cline{8-12} \cline{13-17}
         & & \multicolumn{2}{c|}{\textbf{Vision Tasks}} & \footnotesize{\textbf{FedISIC }} & \footnotesize{\textbf{CIFAR100}} & \footnotesize{\textbf{DermaMNIST}} & \multicolumn{2}{c|}{\textbf{NER Tasks}} & \footnotesize{\textbf{n2c2}} & \footnotesize{\textbf{JNLPBA}}  & \footnotesize{\textbf{BC5-d}} & \multicolumn{2}{c|}{\textbf{RE Tasks}} &  \footnotesize{\textbf{DDI}} & \footnotesize{\textbf{GAD}} & \footnotesize{\textbf{ChemProt}}\\      
    \hline
         BCFL & \multirow{3}*{\ding{55}}  &  \multirow{3}*{100.0\%} & \multirow{5}*{\underline{100.0\%}} & \textbf{100.0\%} & \textbf{100.0\%} & \textbf{100.0\%}&  \multirow{3}*{100.0\%} & \multirow{5}*{\underline{100.0\%}} & \textbf{100.0\%} & \textbf{100.0\%} & \textbf{100.0\%}&  \multirow{3}*{100.0\%} & \multirow{5}*{\underline{100.0\%}} & \textbf{100.0\%} & \textbf{100.0\%} & \textbf{100.0\%}\\
        \, +LoRA & &  & & {97.2\%} & {98.4\%} & 97.5\%  &  & & 96.3\% & 96.0\% & 97.2\% &  & & {96.9\%} & 97.3\% & 94.9\% \\
        \, +BitFit & & &  & 89.9\% & 92.0\% & 90.7\% &  & & 91.0\% & 93.3\% & 87.5\%  &  &  & 90.1\% & 92.3\% & 89.7\%  \\
    \cline{2-3} \cline{5-8} \cline{10-13} \cline{15-17}
        \, +Prune & \multirow{2}*{\ding{51}} & \underline{50.4\%}  &  & 95.9\% & 90.9\% & {99.6\%} & \multirow{2}*{\textbf{\underline{60.9\%}}} &  & {97.4\%} & 98.2\% & 98.5\% & \multirow{2}*{\textbf{\underline{61.1\%}}} &  & 93.3\% & 99.6\% & 78.8\% \\
        \, +KD &  & \textbf{3.3\%} &  & 91.7\% & 80.3\% & 91.2\% &  &  & 97.1\% & {99.0\%} & \underline{99.7\%} & & & 92.8\% & \underline{99.9\%} & {97.6\%} \\
    \hline
        \texttt{DICE} & \ding{51} & \underline{50.4\%} & \textbf{3.5\%} & \underline{98.1\%} & \underline{99.4\%} & \underline{99.8\%} & \textbf{\underline{60.9\%}} & \textbf{6.0\%} & \underline{98.1\%} & \underline{99.2\%} &  {98.8\%} & \textbf{\underline{61.1\%}} & \textbf{6.0\%} & \underline{97.6\%} &  {98.2\%} & \underline{98.7\%} \\
    \bottomrule
    \end{tabular}
    }
    \label{tab:result_of_worker_efficiency1}
\end{table*}

\underline{\textit{Evaluation Metrics.}} The evaluation metrics used in our experiments include 
(i) Participant Workload Metrics: $S_{LM}$ (GB), the minimum size of local models deployed on the workers, 
$|\theta_w^r|$ (GB), the maximum size of parameters to be trained on the minimum local model, 
$S_{VM}$ (GB), the minimum size of validation models deployed on the validators, 
$N_{VM}$, the minimum number of validation models that a validator loads during the complete consensus process, among the honest validators contributing to the global consensus conditions. 
Based on empirical results from previous work \cite{vaswani2017attention,narayanan2021efficient,hoffmann2022training,korthikanti2023reducing,Georgiou2022GreenAD} and Table~\ref{tab:fm_overhead}, 
$S_{LM}$ and $|\theta_w^r|$ jointly characterize the worker's model-processing workload (including forward/backward computation, activation and gradient memory, and communication), while $S_{VM}$ and $N_{VM}$ characterize the minimum model-deployment requirement and the observed validation workload among honest validators contributing to consensus, and their product is a proxy for cumulative validator model-processing workload. 
These metrics enable consistent comparisons across different hardware platforms and thus are selected over hardware-dependent measurements such as peak GPU memory or actual execution time to enhance reproducibility. 
(ii) Model Effectiveness Metrics: $ACC$ and $F1$, the accuracy and F1 score of the customized global model on downstream tasks. 
(iii) System Security Metrics: $RMW$ and $RMV$, the proportion of total rewards successfully claimed by malicious workers and validators, respectively.


Below, we run each experiment three times with different random seeds and report the average of the results of the three runs. 
Participant workload and model effectiveness metrics are normalized by their
values in \textbf{an ideal scenario}, where all participants are honest and unconstrained: each worker trains the full model, and each validator executes the complete VA tasks. 
$RMW$ and $RMV$ are reported directly as reward fractions and are not normalized by their zero values in the ideal scenario.

\subsection{Efficiency Performance} \label{ssec:exp_efficiency}
We first evaluate the efficiency of \texttt{DICE} with honest yet hardware-constrained participants against the ideal scenario without constrained participants. 
To isolate the impact of each participant role's constraints, we conduct two controlled experiments: one with only the workers constrained and another with only the validators constrained. 
Table~\ref{tab:result_of_worker_efficiency1} reports results of the two experiments and reveals the following findings:

(i) Compared to the current BCFL paradigm, \texttt{DICE} reduces worker model size ($S_{LM}$) by 40--50\% and tunable parameters ($|\theta_w^r|$) by nearly 60\%, enabling a substantial reduction in worker-side computational and memory costs. 
It also reduces validator-side model scale ($S_{VM}$) by 40--50\% and the minimum number of validation models that an honest validator must load to satisfy the global consensus conditions ($N_{VM}$) by an order of magnitude. 
Consequently, the product $S_{VM} \times N_{VM}$, which represents validator model-processing workload, becomes substantially smaller. 
Additionally, \texttt{DICE} maintains 96.6\%--99.8\% of ideal performance across tasks under both worker-constrained and validator-constrained experiments. 

(ii) BCFL and its PEFT-enhanced variants (+LoRA, +BitFit) sacrifice over 10\% of ideal performance in most tasks of worker-constrained experiments, as they cannot facilitate workers who are unable to deploy a full model to contribute their task-specific data.  
Under validator-constrained experiments that remove worker constraints to enable usage of all data, these baselines achieve better ACC/F1. 
However, these baselines require not only full model deployment and forward computation but also multi-update verification, excluding honest but hardware-constrained participants from engaging as validators for more robust system security. 

(iii) Compression-enhanced variants (+Prune, +KD), though effective at enabling constrained participants to engage in customization, suffer 4--20\% ACC/F1 drops on complex tasks (e.g., FedISIC, CIFAR100, DDI). 
The root cause is that the final models customized by these baselines have significantly fewer parameters than the original PMs, which limits their representational capacity for tasks with complex data distributions. 

In summary, \texttt{DICE} reduces costs on both workers and validators, lowering the participation barrier while achieving near-ideal model performance. 
On the worker side, \texttt{DICE} addresses the inherent limitations of compared baselines by DPL. 
DPL eliminates the full model deployment requirement on constrained participants without compromising model structure on requesters, and adjusts the training pattern dynamically, thus yielding reliable lightweight training across diverse tasks. 
On the validator side, \texttt{DICE}'s efficiency gain is more pronounced. 
Using MSV, TCC, and EDI, \texttt{DICE} not only reduces per-update verification costs but also resolves a critical scalability bottleneck that prior work cannot address.

\subsection{Comparison with Current BCFL Paradigm} \label{ssec:exp_compare}

We then compare \texttt{DICE} with the current BCFL paradigm and its enhanced variants under a realistic scenario that concurrently incorporates hardware-constrained and malicious participants.

\begin{figure}[h]
    \centering
    \includegraphics[width=0.8\columnwidth]{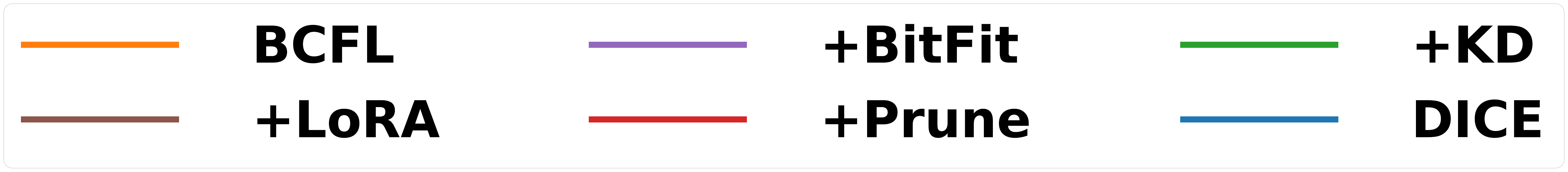}
    \\
    \includegraphics[width=0.342\columnwidth]{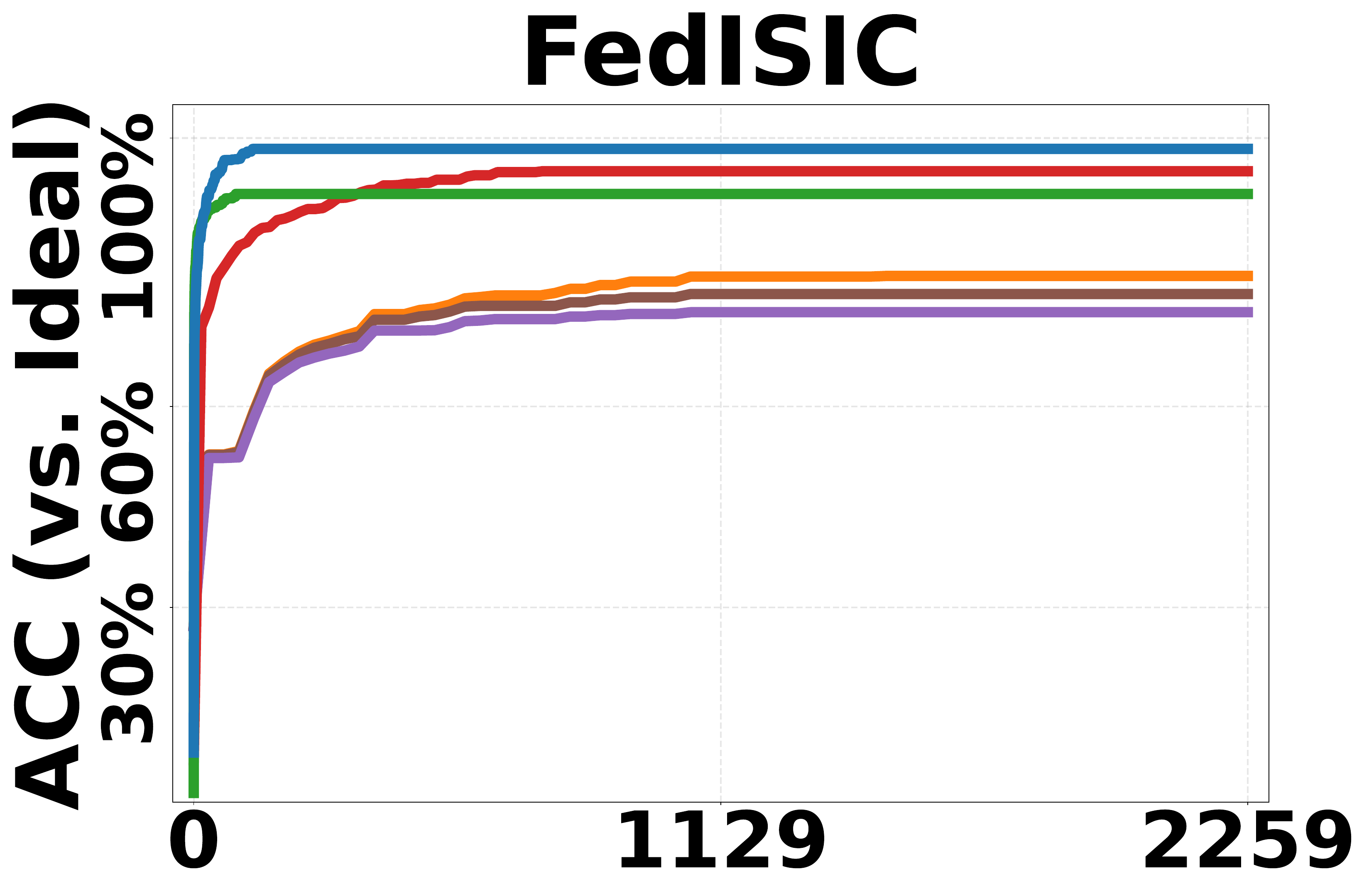}
    \includegraphics[width=0.32\columnwidth]{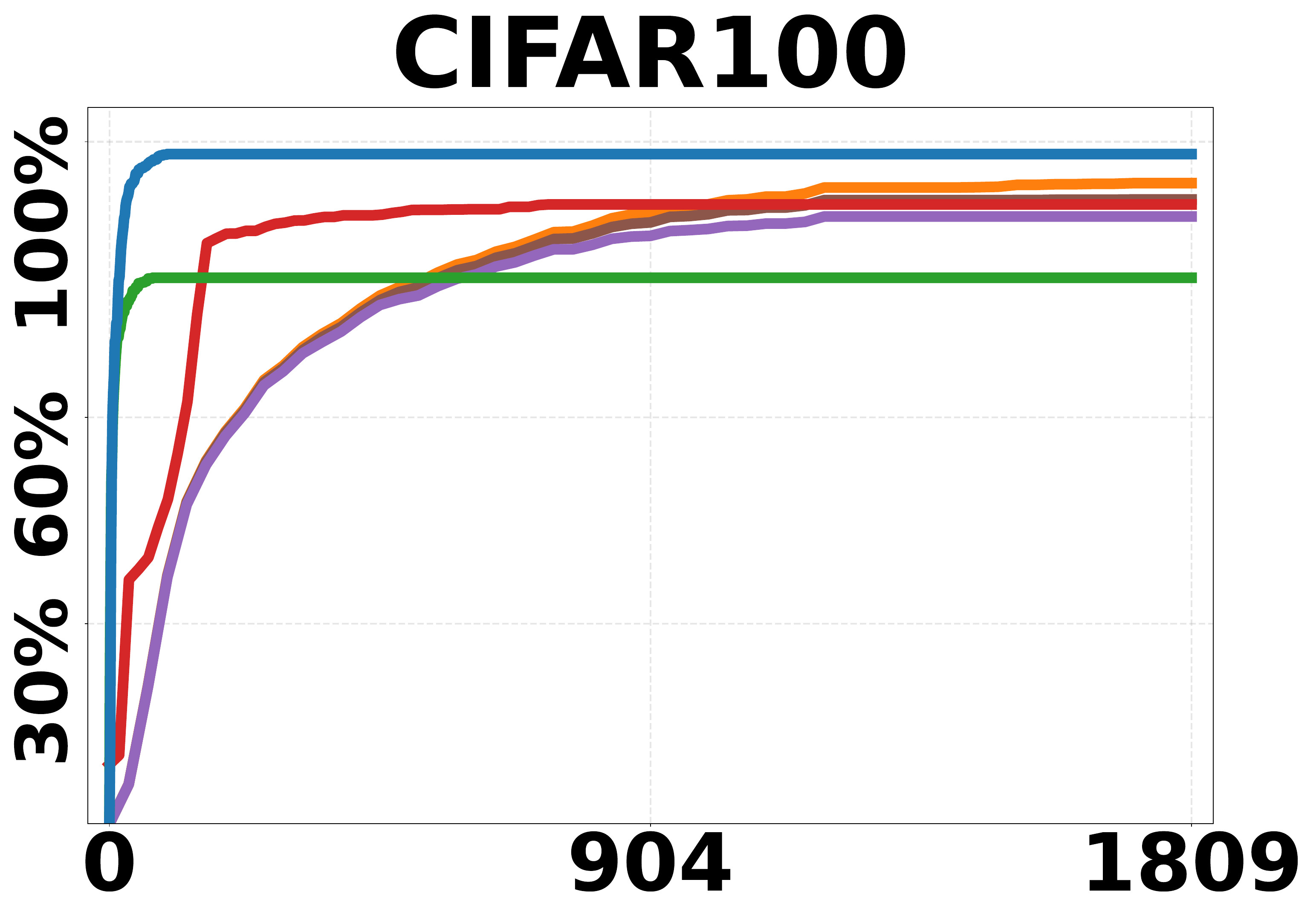}
    \includegraphics[width=0.32\columnwidth]{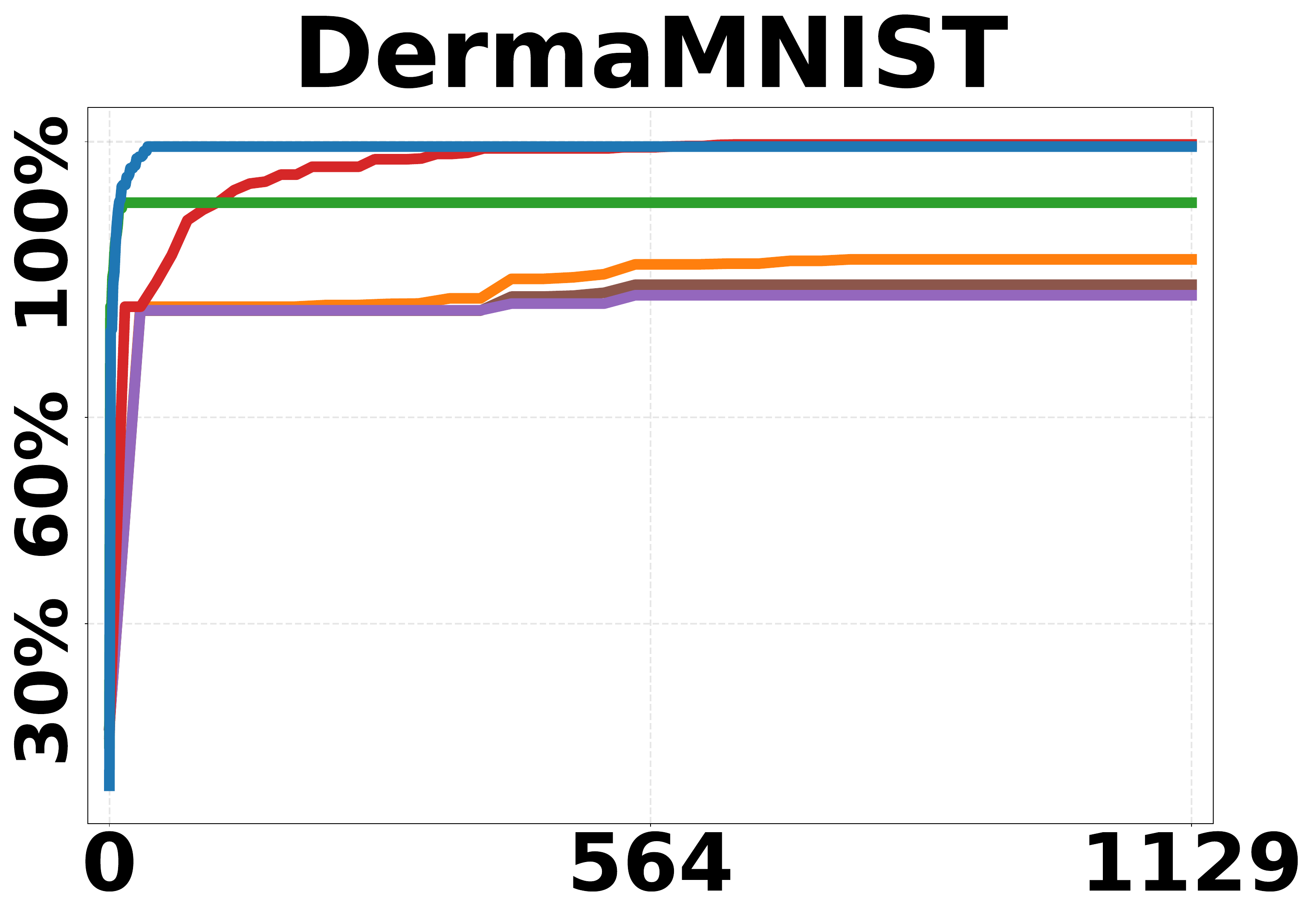}
    \\
    \includegraphics[width=0.342\columnwidth]{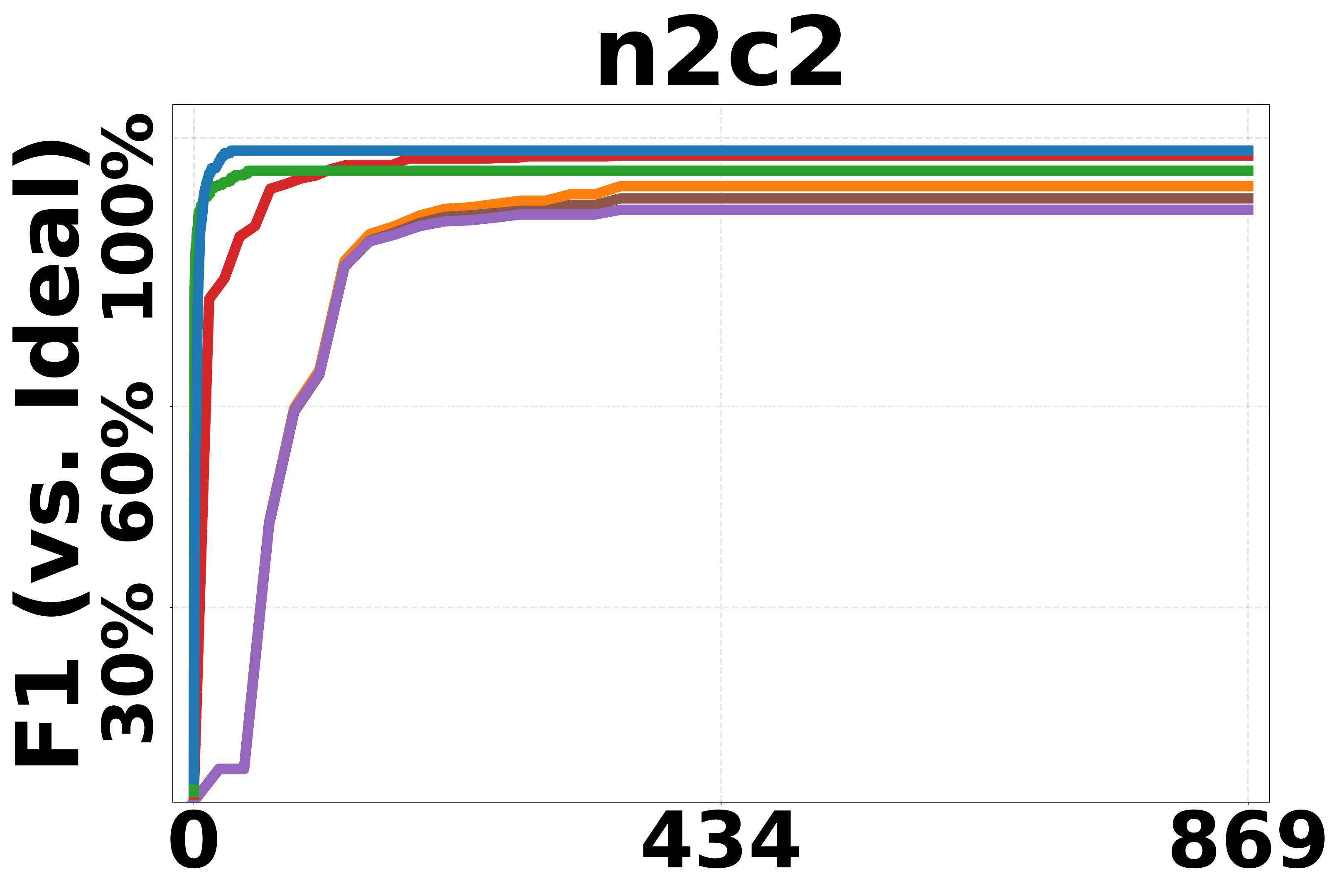}
    \includegraphics[width=0.32\columnwidth]{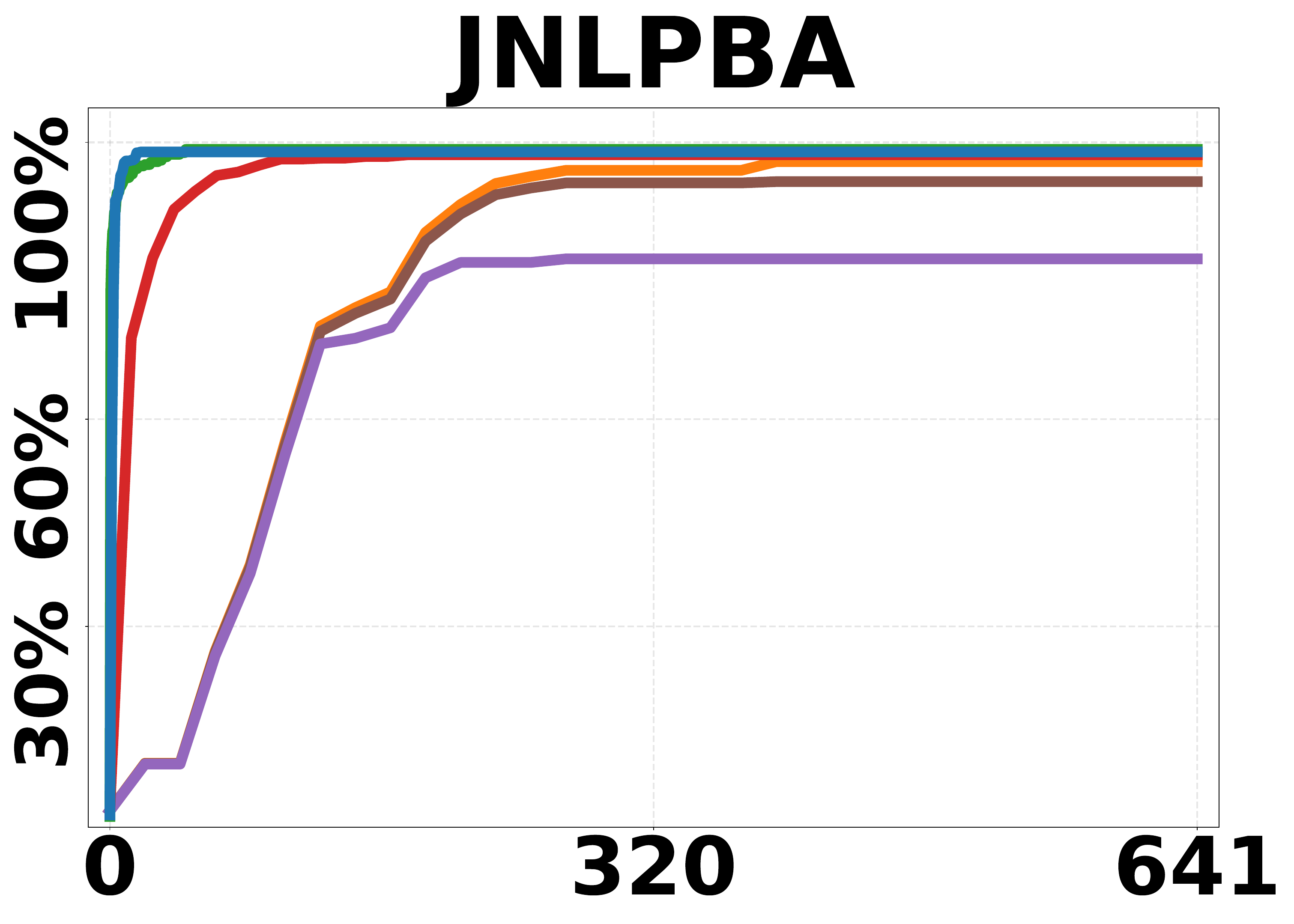}
    \includegraphics[width=0.32\columnwidth]{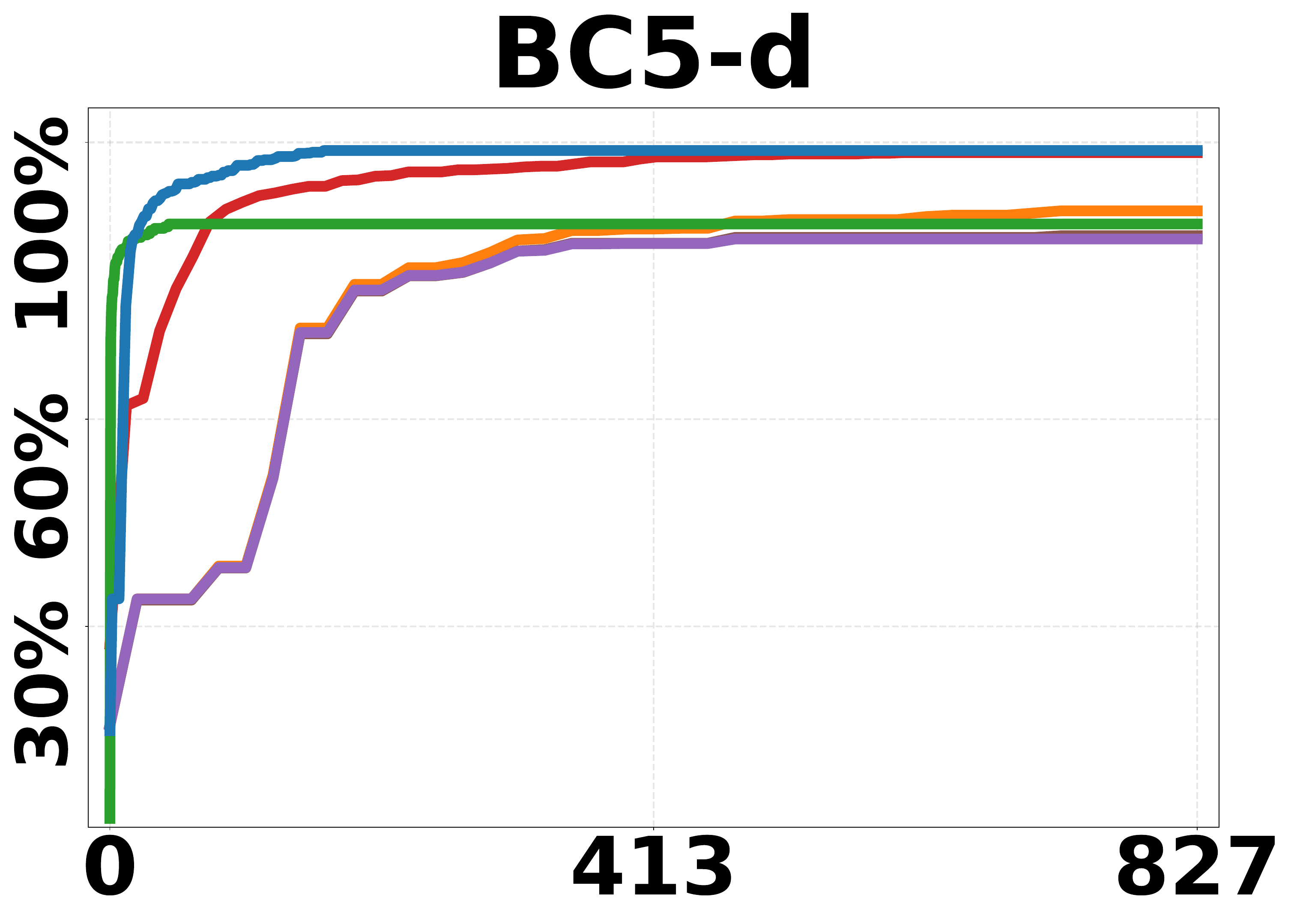}
    \\
    \includegraphics[width=0.342\columnwidth]{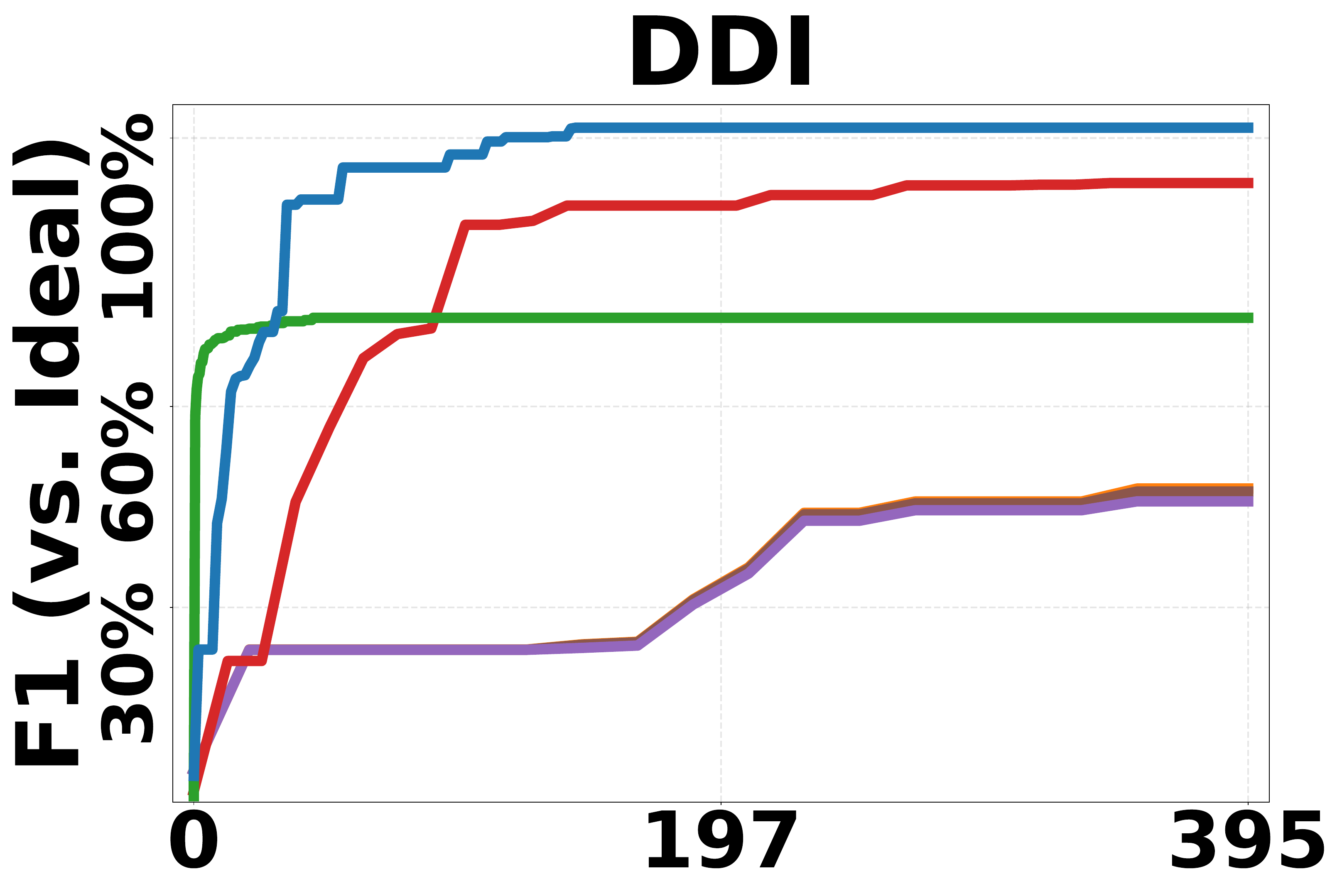}
    \includegraphics[width=0.32\columnwidth]{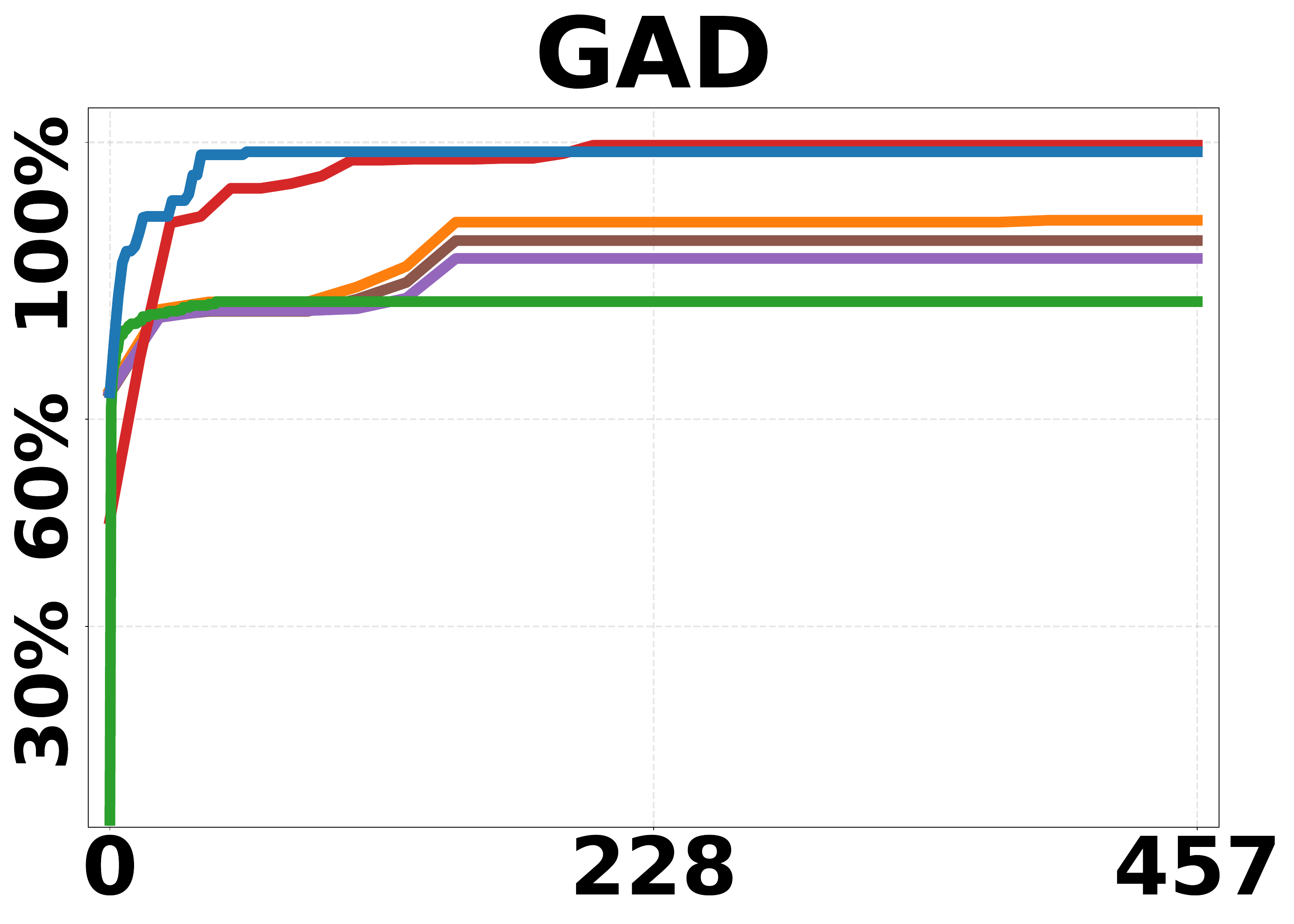}
    \includegraphics[width=0.32\columnwidth]{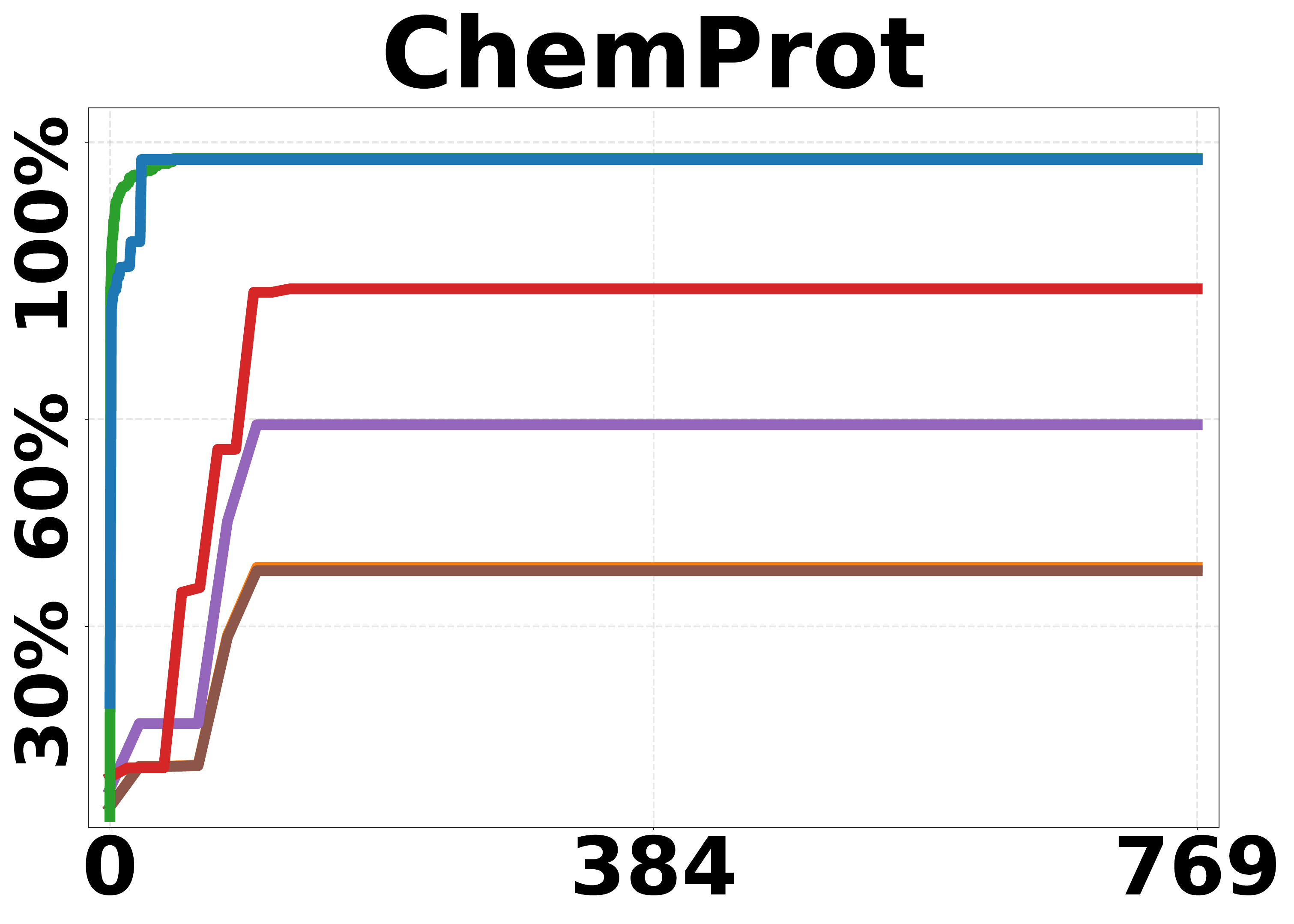}
    \caption{Performance of customized model versus cumulative model-processing workload under different baselines. X-axis: cumulative model-processing workload, measured by accumulated $S_{LM}+N_{VM}S_{VM}$.}\label{fig:result_of_comparison}
    \mbox{ 
        \includegraphics[width=0.322\columnwidth]{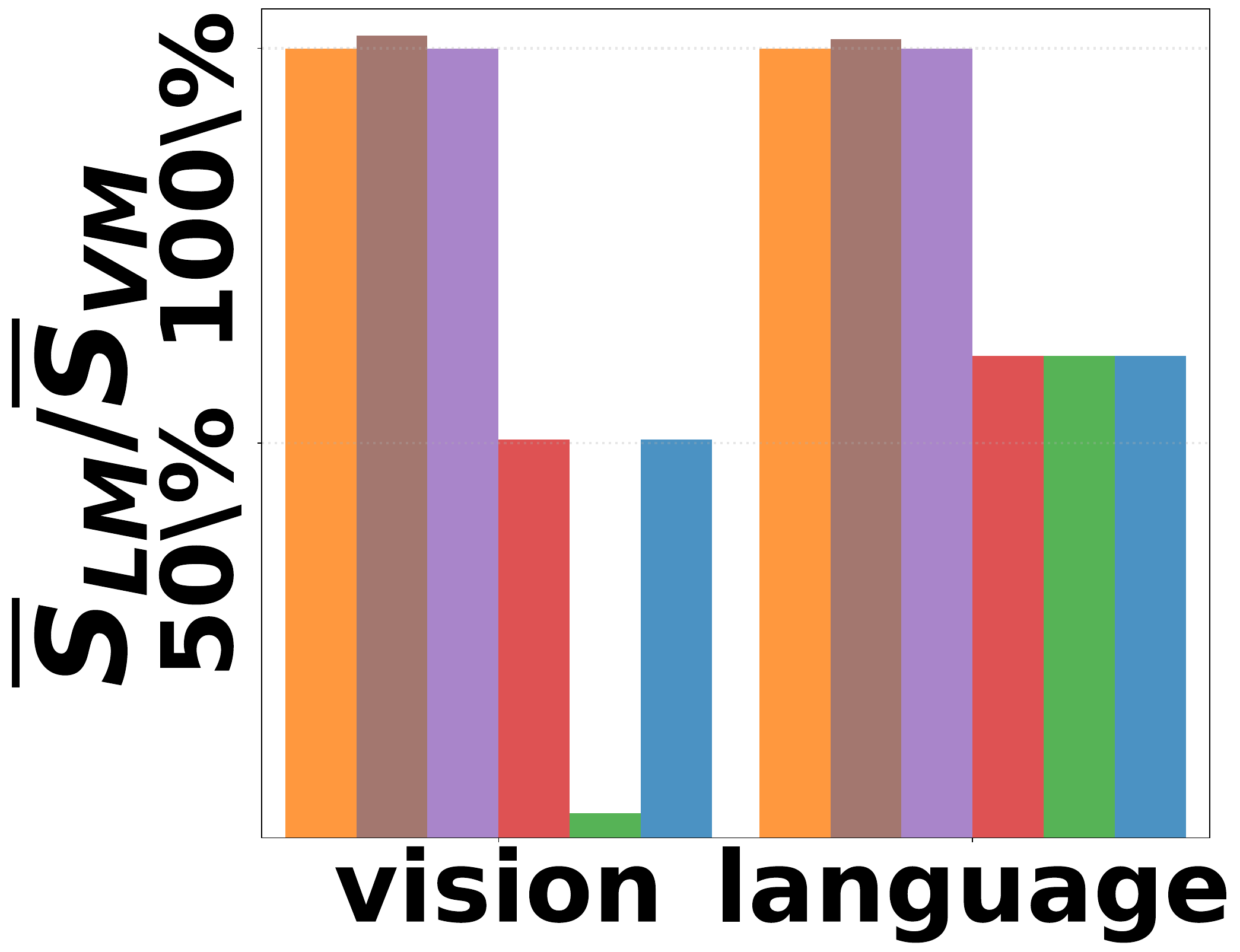}
        \includegraphics[width=0.322\columnwidth]{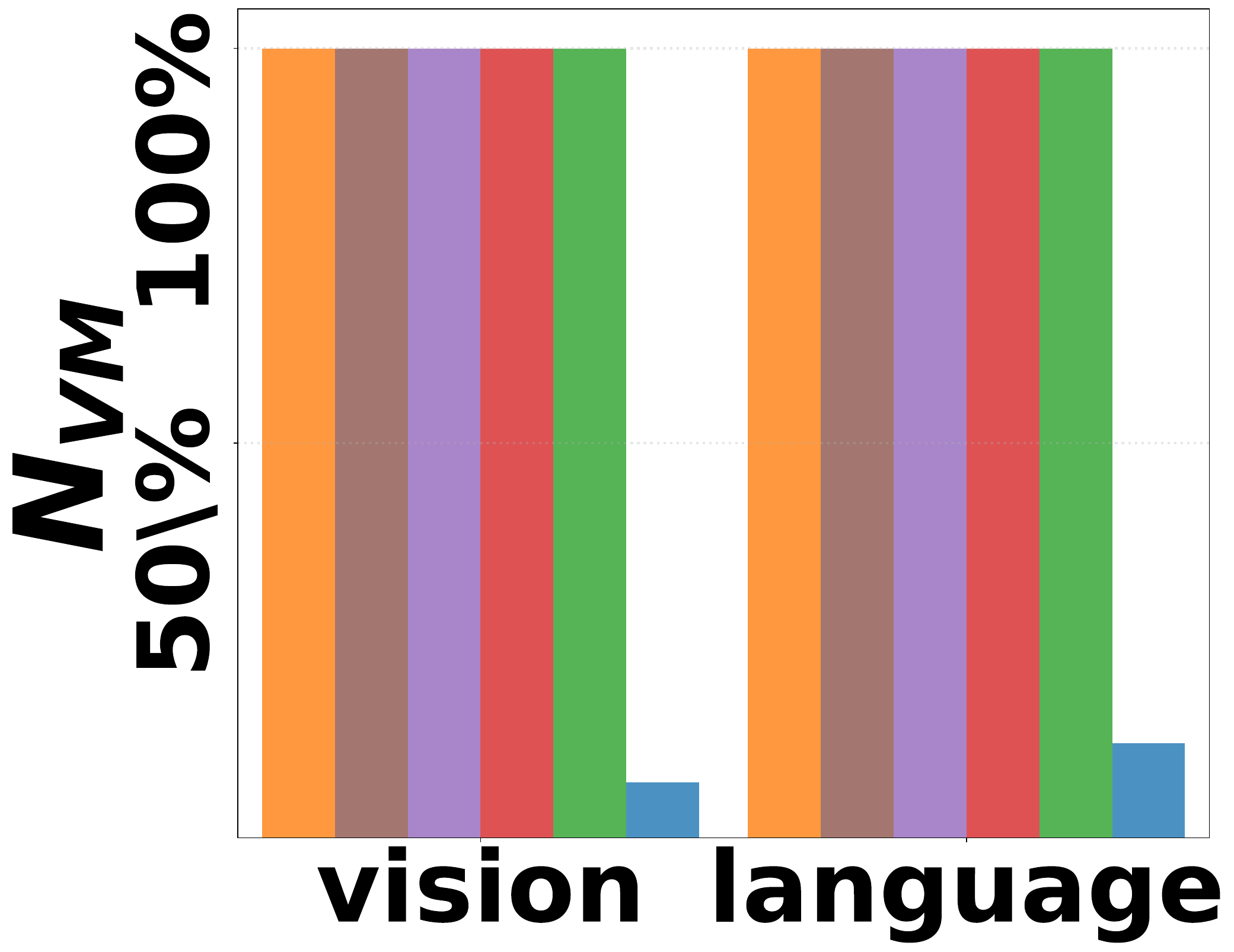}
        \includegraphics[width=0.322\columnwidth]{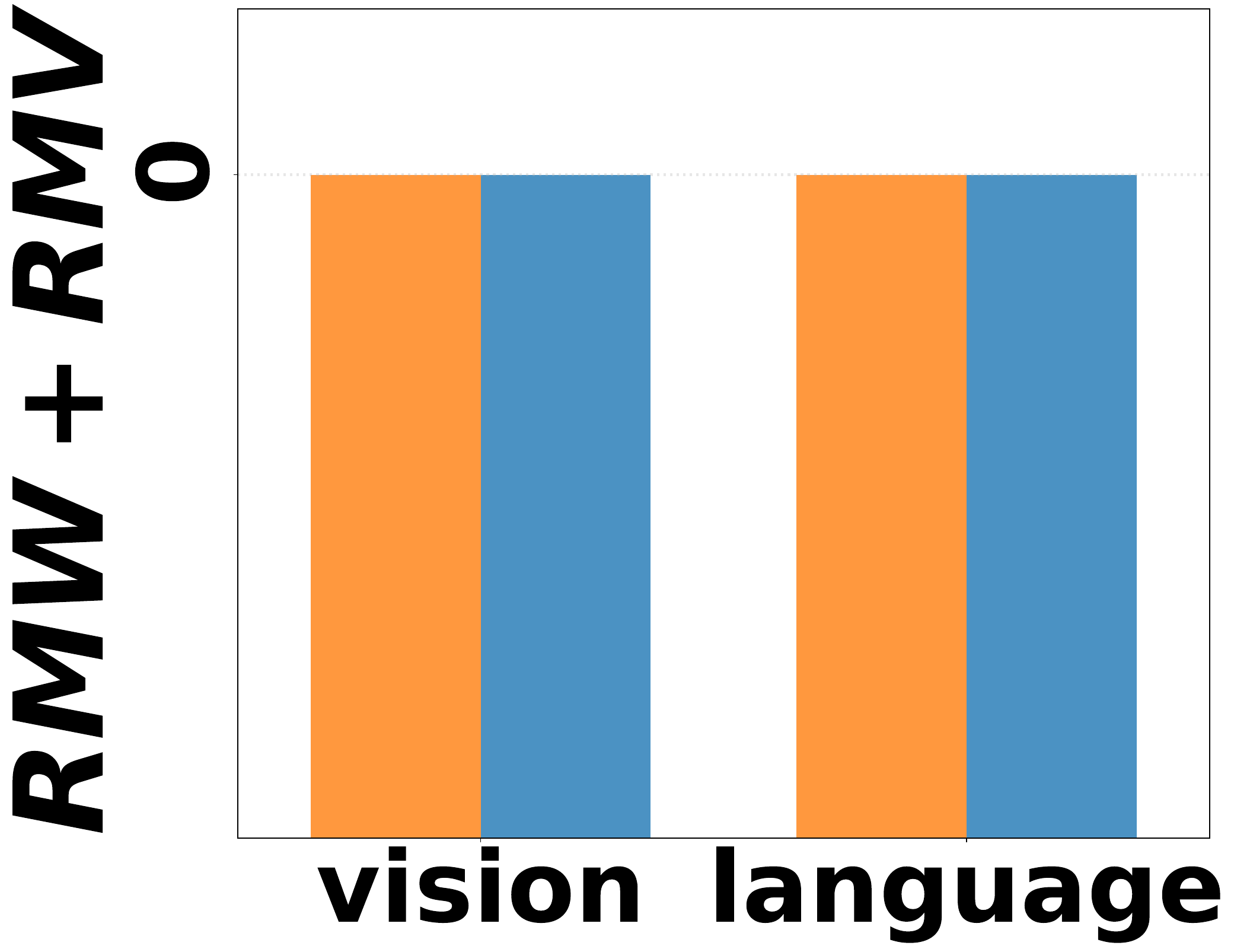}
    }
    \caption{Per-participant workload ($S_{LM}$, $S_{VM}$, $N_{VM}$) and security performance ($RMW+RMV$) under different baselines.}
    \label{fig:result_of_comparison2}
\end{figure}

Fig.~\ref{fig:result_of_comparison} shows how the performance of PMs customized by different baselines varies as participant costs accumulate. 
\texttt{DICE} achieves a superior performance frontier, attaining near-ideal ACC/F1 at significantly lower costs than BCFL and its PEFT-enhanced variants, while avoiding the apparent performance degradation observed in compression-enhanced variants on complex tasks (e.g., +Prune on CIFAR100, DDI, ChemProt; +KD on FedISIC, GAD). 

Fig.~\ref{fig:result_of_comparison2} zooms in on the results to examine how \texttt{DICE} addresses the limitation of baselines. 
The first sub-figure shows that, via DPL and MSV, \texttt{DICE} drastically reduces the size of the model that each worker or validator loads for producing or verifying an update. 
The second sub-figure shows that, by TCC, \texttt{DICE} significantly reduces the number of validation models that an honest validator must load to satisfy the global consensus conditions. 
The complementary role of EDI in preserving these efficiency gains is evaluated in the ablation study in \S\ref{ssec:ablation_study}. 
Together, the above two sub-figures explain \texttt{DICE}'s slower rate of cost accumulation on constrained participants observed in Fig.~\ref{fig:result_of_comparison}. 
Finally, the last sub-figure shows that \texttt{DICE} achieves full reward protection, on par with the BCFL paradigm. 
This indicates that \texttt{DICE}'s near-ideal model performance and efficiency gains observed in Fig.~\ref{fig:result_of_comparison} do not come at the cost of security. 

Collectively, results in Fig.~\ref{fig:result_of_comparison}--\ref{fig:result_of_comparison2} not only further corroborate the findings in \S\ref{ssec:exp_efficiency}, but also confirm that \texttt{DICE} establishes a practical paradigm that simultaneously controls participant expenses, lowers participation barriers, maintains model utility and robust security for customization tasks in real-world Web3 applications.

\subsection{Flexibility Performance} \label{ssec:exp_flexibility}
Previous experiments show that existing efficient customization techniques, despite having limitations, still offer practical value. For example, PEFT reduces the number of parameters to be trained, while compression achieves competitive accuracy with smaller local/validation models on certain tasks.
Therefore, we conduct additional experiments to study \texttt{DICE}'s compatibility with these techniques: (i) using PEFT techniques (LoRA, BitFit) together with DPL, and (ii) replacing DPL with compression techniques (Prune, KD). \texttt{DICE}'s validator-side designs remain unchanged.

Table~\ref{tab:result_of_integration_with_PEFT} shows the results of this experiment and reveals three key findings. 
First, PEFT techniques can further reduce backward computation overhead on workers by reducing the update size $|\theta_w^r|$, but they do not reduce the local or validation model sizes $S_{LM}$ and $S_{VM}$, and thus do not further lower the model-deployment requirements already achieved by DPL and MSV. 
Second, compression techniques provide alternative paths to simultaneously reduce $S_{LM}$, $|\theta_w^r|$, and $S_{VM}$. An extreme reduction is achieved on vision tasks by KD, which uses a much smaller pretrained model, DeiT-Tiny, as a substitute for the requested ViT. 
Finally, the efficiency gains from these techniques come with trade-offs in model performance, which vary across tasks and methods. 


\begin{table}[t] 
    \centering \small
    \caption{Results of \texttt{DICE} integrated with existing efficient customization techniques on different tasks. 
    }
    \label{tab:result_of_integration_with_PEFT}
    \resizebox{\columnwidth}{!}{
    \begin{tabular}{l|c|c|c|c|ccc}
    \toprule
         \multirow{2}*{\textbf{Methods}}  & \multirow{2}*{\textbf{$S_{LM} \downarrow$}}  &  \multirow{2}*{$|\theta_w^r| \downarrow$ } &  \multirow{2}*{\textbf{$S_{VM} \downarrow$}}  & \multirow{2}*{\textbf{$N_{VM} \downarrow$}}  &\multicolumn{3}{c}{\textbf{ACC/F1 $\uparrow$}} \\
         \cline{6-8}
         & & & & &  \footnotesize{\textbf{FedISIC }} & \footnotesize{\textbf{CIFAR100}} & \footnotesize{\textbf{DermaMNIST}} \\
         
    \hline
         \texttt{DICE}  & 50.4\%  & 41.6\%  & 50.4\% & \multirow{5}*{3.5\%}  & 98.3\% & 97.9\% & 99.1\% \\
         \, +LoRA & 51.0\% & 0.6\%   & 50.4\% & & 94.0\% & 94.3\% & 94.6\% \\
         \, +BitFit & 50.4\% & 0.1\%   & 50.4\%  & & 89.1\% & 89.2\% & 89.7\% \\
         \, +Prune & 50.4\% & 50.0\%  & 50.4\% & & 96.1\% & 90.9\% & 98.6\% \\
         \, +KD & 3.3\% & 3.1\% & 3.3\% & & 91.9\% & 80.1\% & 91.2\% \\
    
    \hline\hline
        & & & & & \footnotesize{\textbf{n2c2}} & \footnotesize{\textbf{JNLPBA}}  & \footnotesize{\textbf{BC5-d}}\\
        \hline
         \texttt{DICE} & 60.9\% & 41.7\% & \multirow{5}*{60.9\%} & \multirow{5}*{6.0\%} & 97.9\% & 98.7\% & 98.1\% \\
         \, +LoRA & 61.4\% & 0.6\% & & & 94.1\% & 94.6\% & 95.9\% \\
         \, +BitFit & 60.9\% & 0.1\% & & & 91.4\% & 91.7\% & 93.7\% \\
         \, +Prune & 60.9\% & 50.0\% & & & 97.3\% & 96.9\% & 97.6\% \\
         \, +KD & 60.9\% & 50.0\% & & & 97.1\% & 98.6\% & 98.7\% \\
    \hline\hline
        & & & & & \textbf{DDI } & \textbf{GAD} & \textbf{ChemProt} \\
        \hline
         \texttt{DICE} & 61.1\% & 42.1\% & \multirow{5}*{61.1\% } & \multirow{5}*{6.0\%} & 96.1\% & 98.6\% & 98.3\% \\
         \, +LoRA & 61.6\% & 0.6\% & & & 91.1\% & 94.7\% & 94.0\% \\
         \, +BitFit & 61.1\% & 0.1\% & & & 90.3\% & 91.8\% & 90.8\% \\
        \, +Prune & 61.1\% & 50.4\% & & & 92.4\% & 97.8\% & 79.4\% \\
         \, +KD & 61.1\% & 50.4\% & & & 92.0\% & 98.9\% & 97.1\% \\
    \bottomrule
    \end{tabular}
    }
\end{table}

Based on these findings, \texttt{DICE} offers two configuration options. 
The first option is to use its DPL and MSV schemes with PEFT, extreme compression, or other compatible customization techniques (see \S\ref{sec:related_work}) to further reduce the overhead of producing and verifying an update, at the cost of potential model utility degradation. The second option is to use the default DPL and MSV to balance efficiency and model performance. 
Both options work alongside our core innovations, TCC and EDI, which enable BCFL systems to scale to more workers while maintaining security. 
Together with the adjustable parameter $N_t$ in TCC, these options enable \texttt{DICE} to accommodate diverse device constraints and model performance requirements in practice.


\subsection{Customizing DINOv2 and Qwen2.5} \label{ssec:otherPMs_exp}

This section further evaluates \texttt{DICE}'s generalizability by customizing larger-scale pretrained models than those used in the main experiments, including DINOv2~\cite{oquab2023dinov2} for vision tasks and Qwen2.5~\cite{qwen2.5} for language tasks. 
We also report the results of the current BCFL paradigm for reference. 

\begin{figure}[t]
    \centering
    \includegraphics[width=0.5\columnwidth]{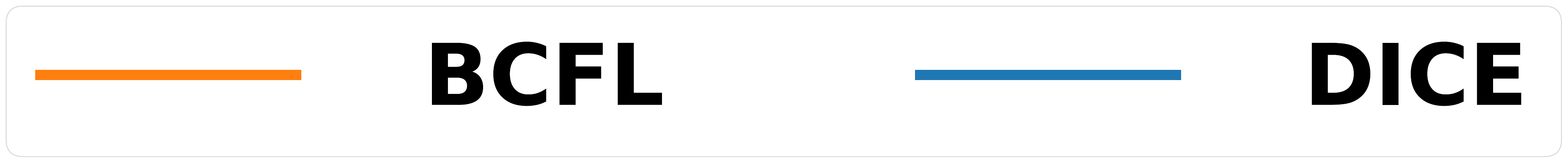}
    \\
    \includegraphics[width=0.342\columnwidth]{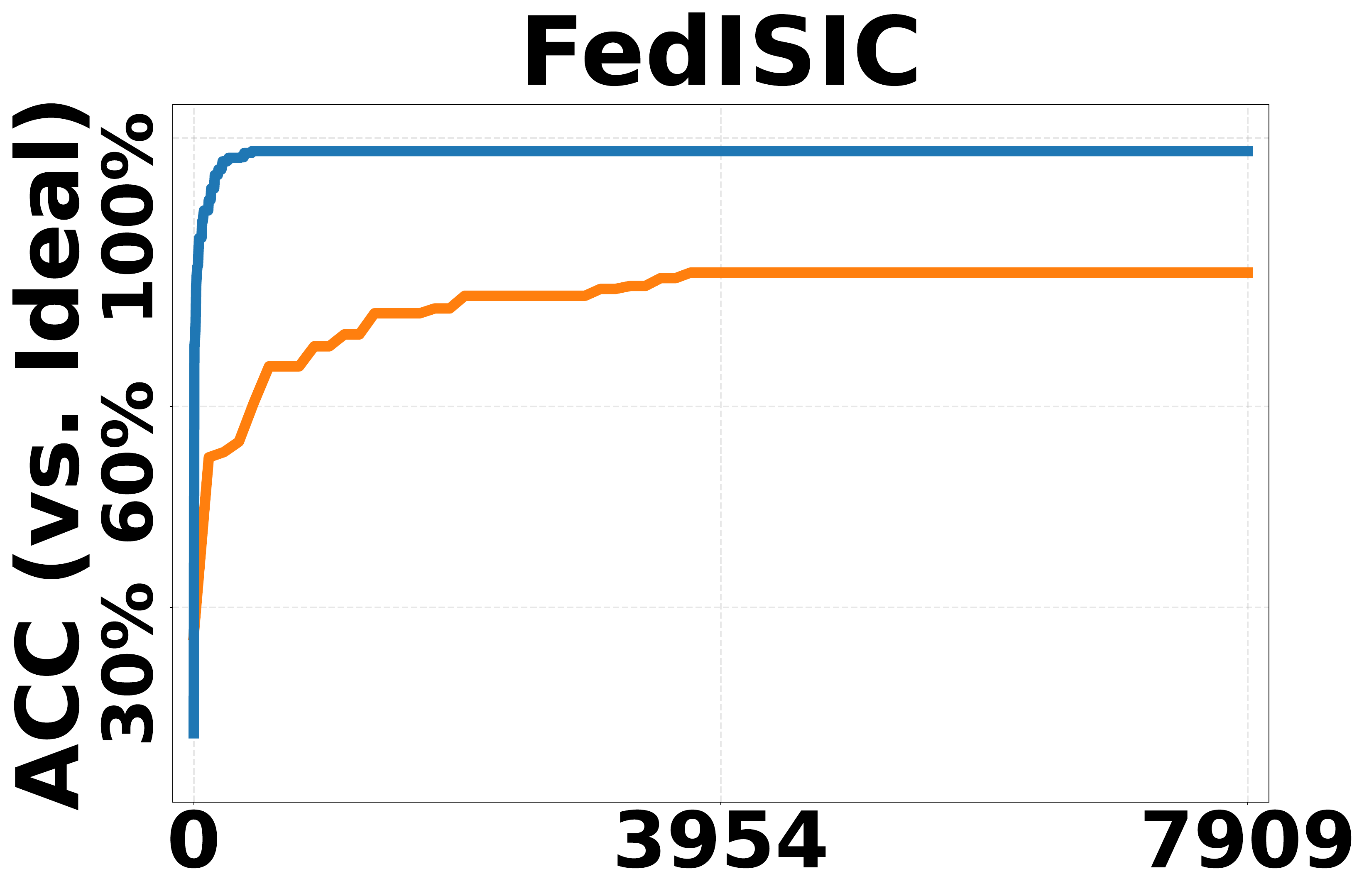}
    \includegraphics[width=0.32\columnwidth]{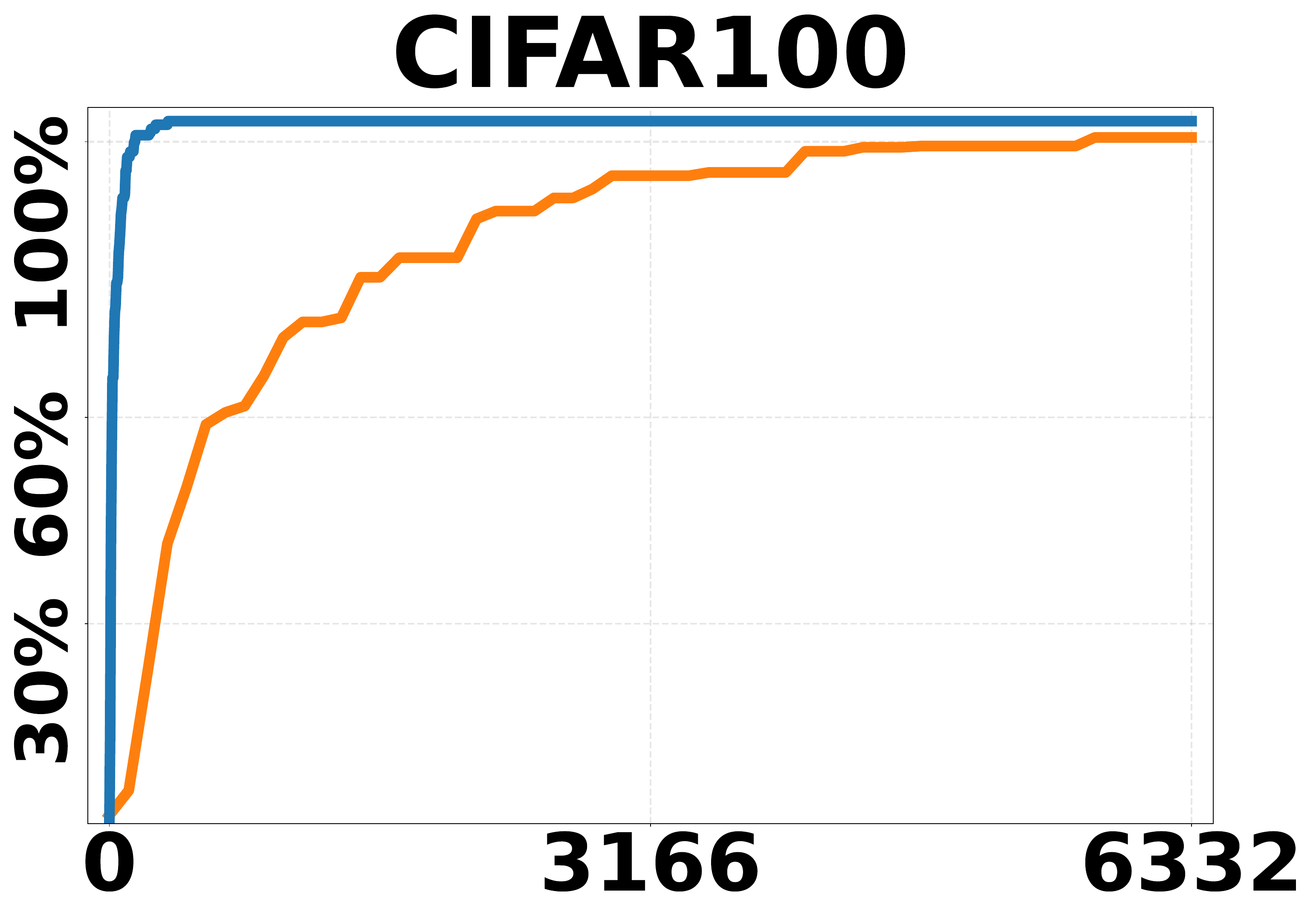}
    \includegraphics[width=0.32\columnwidth]{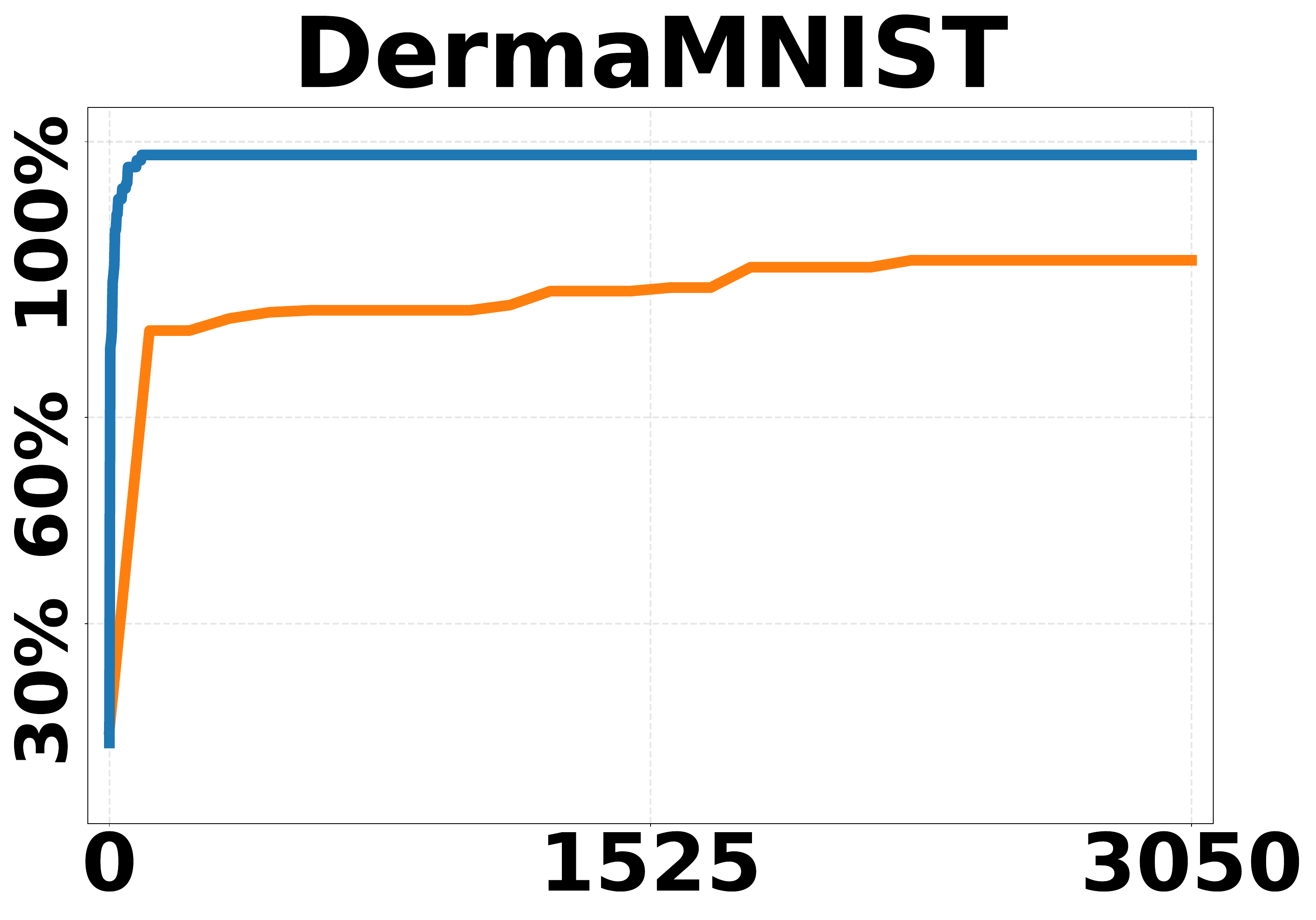}
    \\
    \includegraphics[width=0.342\columnwidth]{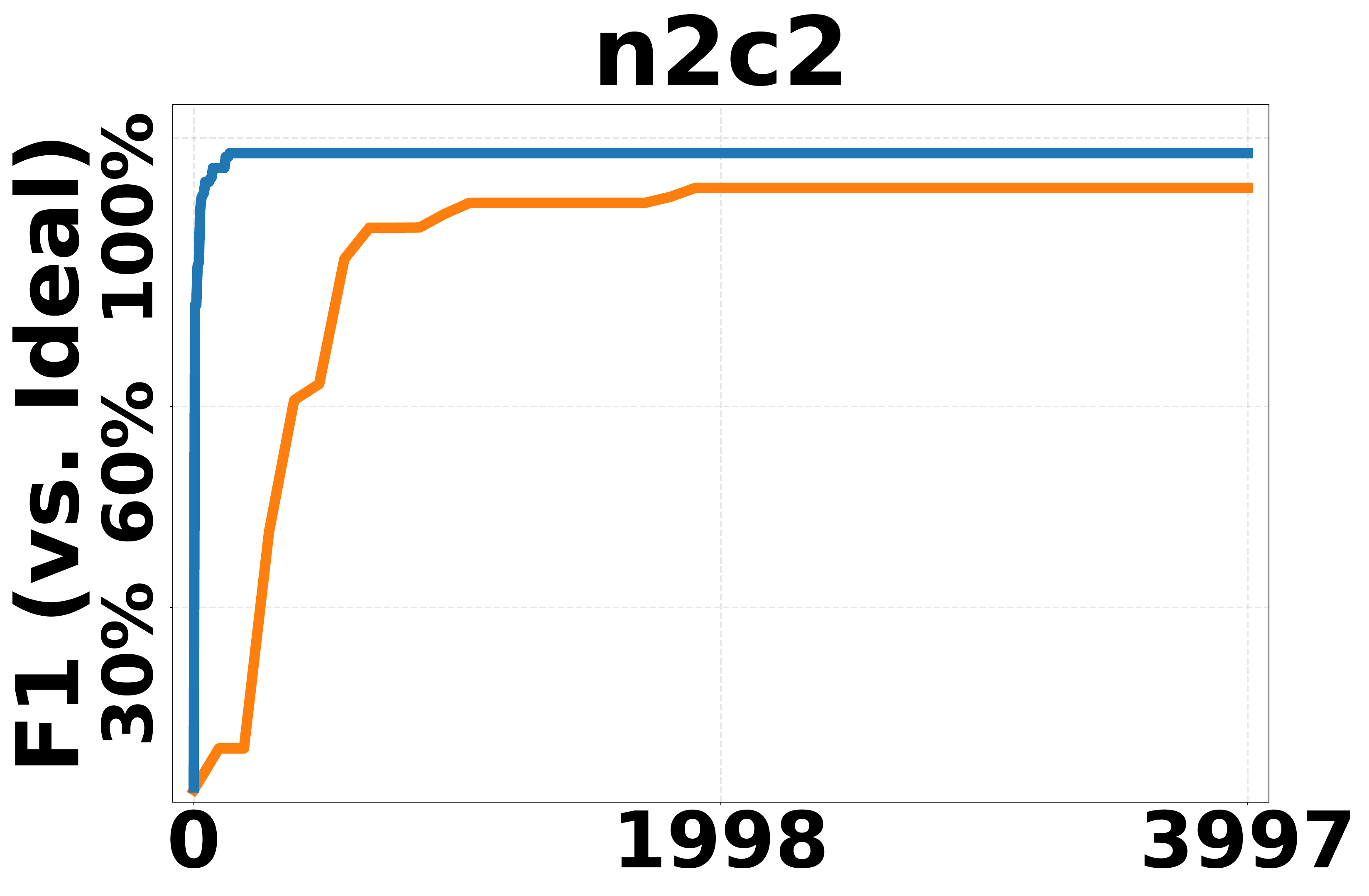}
    \includegraphics[width=0.32\columnwidth]{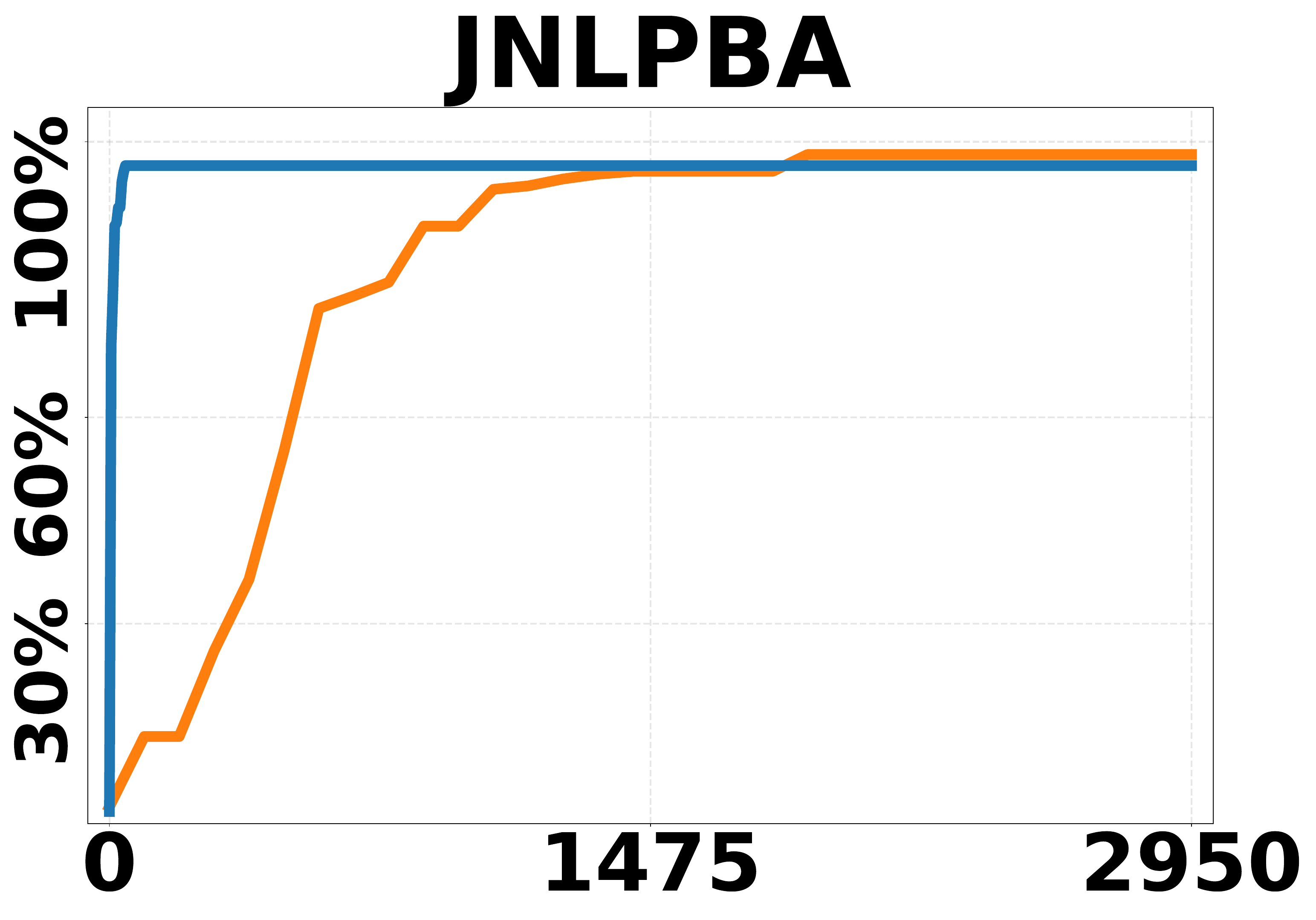}
    \includegraphics[width=0.32\columnwidth]{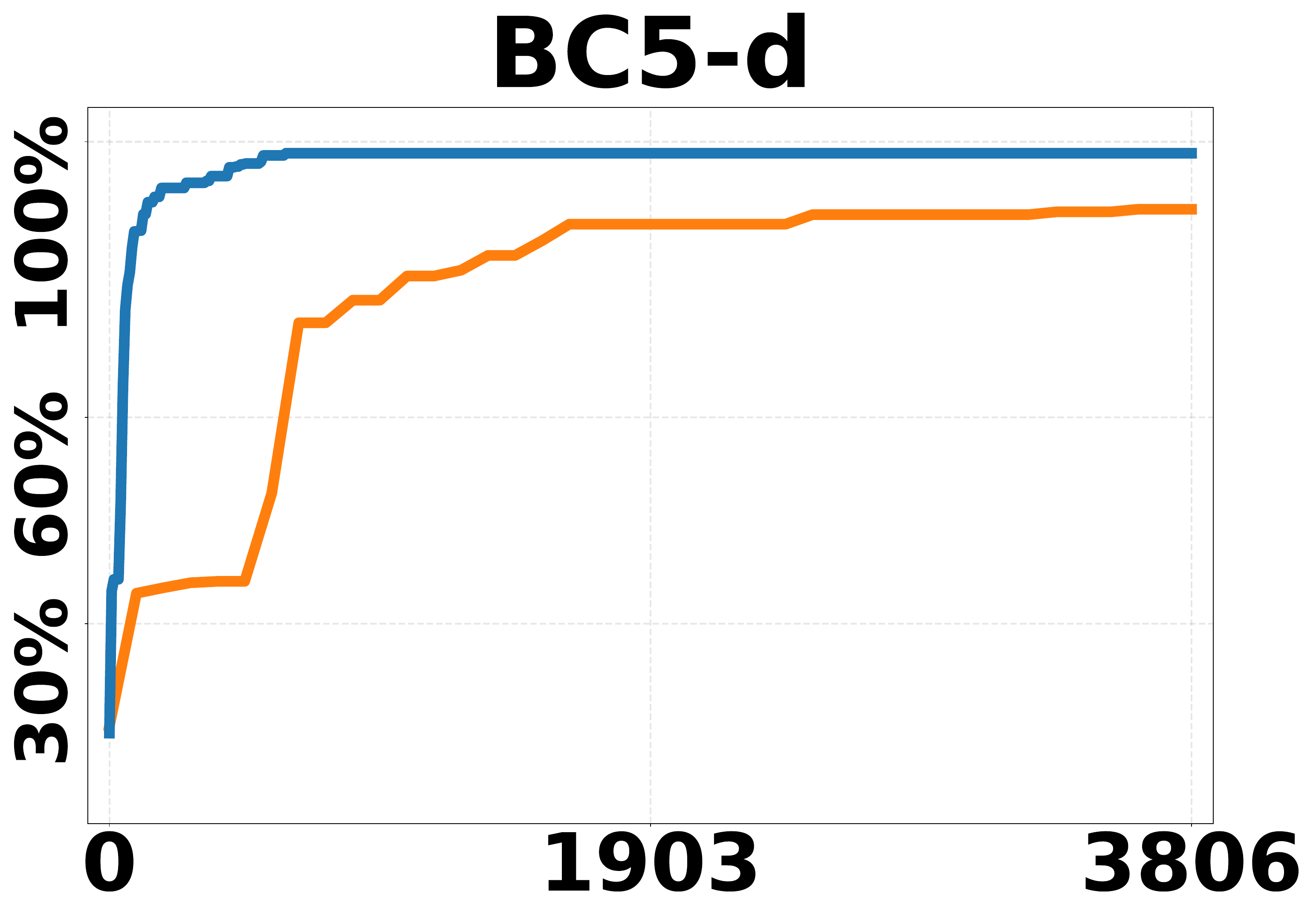}
    \\
    \includegraphics[width=0.342\columnwidth]{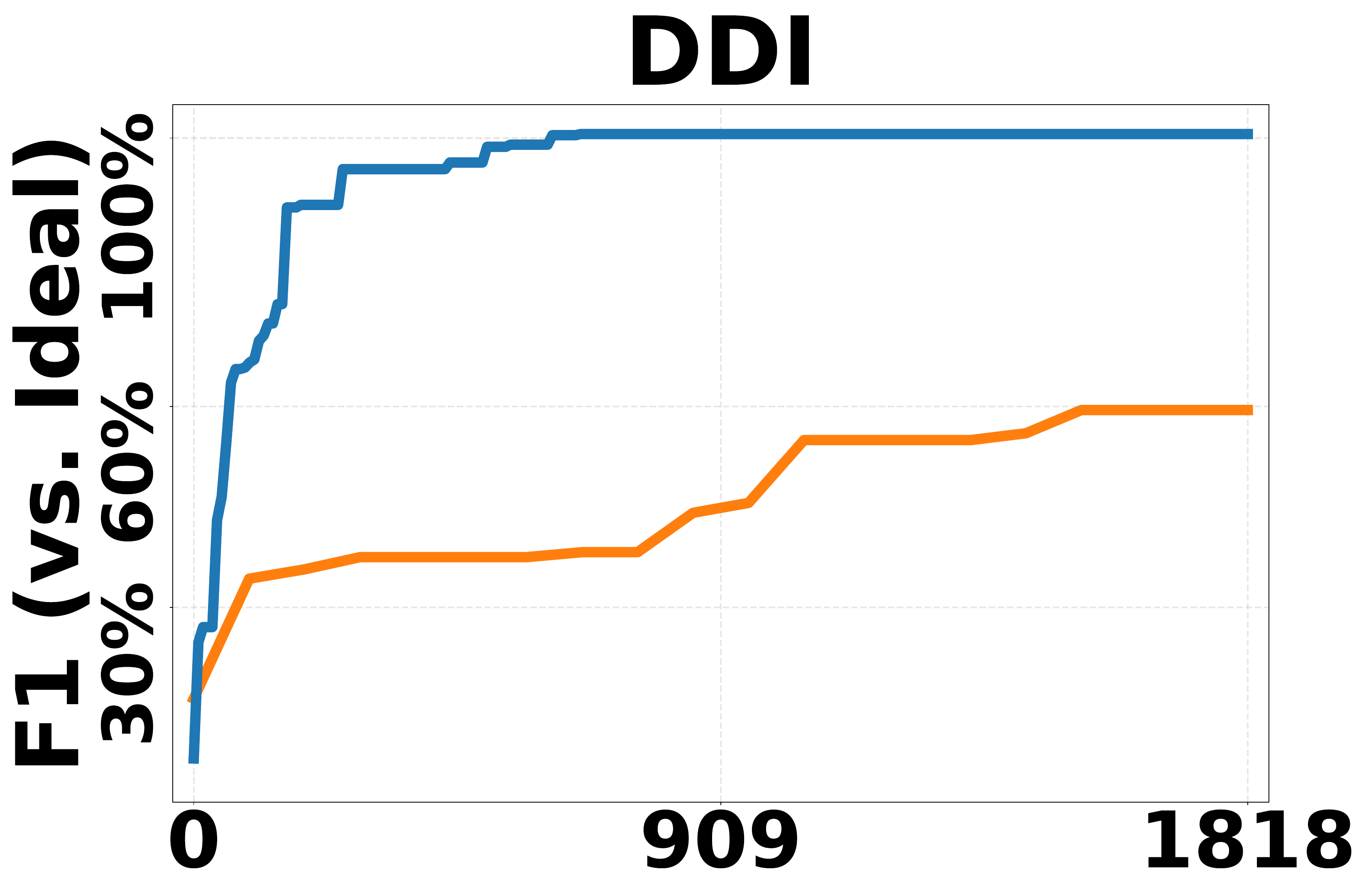}
    \includegraphics[width=0.32\columnwidth]{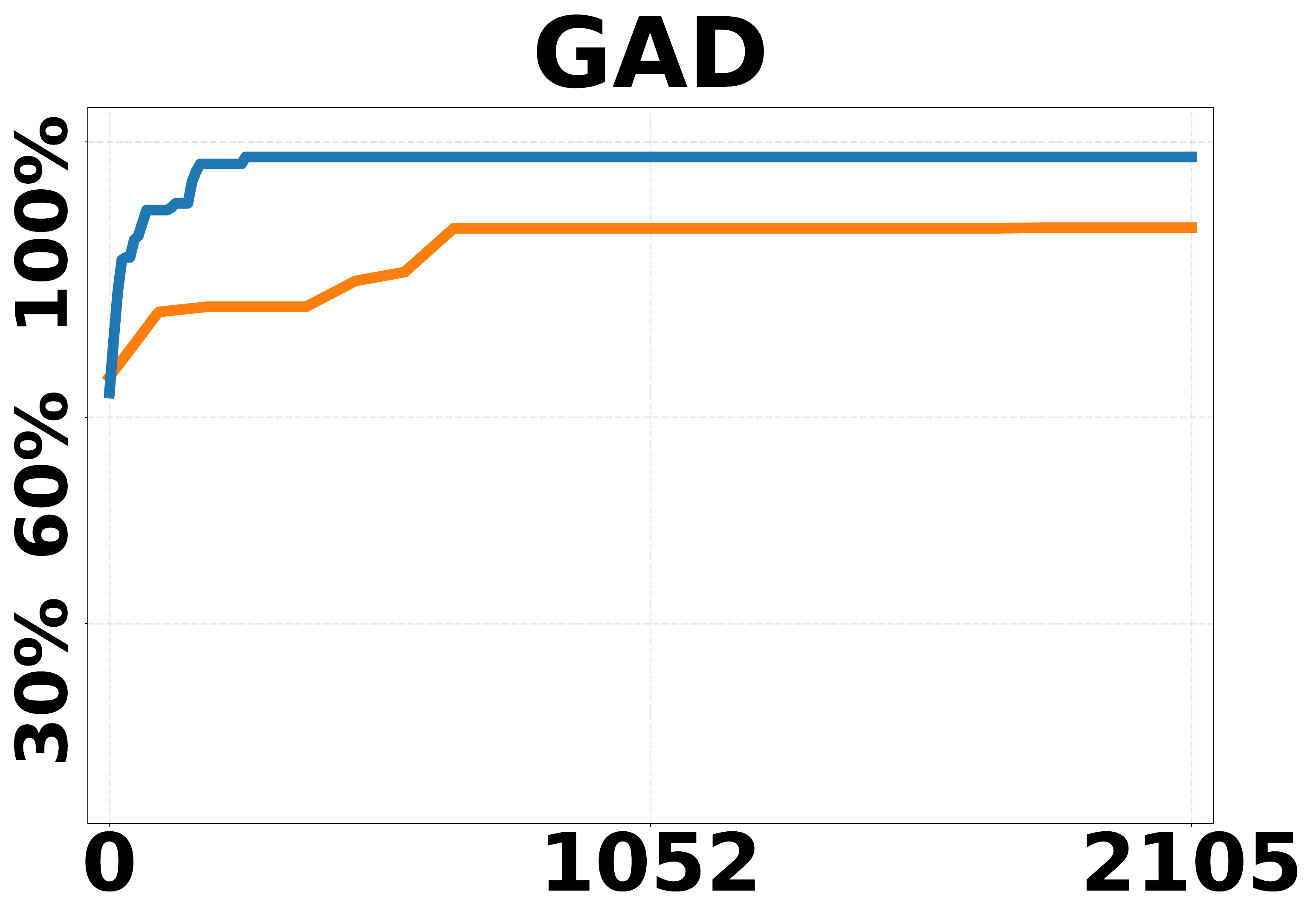}
    \includegraphics[width=0.32\columnwidth]{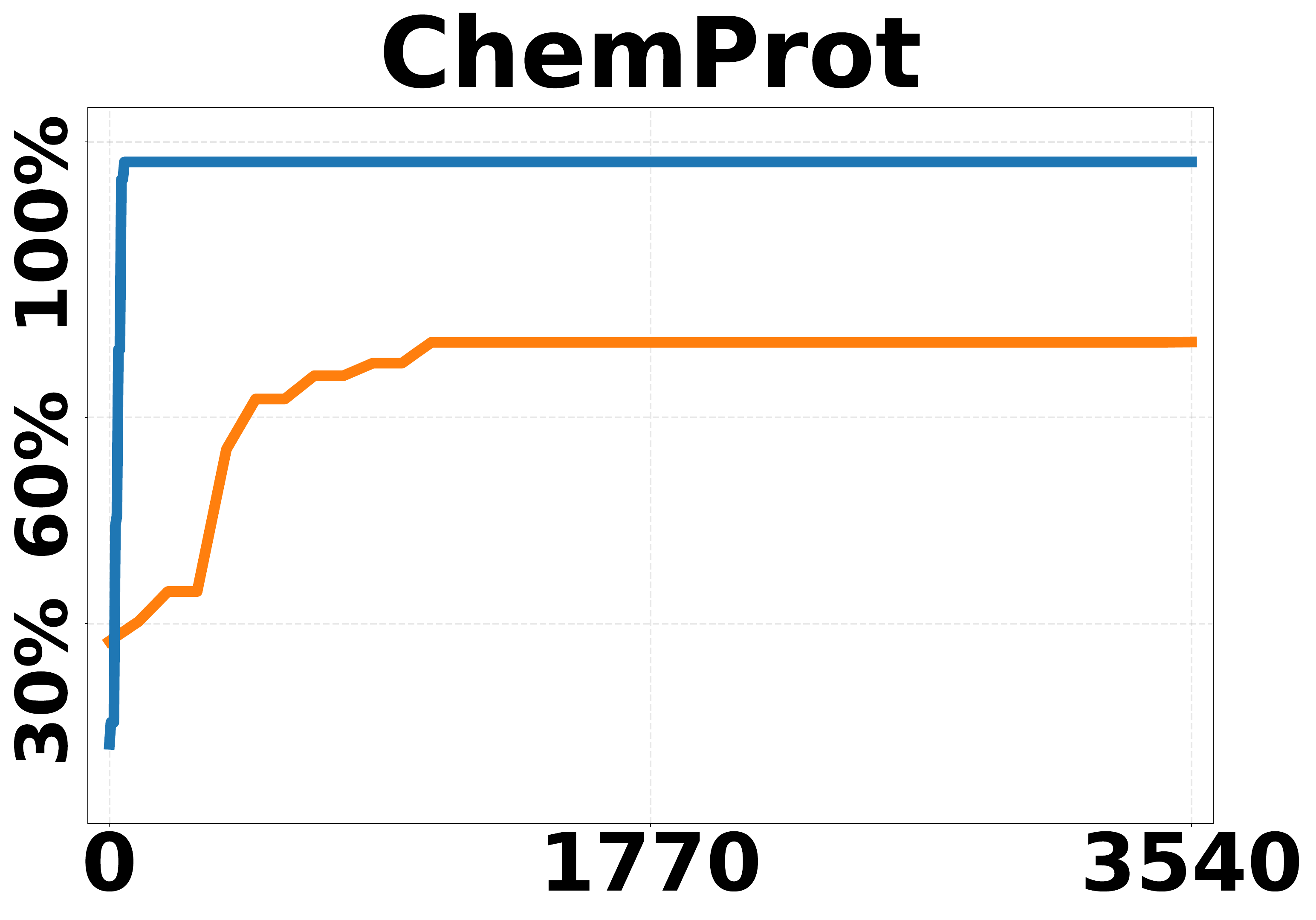}
    \caption{Performance of DINOv2 and Qwen2.5 customized by different baselines. 
    X-axis follows Fig.~\ref{fig:result_of_comparison}.
    }
    \label{fig:result_of_comparison3}
\end{figure}

\begin{figure}[t]
    \centering
    \mbox{
        \includegraphics[width=0.322\columnwidth]{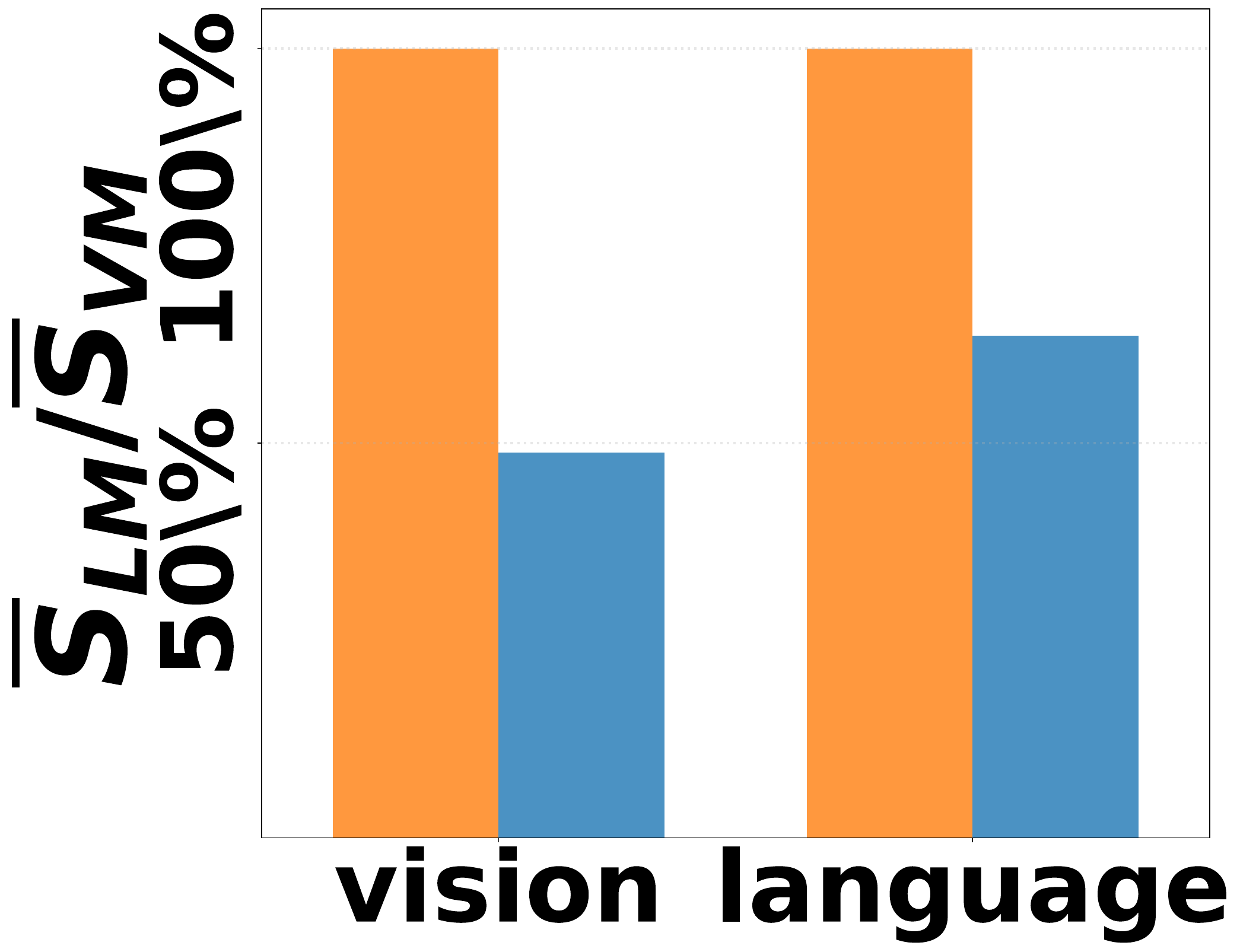}
        \includegraphics[width=0.322\columnwidth]{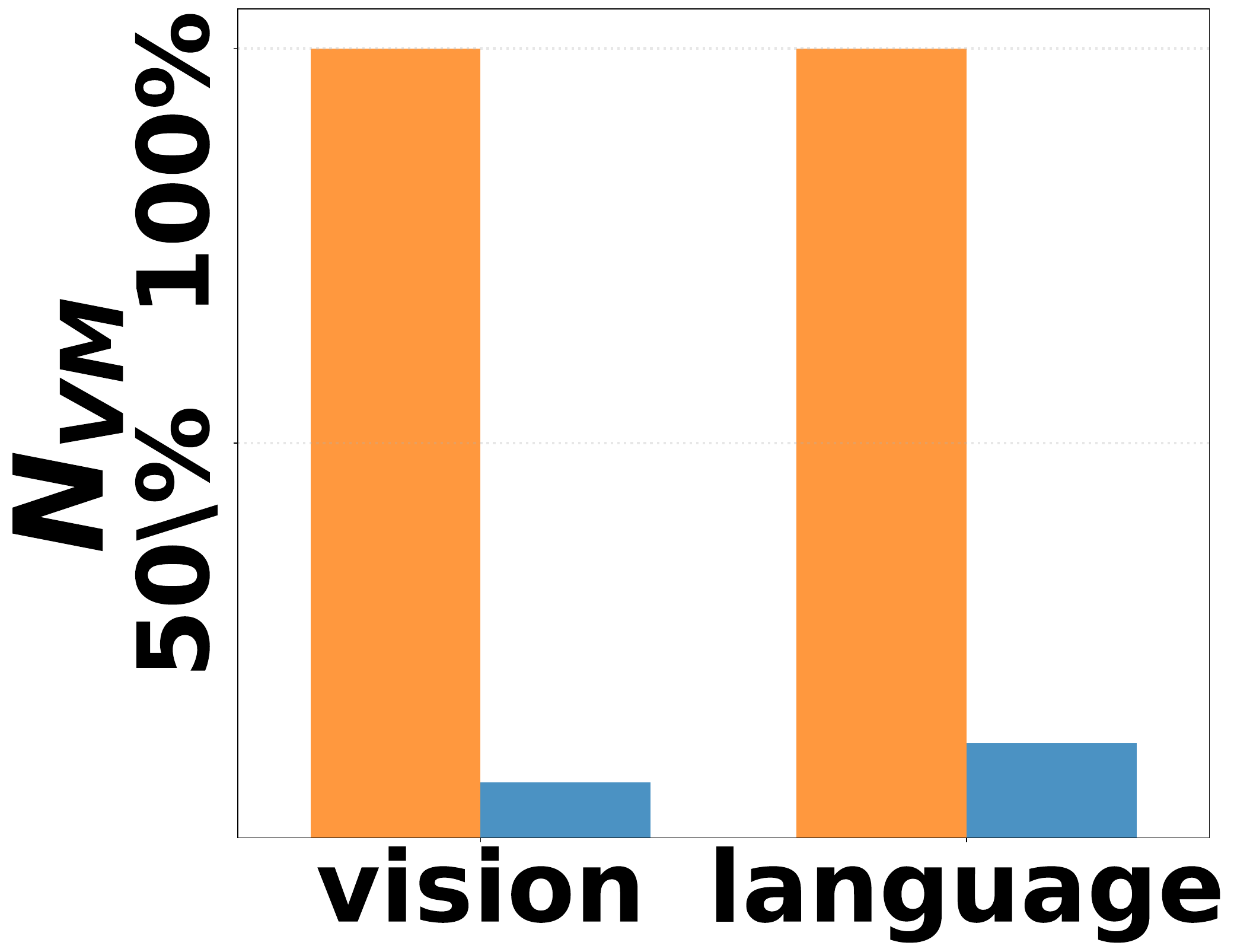}
        \includegraphics[width=0.322\columnwidth]{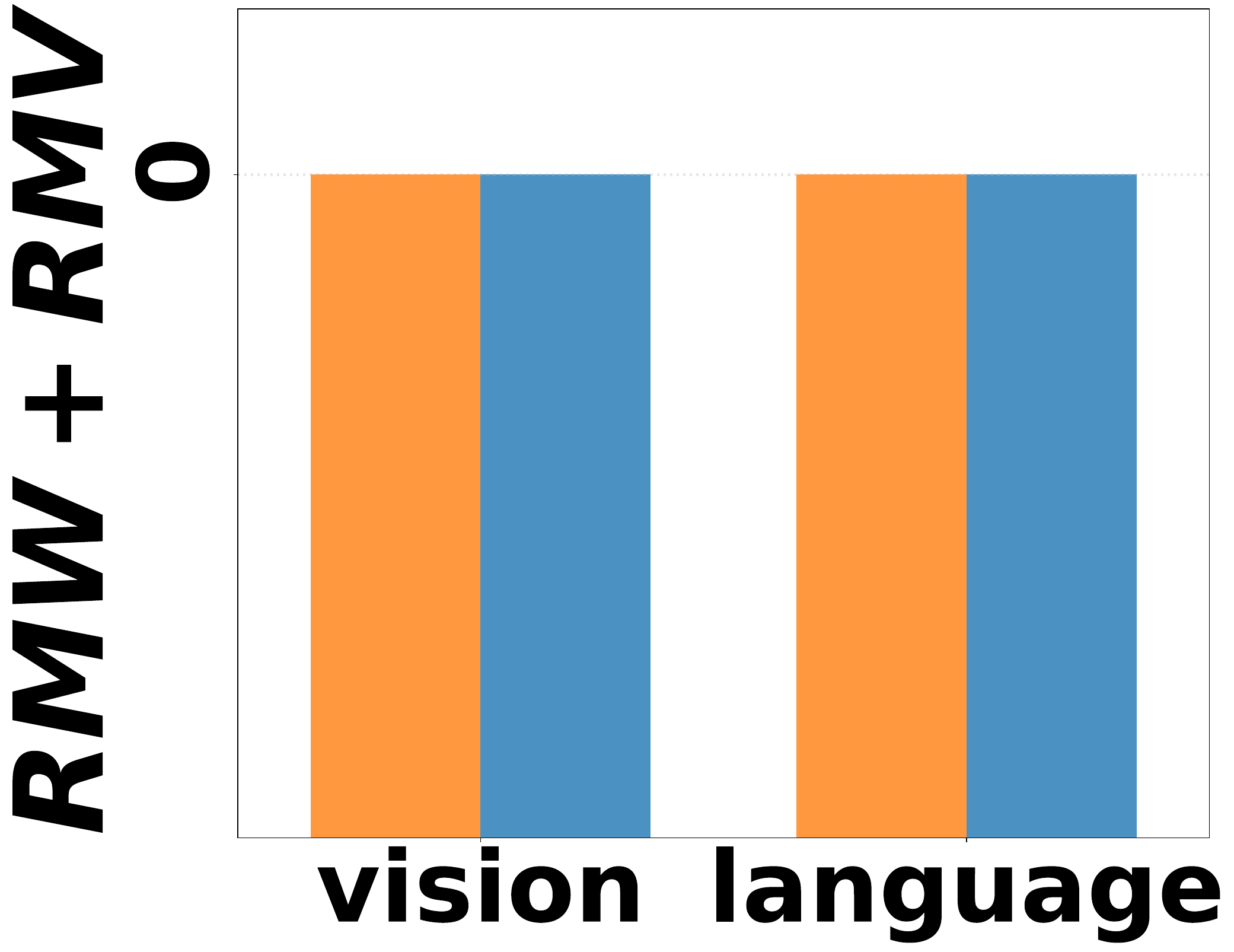}
    }
    \caption{Per-participant workload ($S_{LM}/S_{VM}$, $N_{VM}$) and  security performance ($RMW+RMV$) under different baselines when customizing DINOv2 and Qwen2.5.}
    \label{fig:result_of_comparison4}
\end{figure}

Figs.~\ref{fig:result_of_comparison3}--\ref{fig:result_of_comparison4} present the results of this experiment. 
The results consistently reproduce key findings from \S~\ref{ssec:exp_compare}. 
\texttt{DICE} achieves a superior performance frontier compared to the current BCFL paradigm, attaining higher accuracy/F1 at significantly lower participant workloads across different tasks. 
Additionally, it continues to effectively reduce the model size loaded by workers and validators, decrease per-validator verification workload, and maintain full reward protection. 
These results demonstrate that \texttt{DICE}'s efficiency and security benefits extend across different model families, supporting its applicability to diverse PM customization tasks in Web3 environments.


\subsection{Ablation Study}\label{ssec:ablation_study}
This section conducts five ablation experiments to demonstrate the contribution of DPL and MSV schemes for effective lightweight training and verification and the necessity of using TCC with EDI for efficient consensus. All experiments use FedISIC (vision), n2c2 (NER), and DDI (RE) as representative tasks.

\subsubsection{Effect of DPL's Slice Selection.} 
We first compare DPL's slice selection method against the following alternative methods for determining which H-layers to train in each round: 
(i)~three static methods, namely \textit{T-Only} (training only the top layers), \textit{B-Only} (training only the bottom layers), and \textit{OT} (training both top and bottom layers simultaneously)~\cite{xiao2023offsite}; 
(ii)~four dynamic methods, namely \textit{DyTB} (progressively training from top to bottom layers), \textit{DyBT} (from bottom to top layers), and \textit{DyOT} (from both ends toward the middle layers)~\cite{xiao2023offsite}, and \textit{Random} (randomly selecting a slice to train in each round). 

Table~\ref{tab:ablation_slice_selection} reports the final model performance under different slice selection methods. 
DPL's method consistently outperforms all alternatives across the three tasks. 
In contrast, the alternatives exhibit unstable performance across tasks. 
The static methods (T-Only, B-Only, OT) update only a fixed subset of layers throughout the entire model customization process, failing to explore other potentially informative parts of the model. 
The dynamic methods (DyTB, DyBT, DyOT) progressively cover more layers over time, yet they follow fixed heuristic traversal orders that do not adapt to the actual data distribution and training progress of model customization tasks. 
The random method, while able to explore diverse slices across rounds, lacks any principled guidance to prioritize slices with higher utility. 
DPL's superiority stems from its slice importance scoring, which considers both performance improvement and parameter divergence, enabling the requester to dynamically adjust the training pattern in the actual model customization process. 

\begin{table}[t]
    \centering\small
    \caption{Effect of DPL's importance-score-based slice selection on model performance. }
    \label{tab:ablation_slice_selection}
    \begin{tabular}{l|ccc}
    \toprule
         \textbf{Methods} & \textbf{FedISIC (ACC) $\uparrow$} & \textbf{n2c2 (F1) $\uparrow$} & \textbf{DDI (F1) $\uparrow$} \\
    \hline
         T-Only & 73.8\% & 97.2\% & 81.6\% \\
         B-Only & 95.9\% & 96.1\% & 86.9\% \\
         OT & 87.1\% & 96.9\% & 76.2\% \\
         DyTB & 91.7\% & 97.2\% & 93.1\% \\
         DyBT & 96.0\% & 96.1\% & 86.9\% \\
         DyOT & 91.3\% & 96.9\% & 88.5\% \\
         Random & 89.5\% & 95.8\% & 83.9\% \\
    \hline
         DPL & 98.3\% & 97.9\% & 96.1\% \\
    \bottomrule
    \end{tabular}
\end{table}

\subsubsection{Effect of Emulators.} We then study the effect of the emulators in approximating the input-output mappings of the global H-layers excluded from the compact model, thereby ensuring effective lightweight training and verification.  We vary $L_E$, the number of H-layers in an emulator, from 0 (no emulator) to 2. 

Table~\ref{tab:ablation_emulator} shows the resulting model performance. 
Without an emulator ($L_E=0$), performance drops significantly. In this setting, any update produced or verified on the compact model is derived from incomplete information. Integrating such an update into the full global model disrupts the original cross-layer semantic information flow of that model. 
Increasing $L_E$ from 1 to 2 yields limited benefits or even a slight degradation in the final model performance. 
Moreover, under fixed participant hardware budgets, larger emulators reduce the number of model layers that can be trained or verified in each round.
The above results confirm the effectiveness of emulators in approximating input-output mappings of the global H-layers excluded from the compact model, and demonstrate that our default choice $L_E=1$ balances approximation fidelity and participant-side resource efficiency. 
Exploring more compact and advanced emulator architectures or data-efficient distillation 
methods that reduce the requester's overhead represents a promising direction for future work.

\begin{table}[t]
    \centering\small
    \caption{Effect of emulators (consisting of $L_E$ H-layers) on model performance. }
    \label{tab:ablation_emulator}
    \begin{tabular}{l|ccc}
    \toprule
         \textbf{$L_E$} & \textbf{FedISIC (ACC) $\uparrow$} & \textbf{n2c2 (F1) $\uparrow$} & \textbf{DDI (F1) $\uparrow$} \\
    \hline
         0 & 55.3\% & 67.1\% & 63.8\% \\
         1 (Default) & 98.3\% & 97.9\% & 96.1\% \\
         2 & 98.4\% & 97.1\% & 96.3\% \\
    \bottomrule
    \end{tabular}
\end{table}

\subsubsection{Effect of MSV's Compact-model-based Verification.}
Next, we explore the effectiveness of using a compact validation model for verification in MSV by comparing it against full-model verification (Full-Verify), in which validators evaluate each worker's update using the full global model. 
Table~\ref{tab:ablation_msv_fidelity} reports the performance of the final model and the fraction of the rewards claimed by malicious workers. MSV's results are comparable to those of Full-Verify across all three tasks, indicating the effectiveness of using a compact validation model as a substitute for the full global model for verification. These results empirically justify MSV's design: inserting an emulator into the compact model captures sufficient information about the global H-layers excluded from that model and thus enables effective verification on that model, and using a dynamic threshold compensates for residual bias. Together, MSV enables hardware-constrained validators to participate while retaining model performance and reward protection comparable to Full-Verify.
 
\begin{table}[t]
    \centering\small
    \caption{Effect of MSV's compact-model verification on model performance and security against poisoning attacks from malicious workers.}
    \label{tab:ablation_msv_fidelity}
    \begin{tabular}{l|ccc|ccc}
    \toprule
         \multirow{2}*{\textbf{Methods}} & \multicolumn{3}{c|}{\textbf{ACC/F1 $\uparrow$}} & \multicolumn{3}{c}{\textbf{$RMW \downarrow$}} \\
         \cline{2-7}
         & \textbf{FedISIC} & \textbf{n2c2} & \textbf{DDI} & \textbf{FedISIC} & \textbf{n2c2} & \textbf{DDI} \\
    \hline
         Full-Verify & 98.5\% & 98.5\% & 96.6\% & 0.0\% & 0.0\% & 0.0\%  \\
         MSV & 98.3\% & 97.9\% & 96.1\% & 0.0\% & 0.0\% & 0.0\%  \\
    \bottomrule
    \end{tabular}
\end{table}

\subsubsection{Effect of MSV's Dynamic Validation Threshold.}
We further study the contribution of MSV's dynamic validation threshold by comparing it against a fixed-threshold variant. 
In this variant, the model requester sets $\epsilon_j$ to a constant value at the first round and keeps it unchanged throughout the entire customization process. 

Table~\ref{tab:ablation_threshold} reports the performance of the final model and the fraction of rewards claimed by malicious workers. 
With a fixed threshold, the model 
performance drops below 96\% of the ideal ACC/F1, and malicious workers claim about 10\% of rewards. 
The fixed threshold performs poorly because, as the global model improves over rounds, the fixed threshold becomes increasingly lax, making it easier for malicious workers to produce poisoned updates whose validation performance exceeds the threshold. 
In contrast, the dynamic threshold maintains ACC/F1 above 96\% of the ideal and full reward protection across tasks. 
These results demonstrate that a dynamic validation threshold is crucial for maintaining both model utility and system security, as it tracks the evolving performance of the global model and closes the window that attackers would otherwise exploit. 

\begin{table}[t]
    \centering\small
    \caption{Effect of dynamic validation threshold on model performance and security against poisoning attacks from malicious workers.}
    \label{tab:ablation_threshold}
    \begin{tabular}{l|ccc|ccc}
    \toprule
         \multirow{2}*{\textbf{Methods}} & \multicolumn{3}{c|}{\textbf{ACC/F1 $\uparrow$}} & \multicolumn{3}{c}{\textbf{$RMW \downarrow$}} \\
         \cline{2-7}
         & \textbf{FedISIC} & \textbf{n2c2} & \textbf{DDI} & \textbf{FedISIC} & \textbf{n2c2} & \textbf{DDI} \\
    \hline
         Fixed & 92.7\% & 95.4\% & 92.5\% & 11.2\% & 8.7\% & 9.6\% \\
         Dynamic & 98.3\% & 97.9\% & 96.1\% & 0.0\% & 0.0\% & 0.0\%  \\
    \bottomrule
    \end{tabular}
\end{table}

\subsubsection{Necessity of integrating TCC with EDI.} 
Finally, we examine the role of EDI, particularly its penalty design, in preventing conflict generation and thereby ensuring the efficiency of TCC. 
Unlike the default experimental settings, this experiment enables EDI's complete reward and penalty mechanism and compares it (TCC+EDI) with a variant (TCC-only) that disables penalties under environments with and without malicious validators. 
To highlight the effect of EDI, we configure all malicious validators in this experiment to trigger conflicts via poisoning (i.e., submitting VA results that deviate from the consensus)  when doing so yields positive net gains, as fraud attacks (submitting copied results) do not generate conflicts.

\begin{table}[t]
    \centering\small
    \caption{Effect of EDI on ensuring TCC's efficiency. The reported results are the average proportion of sub-tasks executed by each honest validator out of all VA sub-tasks in each round.}
    \label{tab:ablation_edi}
    \begin{tabular}{l|ccc|ccc}
    \toprule
         \multirow{2}*{\textbf{Methods}} & \multicolumn{3}{c|}{\textbf{All Honest} $\overline{N}_{v,t}/N_{VA} \downarrow$} & \multicolumn{3}{c}{\textbf{20\% Malicious} $\overline{N}_{v,t}/N_{VA} \downarrow$} \\
         \cline{2-7}
         & FedISIC & n2c2 & DDI & FedISIC & n2c2 & DDI \\
    \hline
         TCC-only &  6.3\% & 10.7\% & 10.5\%  & 90.5\% & 88.9\% & 89.9\%  \\
         TCC+EDI & 6.1\% & 10.5\% & 10.6\% & 6.1\% & 10.6\% & 10.5\% \\
    \bottomrule
    \end{tabular}
\end{table}

Table~\ref{tab:ablation_edi} reports the average proportion of VA sub-tasks executed by each honest validator out of all sub-tasks in each round.
When all validators are honest, TCC-only and TCC+EDI perform similarly, as no sub-task becomes controversial in this case and thus no extra mandatory workload is incurred for each honest validator. However, when 20\% of the validators are malicious, the performance of the two variants diverges sharply. 
Under TCC-only, the average proportion of VA sub-tasks executed by each honest validator increases by more than 75 percentage points across the three tasks. 
This is because malicious validators deliberately submit conflicting results to obtain extra time in completing more sub-tasks for more rewards, thereby increasing committee voting procedures for controversial sub-tasks and the prerequisite computation on their descendants.  
In contrast, TCC+EDI performs similarly under both non-adversarial and adversarial environments, with the slight difference attributable to experimental variance. 
With EDI, the penalty for submitting invalid results exceeds the maximum potential reward obtainable from triggering conflicts. Consequently, malicious validators have no incentive to provoke controversies, and no additional sub-tasks require committee voting beyond the root sub-tasks. 
The above results indicate that EDI is important for maintaining TCC's efficiency gains in adversarial environments. 

\section{Related Work} \label{sec:related_work}
The work related to \texttt{DICE} mainly falls into three categories: classical federated learning, blockchain-powered federated learning, and efficient customization techniques.

\textbf{Classical federated learning} aims to train a shared global model from data distributed across different workers while protecting data privacy. The first federated learning algorithm, FedAvg \cite{McMahan17}, and many of its follow-up works \cite{Sahu2018FederatedOF,Karimireddy2020SCAFFOLDSC,Wang2020Federated} proposed an iterative learning scheme that employs a central server to coordinate workers. Data privacy protection is achieved by transmitting only the model parameters, not raw data, between the server and workers. 
Classical federated learning algorithms are designed for scenarios where at least one participant, e.g., the server, is trusted. In contrast, our work is designed for Web3 environments where any participant is untrusted.

\textbf{Blockchain-powered federated learning} aims to enable untrusted Web3 participants to contribute to training models through blockchain techniques \cite{9293091,10.1145/3524104,9833437}. Considering the untrusted yet rational and self-interested characteristics of Web3 participants, many studies have proposed (i) well-designed malicious behavior handling schemes \cite{lin2024refiner,10.14778/3476311.3476313} that rely on a cross-node consensus algorithm to produce correct global models, and (ii) incentive schemes \cite{lin2024refiner,10.14778/3476311.3476313,10.1145/3578581,9369019} that incentivize self-interested entities to participate and behave honestly by providing monetary rewards (e.g., cryptocurrency) proportional to their contributions, rather than relying solely on non-monetary incentives such as reputation or service access \cite{10.1145/3578581}. 
The difference between existing BCFL studies and ours is that these prior studies are designed for small models with at most tens of millions of parameters, such as CNN \cite{lin2024refiner,manoj2025blockchain}, LSTM \cite{lin2024refiner,manoj2025blockchain}, ResNet \cite{bian2023blockchain,manoj2025blockchain}, 
while \texttt{DICE} is designed for large pretrained models having at least 100 million parameters, e.g., BERT \cite{DBLP:journals/corr/abs-1810-04805}, ViT \cite{wu2022tinyvit,dosovitskiy2020vit}, DINOv2 \cite{oquab2023dinov2}, Qwen \cite{qwen2.5,qwen2}, LLaMa \cite{touvron2023llamaopenefficientfoundation,zhang2024tinyllama}, CLIP \cite{Radford2021LearningTV}.  
\texttt{DICE} lowers the participant barrier and breaks the scalability bottleneck of the current BCFL paradigm, enabling the system to scale to more workers for better model performance and to more validators for more robust security. 

\textbf{Efficient customization techniques} aim to reduce the computational and memory overhead of further pretraining or fine-tuning PMs. These techniques typically follow two paradigms: (i) parameter-efficient fine-tuning \cite{zhang2023fedpetuning}, such as LoRA \cite{hu2022lora}, Adapter \cite{houlsby2019parameter}, Prompt Tuning \cite{lester-etal-2021-power}, and BitFit \cite{ben-zaken-etal-2022-bitfit}, which retain the pretrained backbone without reducing its size or parameter bit-width; (ii) model compression \cite{xu2025resource}, such as pruning \cite{lu2024reassessing,xiao2023offsite,hu2025trimllm}, knowledge distillation \cite{hinton2015distilling,sanh2019distilbert,wu2022tinyvit}, and quantization \cite{han2015deep,liang2021pruning}, which alter the model structure or reduce parameter bit-width.  
Many studies have integrated these techniques into federated learning to enable hardware-constrained participants to contribute \cite{fan2025ten}. 
However, most of them focus on centralized settings, where a powerful trusted server is adopted to verify or aggregate all worker updates  \cite{Wang2023FedBFPTAE,Tupper2025PersonalizedFF,xu2025dp}. 
In contrast, \texttt{DICE} is designed for decentralized BCFL systems \cite{lin2024refiner,desai2021blockfla,Liu2020,9284684,10.1109/TDSC.2019.2952332,9293091}, where verification and aggregation are performed by validators, each of whom is untrusted, potentially hardware-constrained, and required to redundantly execute the same VA tasks to ensure the validity of cross-validator consensus. 
For existing BCFL systems, the per-validator workload that grows linearly with worker scale becomes the dominant bottleneck hindering them from scaling to a larger number of data holders. Moreover, this cannot be simply resolved by integrating efficient model customization methods. 

\section{Conclusion} \label{sec:conclusion}
This paper presented \texttt{DICE}, a novel BCFL algorithm designed for efficient and scalable customization of pretrained models in Web3. 
We systematically analyzed the efficiency and scalability challenges that prevent existing BCFL systems from being directly applied to customize large PMs. 
To overcome these limitations, we introduced two sets of novel techniques, namely DPL with MSV to support lightweight training and verification of large PMs, and TCC with EDI to resolve the scalability bottleneck that prior BCFL systems cannot address even when enhanced by existing efficient customization methods.
We presented a rigorous theoretical analysis of \texttt{DICE}'s efficiency, scalability, and security. 
Extensive experiments on customizing vision and language PMs corroborated our analysis, showing that compared to the current BCFL paradigm, \texttt{DICE} significantly reduces costs for both workers and validators while maintaining competitive model performance and robust security. 
Moreover, \texttt{DICE} is compatible with existing efficient customization solutions, accommodating diverse device constraints and model performance requirements without compromising security. 

\begin{acks}
 This work was funded by the Fundamental and Interdisciplinary Disciplines Breakthrough Plan of the Ministry of Education of China (No. JYB2025XDXM103) and the Leading Talent of Technological Innovation Program (No. 2023R5214) of Zhejiang Province. We also acknowledge anonymous reviewers for their valuable feedback. 
\end{acks}

\bibliographystyle{ACM-Reference-Format}
\balance
\bibliography{sample-base}

\end{document}